\documentclass[10pt,twocolumn,letterpaper]{article}

\usepackage{iccv}              %

\usepackage[dvipsnames]{xcolor}
\usepackage{multirow}
\usepackage{soul}
\usepackage{tabularx}
\usepackage{adjustbox}
\usepackage{tabularx} 
\usepackage{algorithm}
\usepackage{algpseudocode}
\usepackage{graphicx}
\usepackage{caption}

\newbox\jsavebox
\newcommand{\jsubfig}[2]{%
	\sbox\jsavebox{#1}%
	\parbox[t]{\wd\jsavebox}{\centering\usebox\jsavebox\\#2}%
	}

\newcommand{\whitetxt}[1]{{\color{white}#1}\normalfont}
\newcommand{\ignorethis}[1]{}

\def\methodName{BlendedPC}

\def\inpaintingModel{Inpaint-E}

\newcommand{\norm}[1]{\left\Vert #1 \right\Vert_2}

\definecolor{iccvblue}{rgb}{0.21,0.49,0.74}
\usepackage[pagebackref,breaklinks,colorlinks,allcolors=iccvblue]{hyperref}

\def\paperID{5344} %
\def\confName{ICCV}
\def\confYear{2025}

\title{Blended Point Cloud Diffusion for Localized Text-guided Shape Editing}

\author{
Etai Sella$^{1*}$ \ \ \
Noam Atia$^{1*}$ \ \ \
Ron Mokady$^{2}$ \ \ \
Hadar Averbuch-Elor$^{3}$ 
\\[2mm]
\vspace{1em}
$^1$Tel Aviv University \ \ \
$^2$BRIA AI \ \ \
$^3$Cornell University
\\[0.5mm]
\vspace{-1mm}
\centering
\small{\url{https://tau-vailab.github.io/BlendedPC/}}  
}

\begin{document}
 \maketitle
\begin{abstract}
Natural language offers a highly intuitive interface for enabling localized fine-grained edits of 3D shapes. However, prior works face challenges in preserving global coherence while locally modifying the input 3D shape. In this work, we introduce an inpainting-based framework for editing shapes represented as point clouds. Our approach leverages foundation 3D diffusion models for achieving localized shape edits, adding structural guidance in the form of a partial conditional shape, ensuring that other regions correctly preserve the shape's identity. Furthermore, to encourage identity preservation also within the local edited region, we propose an inference-time coordinate blending algorithm which balances reconstruction of the full shape with inpainting at a progression of noise levels during the inference process. Our coordinate blending algorithm seamlessly blends the original shape with its edited version, enabling a fine-grained editing of 3D shapes, all while circumventing the need for computationally expensive and often inaccurate inversion.
Extensive experiments show that our method outperforms alternative techniques across a wide range of metrics that evaluate both fidelity to the original shape and also adherence to the textual description. %
\end{abstract}
  
\begin{figure} %
\centering
\jsubfig{\includegraphics[width=0.49\textwidth]{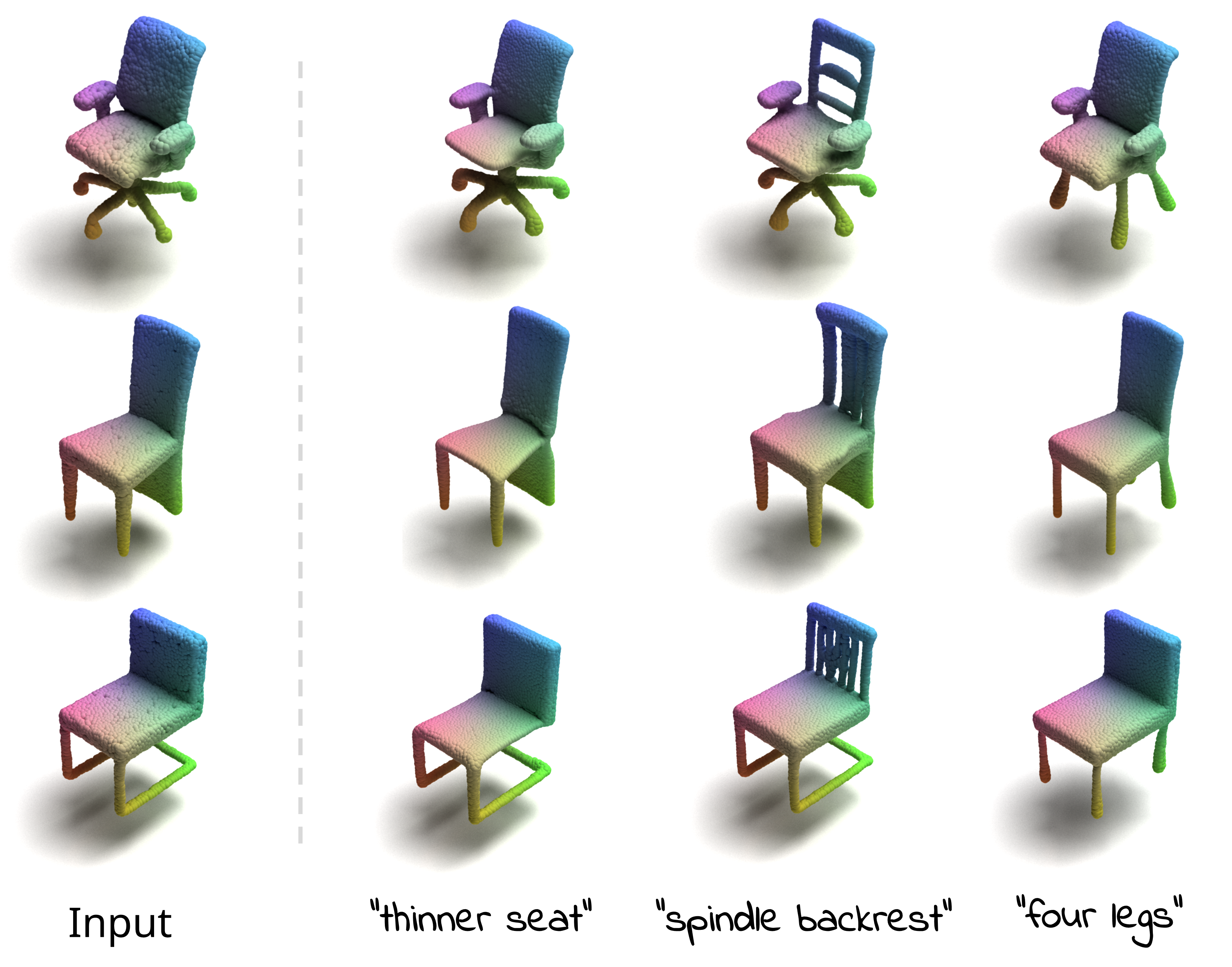}}{}
\vspace{-17pt}
\caption{
Given an input point cloud and a target text prompt, our localized editing technique produces an output point cloud that accurately conveys the text while faithfully preserving the input's structural identity, particularly outside of the desired edit region.
}
\label{fig:teaser}
\end{figure}

\def\thefootnote{*}\footnotetext{Denotes equal contribution}
\section{Introduction}
We are witnessing tremendous success in text-guided 3D generation---a task that seemed nearly impossible until recently, and nowadays enables the creation of high-fidelity 3D assets through the use of text~\cite{poole2022dreamfusion,lin2023magic3d}. %
This generative power has also fueled increasing interests in text-guided editing of existing 3D shapes~\cite{sella2023vox,chen2023fantasia3d,sella2024spice}. However, most existing text-guided editing techniques~\cite{achlioptas2022changeit3d,sella2024spice} manipulate input shapes globally, without \emph{explicit} guidance over which regions should be edited and, conversely, which regions should be preserved. While these methods can perform semantic fine-grained edits conveying the target text prompt, they often do not sufficiently preserve the shape's structure, creating additional unintended and undesirable changes. 

Inspired by localized text-guided image editing techniques~\cite{rombach2022high,avrahami2022blended,avrahami2023blended}, we introduce a framework that reframes the localized shape editing task as a semantic inpainting problem. Our approach, coined \emph{Blended Point Clouds} (\methodName{}), leverages a foundation point cloud diffusion model~\cite{nichol2022point}, which is tuned to enable text-guided semantic shape inpainting while preserving the model's generative capabilities. 
In particular, we repurpose an existing small-scale dataset linking 3D shape pairs with textual descriptions of their pairwise geometrics differences~\cite{achlioptas2023shapetalk}. By leveraging off-the-shelf language and point cloud segmentation models, we identify textual descriptions describing part-specific differences (i.e. ``the target has shorter \emph{legs}'') and construct a set of \{\emph{partial shape, complete shape, textual description}\} triplets associated with these localized differences. These triplets serve as training data for our inpainting model that learns to generate complete shapes from partial shapes and textual descriptions.

While our inpainting model can indeed inpaint missing regions in a manner that faithfully adheres to the target text prompt---similarly to shape completion models~\cite{wang2017shape,yu2021pointr,chu2024diffcomplete}---models conditioned only on partial shapes are by design \emph{blind} to the appearance of the edited region. Consider, for instance, the examples in Figure \ref{fig:teaser}. To precisely edit the chair's back and add spindles to it (third column), it is crucial to provide the full input shape and not only a partial conditional (\emph{i.e.}, back-less) chair. 

Hence, we propose an inversion-free \emph{coordinate blending} technique that locally blends noised versions of the input shape with the inpainted shape at a progression of diffusion steps. Our key insight is that we can bypass artifact-prone inversion-based techniques and enforce identity preservation also within the local edited region by leveraging the model's inherent ability to reconstruct complete input shapes. Our coordinate blending mechanism is applied at inference-time, and allows for achieving a seamless blending of the edited region with the rest of the shape. 

We conduct extensive experiments, comparing our method to existing techniques performing semantic fine-grained edits of 3D shapes using natural language~\cite{achlioptas2022changeit3d,sella2024spice}, evaluating performance both in terms of preservation of the shape's structural identity as well as the extent to which the target shape conforms to the text prompts.

\section{Related Work}

\subsection{Fine-grained Shape Editing}
Shape editing, particularly edits that manipulate the fine-grained structure of the shape, is a longstanding problem in computer vision and graphics, initially applied for character animation~\cite{magnenat1988joint,badler1990making}. Below we cover only works most closely related to our work.

To achieve a high level of control for enabling editability of shapes, methods such as DualSDF~\cite{hao2020dualsdf} and CNS-Edit~\cite{hu2024cns} construct a dual representation which couples an easily editable, low-level representation with a more intricate (high-quality) representation whose structure aligns with the low-level representation. EXIM~\cite{liu2023exim} also introduces a hybrid representation, specifically composed of an explicit part that enables coarse localization and an implicit part that enables fine global geometric editing and color modifications. NeuMesh~\cite{yang2022neumesh} couples a mesh representation with a neural representation by attaching neural features to mesh vertices, allowing mesh based geometry editing of detailed shapes. iShapEditing~\cite{li2024ishapediting} leverages the explicit qualities of the triplane representation to enable point drag based editing. SPAGHETTI~\cite{hertz2022spaghetti} proposes a shape representation composed of Gaussian Mixture Models which allowing part-level control. While these methods provide fine-grained control over shape editing, they typically rely on less intuitive user inputs compared to textual descriptions. Furthermore, they are often limited to relatively simple geometric operations such as creating holes, elongating parts or expanding them.

Recent advancements in text–image representations, most notably CLIP \cite{radford2021learning}, has led to a surge of text-guided shape editing techniques, which aim to leverage the flexibility and expressiveness of natural language for enabling more diverse and complex shape modifications. Several methods~\cite{chen2022tango, michel2022text2mesh, gao2023textdeformer} use CLIP to stylize or deform meshes by matching their 2D projections with a target text prompt. CLIP has also been widely used to optimize Neural Radiance Fields (NeRFs)~\cite{mildenhall2021nerf} to resemble objects described in text prompts~\cite{jain2022zero, lee2022understanding, wang2022clip}. Likewise, optimization-based methods---first powered by CLIP, and later by Score Distillation Sampling~\cite{poole2022dreamfusion}---have also shown promise when applied to a 3D Gaussian Splatting representation \cite{wu2024gaussctrl,wang2024gaussianeditor,palandra2024gsedit}, a new widely-adopted 3D representation that bears some resemblance to point clouds (which we focus on). However, while unsupervised optimization-based methods can  often convey complex semantic concepts, they struggle to perform fine-grained text-guided semantic editing, as we also demonstrate in the supplementary material.

To this end, supervised methods leveraging datasets of source--target shape pairs and a target prompt, which describes the edit that connects the shapes, have been proposed for addressing the task of fine-grained shape editing. Methods such as InstructP2P~\cite{xu2023instructp2p} and ShapeWalk~\cite{slim2024shapewalk} propose using GeoCode \cite{pearl2022geocode}, a method that maps input shapes to an editable parameter space, in conjunction with an LLM to create a datasets containing source--target shape pairs and target text descriptions. They then use their curated datasets to train text-guided shape editing models.

ShapeTalk~\cite{achlioptas2023shapetalk} constructs a dataset using randomly paired shapes from existing shape datasets and using human annotators whom are asked to provide text prompts describing the differences between two input shapes. Their work also introduces ChangeIt3D, an editing model trained on ShapeTalk. Other methods such as LADIS~\cite{huang2022ladis} and Spice-E~\cite{sella2024spice} also use ShapeTalk to facilitate editing, the latter also harnessing a pre-trained text-to-shape diffusion model, enabling it to tackle out-of-distribution editing instructions. These supervised methods are most closely related to work. However, as we extensively show in our experiments, they struggle with localized editing instructions which constrains the edit to a specific region, and requires maintaining high structural fidelity in other regions.

\subsection{Local Editing via Inpainting}
A key concept guiding our work is that localized editing can be approached as a inpainting problem. In 2D images, the emergence of powerful image synthesis diffusion models \cite{dhariwal2021diffusion, ho2020denoising, rombach2022high, nichol2021glide} led to a proliferation of image inpainting methods, as these foundational models proved to be very well suited for this task. Rombach et al. \cite{rombach2022high} and GLIDE \cite{nichol2021glide} introduce an inpainting model as a possible application of their respective diffusion models. These models approach this task by concatenating an incomplete image and binary mask to the noisy latent propagating through the diffusion network and training the diffusion model to complete the missing regions. Similarly, SmartBrush \cite{xie2023smartbrush} introduces a model trained to denoise the incomplete image directly while also predicting a more accurate binary mask for the desired edit. While training an inpainting task-specific diffusion model has become somewhat of a go-to approach for image inpainting, it requires training a diffusion model which can be very computationally demanding. Blended diffusion \cite{avrahami2022blended} thus takes a different approach and uses CLIP embeddings of the incomplete image and guiding text to steer the sampling process of a pretrained unconditional diffusion model towards completing the inpainted region in a manner that aligns with the text condition.
RePaint \cite{lugmayr2022repaint} introduced a free form inpainting algorithm blending noisy versions of the ``known'' regions of an image with the inpainted regions during the inference process of a diffusion model given an input binary mask. Later, Blended Latent Diffusion \cite{avrahami2023blended} introduced a similar approach which additionally incorporated textual guidance and operated in the latent space of a latent diffusion model (LDM) \cite{rombach2022high}.
Our work is inspired by these insights, which we utilize for inpainting 3D shapes represented as point clouds. 

For 3D shapes, shape inpainting, typically referred to as shape completion, has been extensively studied, as it has many real world applications, including for robotics and autonomous driving. The explicit and simplistic nature of point clouds, coupled with the fact that point clouds are a natural output of 3D scanning, make point clouds a natural representation for this task. With the introduction of GANs \cite{goodfellow2020generative}, many works applied the GAN framework for generating complete point clouds from partial, incomplete ones \cite{wang2017shape, zhang2021unsupervised, wang2020cascaded}. Other approaches built upon the success of transformer encoder decoder networks in Natural Language Processing and applied transformer architectures for point cloud shape completion \cite{yu2021pointr, zhang2022point, xiang2021snowflakenet, yan2022shapeformer}. More recently, point cloud completion and inpainting methods have also utilized diffusion models to great effect \cite{chu2024diffcomplete, galvis2024sc, Zhou_2021_ICCV}. While these methods have shown to be very apt in reconstructing partial shapes, they are limited to reconstruction and do not accept additional text guidance. Conversely, SALAD \cite{koo2023salad} and SDFusion \cite{cheng2023sdfusion} introduce techniques capable of performing text-guided part completion. However, as part \textit{completion} methods they are inherently ``blind'' to the original part-to-edit, which makes these methods ill-posed for fine grained shape \textit{editing}.

\section{Method}
\begin{figure*} %
\centering
\jsubfig{\includegraphics[width=0.9\textwidth]{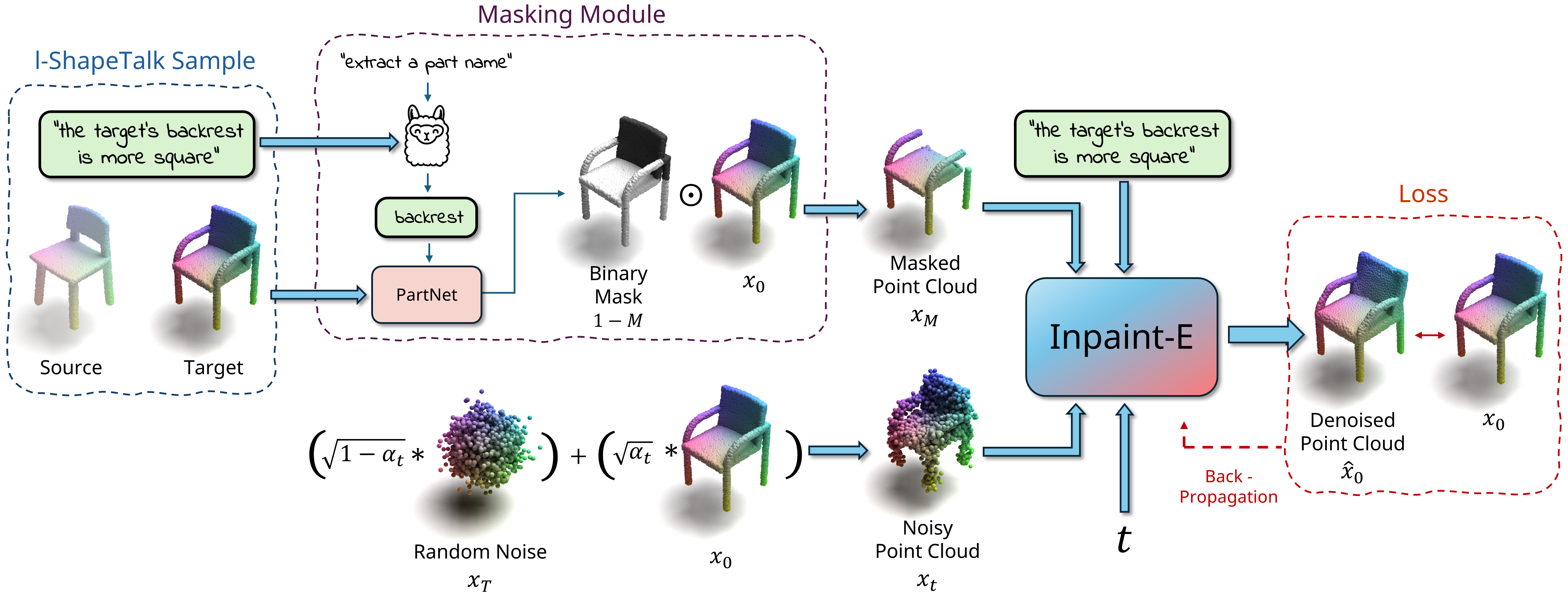}}{}
\vspace{-7pt}
\caption{\textbf{An overview of \inpaintingModel{}'s training procedure.} At each iteration we sample a target shape $x_0$ and a text prompt from l-ShapeTalk (Section \ref{sec:data}) and use Llama \cite{dubey2024llama} and PartNet \cite{mo2019partnet} to extract a binary mask $M$ that indicates the part that requires editing. We then use the binary mask to produce a partial point cloud that is given to \inpaintingModel{} as guidance in addition to the text prompt and timestep $t$. We then add random noise to the target point cloud $x_0$ and feed it to \inpaintingModel{} for denoising. As further detailed in Section \ref{sec:inpaint_e}, \inpaintingModel{}'s training objective is to minimize the difference between the output and target point cloud's 3D coordinates. 
}
\label{fig:method_training}
\end{figure*}

In this section, we introduce our approach for performing precise localized edits on point clouds given textual instructions. We begin by providing background on point cloud diffusion models (Section \ref{sec:prelim}). We then explain and justify our inpainting-based approach for achieving localized fine-grained editing (Section \ref{sec:inpainting_based_editing}), introducing our 3D inpainting model \inpaintingModel{}---one of the central components of our work. Similar to prior inpainting models, it lacks the prior knowledge of the edited part. To bridge this gap, we introduce our novel \emph{inversion-free coordinate blending} mechanism, which allows for better preserving identity (Section \ref{sec:inference_time}). Finally, The data used to train \inpaintingModel{} and evaluate our method is covered in Section \ref{sec:data}. 

\subsection{Preliminaries}
\label{sec:prelim}
Text-guided point cloud diffusion models aim to map a random noise vector $x_T$ and textual condition $\mathcal{C}$ to an output point cloud $x_0$, which corresponds to the given conditioning prompt. A point cloud sample $x \in \mathbb{R}^{K \times d}$ consists of $K$ points, each described by $d$ coordinates. In order to perform sequential denoising, the model $\varepsilon_\theta$ is trained to predict artificial noise. The noisy sample is defined by
\begin{equation}
x_t = \sqrt{\bar{\alpha}_t} x + \sqrt{1 - \bar{\alpha}_t} \varepsilon,
\end{equation}
where ${\alpha}_t$ is determined by a noise schedule and $\varepsilon \sim \mathcal{N}(0, I)$.
The model is trained to optimize: 
\vspace{-1mm}
\begin{equation}
\min_\theta \mathbb{E}_{x_0,\varepsilon\sim N(0,I),t\sim \mathcal{U}(1,T)} \norm{\varepsilon-\varepsilon_\theta(x_t,t,\mathcal{C})}^2.
\end{equation}

At inference, given a noise vector $x_T$, the noise is gradually removed by sequentially predicting it using the trained model for $T$ steps, refining it into a coherent sample from the learned data distribution.

\subsection{Inpainting-based Editing}
\label{sec:inpainting_based_editing}
Our method enables fine-grained, localized editing of point clouds based on textual instructions. While generic editing methods might result in a global change, we focus on editing a specific region, allowing more controllability for the user. For example, given a chair, a user may want to edit only the legs while keeping the backrest and seat intact. 

Given a shape $x$ and a textual instruction $\mathcal{C}$, we produce an edited shape $x_e$ such that the editing is limited to a local region, represented by a binary mask $M$. The structural text-based editing from $x$ to $x_e$ is then confined only to the masked region, i.e., $(1-M) \odot x = (1-M) \odot x_e$, where $\odot$ represents element-wise multiplication.

We recognize that our two objectives—performing meaningful text-based editing and restricting changes to a local region—could be conflicting. This challenge partially arises from a global bias in the dataset; for example, a chair with a thin back often has thin legs as well, meaning that altering one feature may unintentionally affect others. To address this, we developed a two-step approach. First, we extract a mask based on the textual description to identify the target region. Then, we perform localized editing using three inputs: the original shape, the local mask, and the textual instruction.

In this work, we acquire the mask using an off-the-shelf segmentation model. Specifically, we instruct an LLM (llama3 \cite{dubey2024llama}) to extract a part name from a given instruction prompt. For example given the prompt \textit{``the target has a thicker seat''} the language model should return the word ``seat''. We then give the part name and the complete point cloud as input to PointNet \cite{qi2017pointnet}, a neural network trained on the PartNet~\cite{mo2019partnet} dataset to perform semantic segmentation of the input point cloud. The model then produces a binary mask indicating the points belonging to the part referred to in the editing prompt. 

Next, we observe that our local editing setting can be formulated as a text-guided inpainting problem, where we 
``discard'' the original edit region and in-paint it according to a textual edit description.
While text-based editing training usually requires supervised pairs, text-based inpainting models are trained in self-supervised (i.e., no pairs) fashion, where the model learns to complete a local region guided by the textual condition. Accordingly, we construct a text-guided point cloud inpainting model by making minimal architectural changes to an existing text-to-point-cloud generator and fine-tune it on a small-scale dataset.

\subsubsection{Training the \inpaintingModel{} Model}
\label{sec:inpaint_e}
Our point cloud inpainting diffusion model, coined \inpaintingModel{}, is provided with a masked point cloud $x_M = x \odot (1-M)$ and a text condition $\mathcal{C}$ as input. That is, we set the masked coordinates in $x$ (i.e., region to inpaint) to $(0, 0, 0)$. Additionally, to avoid ambiguity between masked points and points near the origin and clearly indicate the masked region, we also set the auxiliary $RGB$ color channels of the masked points to $(1, 1, 1)$, and $(0, 0, 0)$ for all other points. The model is then trained to complete the masked area guided by the textual description and the masked point cloud with the target being the ground truth, non-masked, point cloud $x$. %
As there is no existing inpainting model specifically designed for our tasks, we use Point-E \cite{nichol2022point}, a foundational point cloud diffusion model, as our backbone. We modify the main generator module, which generates samples of $1024$ points. Since the structure is determined in the generator stage, we do not modify the upsampler module, which upsamples the outputs to $4096$-points.

We require guidance of both text prompt and a masked point cloud, while Point-E only accepts a text prompt as guidance. Therefore, we follow Spice-E \cite{sella2024spice} and modify the Point-E generator's architecture to use \emph{Cross-Entity} attention blocks instead of self-attention, enabling the propagation of guidance information from the partial point cloud into the de-noising process of the complete point cloud. By following this procedure we introduce the masked point clouds into the generation process as structural priors while preserving Point-E's generation capabilities. 

We finetune our model using the loss and training procedure used in Point-E. At each training iteration we aim to predict the noise added to the target point cloud, the only difference with Point-E being the use of the masked point cloud as additional guidance:
\begin{equation}
    \mathcal{L}_{\inpaintingModel{}} =  \norm{\varepsilon-\varepsilon_\theta(x_t,t,\mathcal{C},x_M)}^2
\end{equation}

As previously mentioned, we obtain the editing masked $M$ by extracting a part name from the textual condition $\mathcal{C}$ using an LLM and segmenting that part in  using an off-the-shelf segmentation model. \inpaintingModel{}'s training procedure is outlined in Figure \ref{fig:method_training}. 

\begin{figure*} %
\centering
\jsubfig{\includegraphics[width=0.85\textwidth]{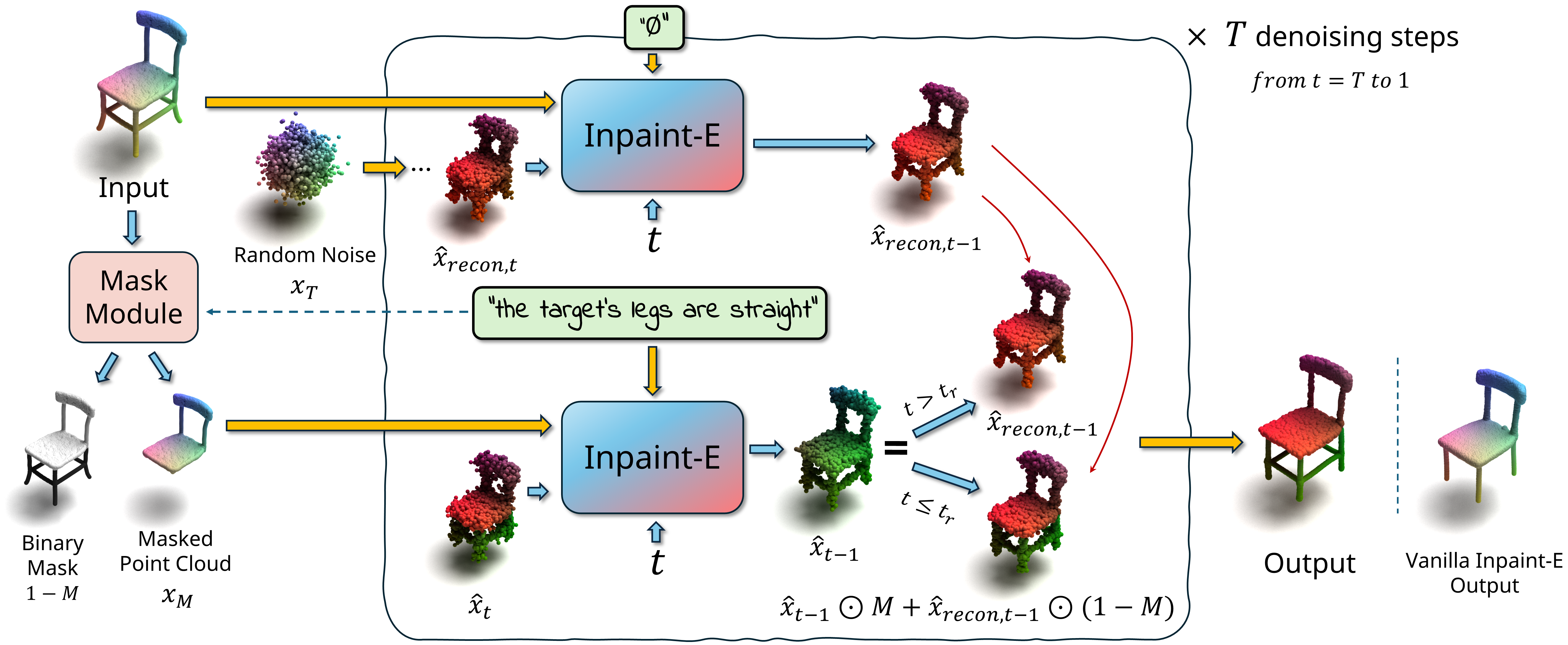}}{}
\vspace{-7pt}
\caption{\textbf{Inversion-Free Inference Time Coordinate Blending}. A diagram illustrating our proposed inference-time mechanism, as described in Algorithm \ref{alg:inference} and Section \ref{sec:inference_time}. The red points represent points generated by \inpaintingModel{} using the empty string and input point cloud as guidance (i.e., \emph{reconstruction denoising}), and the green represent points generated by \inpaintingModel{} using the full editing prompt and the partial point cloud $x_M$ as guidance (i.e. \emph{inpainting}), blended together to achieve a fine-grained shape editing.}%
\label{fig:method_inference}
\end{figure*}

\subsubsection{Inversion-Free Coordinate Blending}
\label{sec:inference_time}
To enable \emph{fine-grained} editing, our method requires prior knowledge of the edited part's shape. For example, if we want to make a chair’s legs longer, we first need to capture the shape of the original legs. However, our inpainting model is inherently blind to the masked region. Therefore, we propose to modify the inference process of \inpaintingModel{} to balance both inpainting and reconstruction of the \emph{full} shape.

Reconstructing a shape or an image during the inference process of a diffusion model usually involves inversion---the process of finding an initial noise vector that, when passed through the diffusion process, reconstructs the input while retaining editability. Unfortunately, inversion methods are computationally expensive and often inaccurate, especially in conditional models \cite{hertz2022prompt, mokady2023null}. Therefore, we design our inference time algorithm to avoid inversion altogether by leveraging \inpaintingModel{}'s enhanced reconstruction capabilities. We observe that \inpaintingModel{} is capable of reconstructing the non-edit region with high accuracy for any text prompt and noise vector, thus by extending the non-edit region to encompass the entire shape \inpaintingModel{} is able to produce gradually de-noised versions of the input shape for any initial noise vector. With this in mind, we propose a \emph{reconstruction denoising} mechanism, where  the model is conditioned on the complete point cloud $x$ and a null text prompt $\mathcal{C}_0 =$ ``'' and iteratively produces a sequence of progressively denoised point clouds $\{\hat{x}_{recon, i}\}_{i=0}^T$ from an inital noise vector $x_T$ such that the final output $\hat{x}_{recon, 0}$ closely resembles the original point cloud $x$. %
To support this, we occasionally replace during training the masked point cloud $x_{M}$ and the text condition $\mathcal{C}$ with the full point cloud $x$ and an empty prompt $\mathcal{C}_{0} = ``\,"$.

Our inference process then begins by reconstructing the full shape as previously described from timestep $t = T$ down to a transition timestep $t = t_r$. From $t = t_r$ down to $t = 0$ we perform \emph{coordinate blending}, a process by which we both perform reconstruction denoising and text conditioned inpainting denoising and blend the two intermediate outputs into one according to the edit mask. Formally, we use \inpaintingModel{} with $\mathcal{C}_{0} = ``\,"$ and $x$ as guidance inputs to denoise $\hat{x}_{recon, t}$ into $\hat{x}_{recon, t - 1}$ and use \inpaintingModel{} a second time with $\mathcal{C}$ and $x_M$ as guidance inputs to denoise $\hat{x}_t$ into $\hat{x}_{t-1}$. We then combine these two noisy point clouds into one according to the following blending operation:   \[\hat{x}_{t-1} \gets \hat{x}_{t-1} \odot M + \hat{x}_{recon, t-1} \odot (1-M),\] ensuring that points outside \( M \) in \( \hat{x}_{t-1} \) remain unchanged from the reconstruction \( \hat{x}_{recon, t-1} \). In this way, we guarantee that the unedited points in the output \( \hat{x}_0 \) are nearly identical to the  original point cloud \( x \), preserving identity across the unmasked regions.

By only beginning inpainting at \( t = t_r \), we allow the model to work from an initial, approximate representation of the original masked part. This ensures that the inpainted content in \( \hat{x}_0 \) retains strong resemblance to the original, enhancing the model’s ability to perform identity-preserving edits. We empirically set $t_r = 20$ (of $T=64$ total steps) in our experiments, as we found it to strike a good balance between identity preservation and edit flexibility. Our full algorithm is presented in Algorithm \ref{alg:inference} and Figure \ref{fig:method_inference}. 

\begin{algorithm}
\caption{Inference-time coordinate blending \ignorethis{mechanism} }
\label{alg:inference}
\begin{algorithmic}
\Require $x, t_r, \mathcal{C}, M$
\State Set $x_{M} \gets x \odot (1-M)$
\State Set $\mathcal{C}_{0} \gets ``\,"$ 
\State  Initialize $\hat{x}_{recon, T} \gets \mathcal{N}(0, I)$ 

\For{$t = T, T-1, \ldots, 1$} 
    \State $\hat{x}_{recon, t-1} \gets  InpaintE(\hat{x}_{recon, t}; x, \mathcal{C}_{0}, t)$
    \If{$t > t_r$} \Comment{{\footnotesize ``Reconstruction Denoising''}}
    \State $\hat{x}_{t-1} \gets  \{\hat{x}_{recon, t-1}\}$
    \Else  \Comment{{\footnotesize ``Coordinate Blending''}}
    \State $\hat{x}_{t-1} \gets InpaintE(\hat{x}_{t}; x_{M}, \mathcal{C}, t)$
    \State $\hat{x}_{t-1} \gets \hat{x}_{t-1} \odot M + \hat{x}_{recon, t-1} \odot (1-M)$
    \EndIf
\EndFor
\Ensure $\hat{x}_0$
\end{algorithmic}
\end{algorithm}
\vspace{-4mm}
\subsection{Training Data}
\label{sec:data}
Training our point cloud inpainting model requires triplets of a point cloud \( x \), a binary mask \( M \), and a text condition \( \mathcal{C} \). Previous image inpainting models often used random masks \cite{rombach2022high}, but recent works \cite{nitzan2024lazy, xie2023smartbrush} have shown that using accurate masks and descriptive text prompts enhances quality. These accurate masks are typically obtained with off-the-shelf segmentation models. 

Accordingly, we train and evaluate our model on ShapeTalk \cite{achlioptas2023shapetalk}, a dataset comprising shape pairs from ShapeNet \cite{chang2015shapenet}, PartNet \cite{mo2019partnet}, and ModelNet \cite{vishwanath2009modelnet}, accompanied by human-written text prompts that describe shape differences. Although these prompts make ShapeTalk valuable for training text-guided editing models, the dataset has two main limitations for local editing. First, many prompts describe global attributes (e.g., ``the target looks more comfortable''), and second, because pairs are randomly selected, shape differences are not localized to specific parts. 

To enable local editing, we created l-ShapeTalk, a subset containing prompts that refer to specific parts (e.g., ``the target has a shorter backrest''). We used a large language model (LLM) to identify part names within prompts, which then guided our segmentation model to generate edit masks, discarding samples where segmentation failed. For evaluation, we use both l-ShapeTalk and the full ShapeTalk dataset, selecting the most relevant part for editing from each prompt. During training, we use masked versions of ShapeTalk ``target'' shapes, while for evaluation, we mask the ``source'' shapes to reflect our training objective (inpainting) versus our end goal (editing). We follow the original train/test splits in ShapeTalk for creating the localized test subset. Our dataset will be released, and additional details are provided in the supplementary material.

\begin{table*}[t]
\centering
\setlength{\tabcolsep}{3.0pt}
\def\arraystretch{1.0}
\begin{tabularx}{\textwidth}{lcccccccccccccc}
\toprule
 &           & \multicolumn{6}{c}{Shapetalk} &  & \multicolumn{6}{c}{l-Shapetalk} 
                          \\ 
\cmidrule(lr){3-8} \cmidrule(lr){10-15} 
Metric && $\text{CLIP}_{Sim}\uparrow$& $\text{CLIP}_{Dir}\downarrow$ & 
GD$\downarrow$ & CD$\downarrow$ & FPD$\downarrow$ & l-GD$\downarrow$ && $\text{CLIP}_{Sim}\uparrow$& $\text{CLIP}_{Dir}\downarrow$ &  GD$\downarrow$ & CD$\downarrow$ & FPD$\downarrow$ & l-GD$\downarrow$ \\ \midrule
Changeit3D && 0.21&	1.02 &	0.65&	0.18&	183.02&	0.19
 && 0.21&	1.02 &	0.87&	0.19	&217.28&	0.20
\\ 
Spice-E && 0.25&	1.01 &	1.84&	0.24	&390.02&	0.31
 && 0.25&	1.01 &	2.26&	0.26&	487.71&	0.39
\\  
Ours && \textbf{0.26}&	\textbf{0.99}	&\textbf{0.34}	&\textbf{0.05}&	\textbf{33.64}	& \textbf{0.07}
 && \textbf{0.27}	& \textbf{0.99} &	\textbf{0.29}&	\textbf{0.04}	&\textbf{13.51}	&\textbf{0.05}
\\ 
\bottomrule
\end{tabularx}
\vspace{-8pt}
    \caption{\textbf{Quantitative Evaluation}. We compare performance against Changeit3D~\cite{achlioptas2022changeit3d} and Spice-E~\cite{sella2024spice} across several metrics, which are further detailed in Section \ref{sec:metrics}. As illustrated above, our method outperforms prior work across all metrics, considering all test samples (Shapetalk) as well as the localized subset we extracted (l-Shapetalk).}
\label{tab:metrics}
\end{table*}

\section{Experiments}
\label{sec:experiments}

In this section, we present our main results and comparisons. The data used throughout our experiments is detailed in Section \ref{sec:data}. Following prior work~\cite{sella2024spice}, results are reported over the \emph{Chair}, \emph{Table} and \emph{Lamp} categories. %
Additional experimental details, comparisons and many more results from various shape categories, including fly-through visualizations demonstrating the quality of our results from multiple views, are provided in the supplementary material.

\subsection{Evaluation Metrics}
\label{sec:metrics}

We adopt a wide range of metrics proposed by prior work~\cite{achlioptas2022changeit3d,sella2023vox} to evaluate performance over multiple aspects.

\smallskip \noindent \textbf{Identity Preservation.} We evaluate structural identity preservation  using the following two metrics:  
 \newline \emph{Geometric Distance} (GD) measures the geometric distance between the input and output point clouds using the standard Chamfer distance measure.
 \newline \emph{\textit{localized}-Geometric Distance} (l-GD). Given a binary editing mask marking the points that need to be edited according to the text prompt, l-GD measures the geometric distance (GD) of the points outside of the mask between the input and output shapes. The distance in this metric, similarly to GD, is measured using Chamfer distance. 

\smallskip \noindent \textbf{Structural Fidelity.}  We evaluate structural-similarity to the input shapes using the following two metrics:
 \newline \emph{Fréchet Point Distance} (FPD) \cite{shu20193d}. Similarly to Fréchet Inception Distance \cite{heusel2017gans} in 2D, it measures the quality of the generated outputs by comparing the distribution of output features to those of the inputs using a pretrained network.
 \newline \emph{Class Distortion} (CD) Measures the quality of the outputs by using a pretrained classifier to predict their shape category (chair, airplane, etc.). The assumption is that the more the model is confident of the tested output being of its intended category the more it is structurally sound. Thus, we calculate the absolute difference of the probability assigned to the underlying class between the input/output shapes.

\smallskip \noindent \textbf{Edit Fidelity.} We evaluate how well the generated results capture the target text prompt using the following metrics:  
\newline \emph{CLIP Similarity} ($\text{CLIP}_{Sim}$) measures the semantic similarity between the output objects and the target text prompts. We encode both the prompt and images rendered from our point cloud outputs using CLIP's \cite{radford2021learning} text and image encoders, measuring the cosine-distance between them. %
 \newline \emph{CLIP Direction Similarity} ($\text{CLIP}_{Dir}$) evaluates the quality of the edit in regards to the input by measuring the directional CLIP similarity. This metric measures the distance between the direction of the change from the input and output rendered images and the direction of the change from an input prompt %
 to the one describing the edit.  %

\begin{figure} %

\rotatebox{90}{\hspace{0.2cm}\footnotesize{\emph{It has more}}}
\rotatebox{90}{\hspace{0.3  cm}\footnotesize{\emph{pointy top}}}
\jsubfig{\includegraphics[height=1.75cm, trim={4.0cm 4.0cm 4.0cm 4.0cm}, clip]{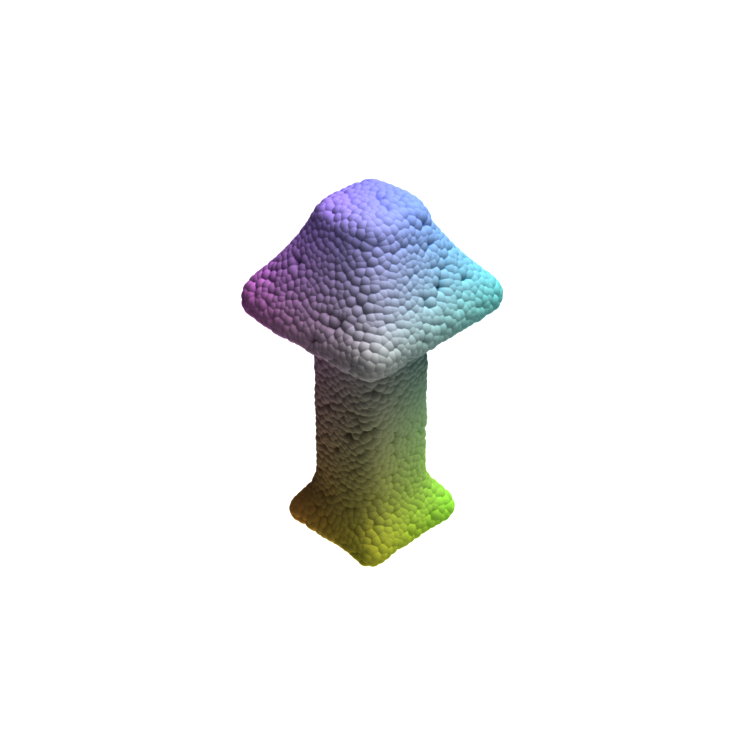}}{}\hfill
\jsubfig{\includegraphics[height=1.75cm, trim={2.0cm 2.0cm 2.0cm 2.0cm}, clip]{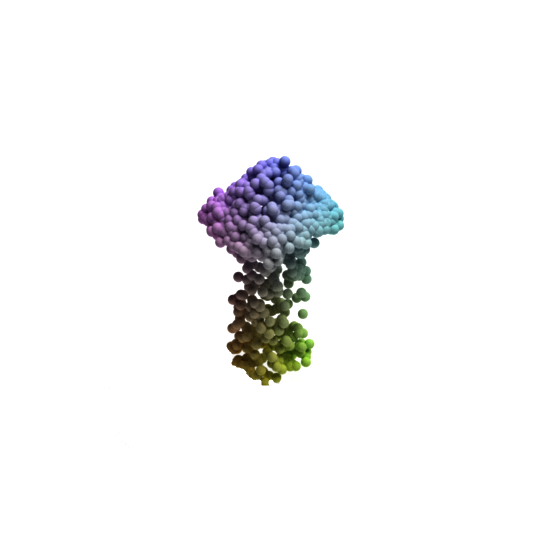}}{} \hfill
\jsubfig{\includegraphics[height=1.75cm, trim={2.5cm 2.5cm 2.5cm 2.5cm}, clip]{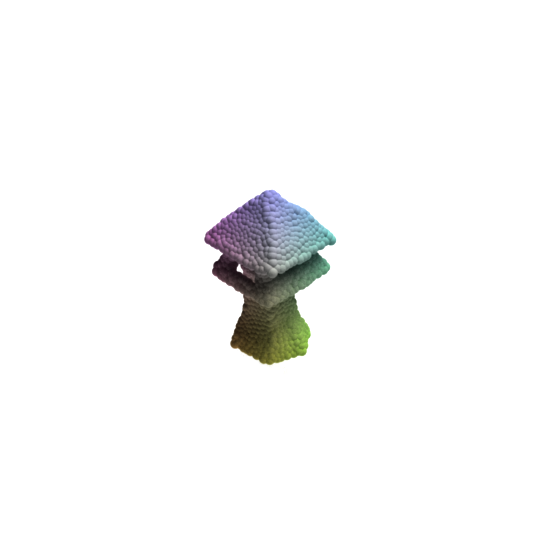}}{}\hfill
\jsubfig{\includegraphics[height=1.75cm, trim={4.0cm 4.0cm 4.0cm 4.0cm}, clip]{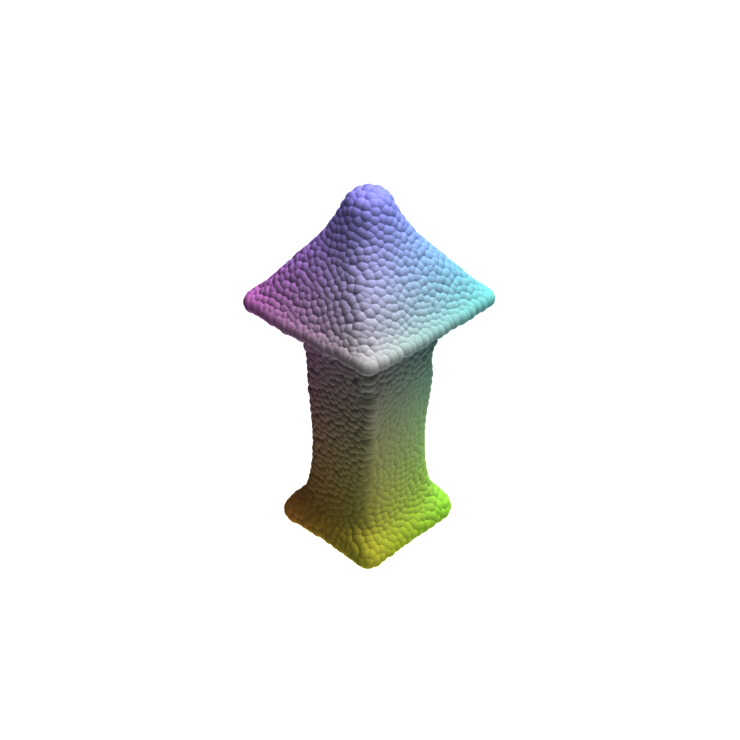}}{}  \vspace{2.0pt}

\rotatebox{90}{\hspace{0.4cm}\footnotesize{\emph{Shade is}}}
\rotatebox{90}{\hspace{0.1cm}\footnotesize{\emph{more rounded}}}
\jsubfig{\includegraphics[height=1.75cm, trim={7.0cm 7.0cm 7.0cm 7.0cm}, clip]{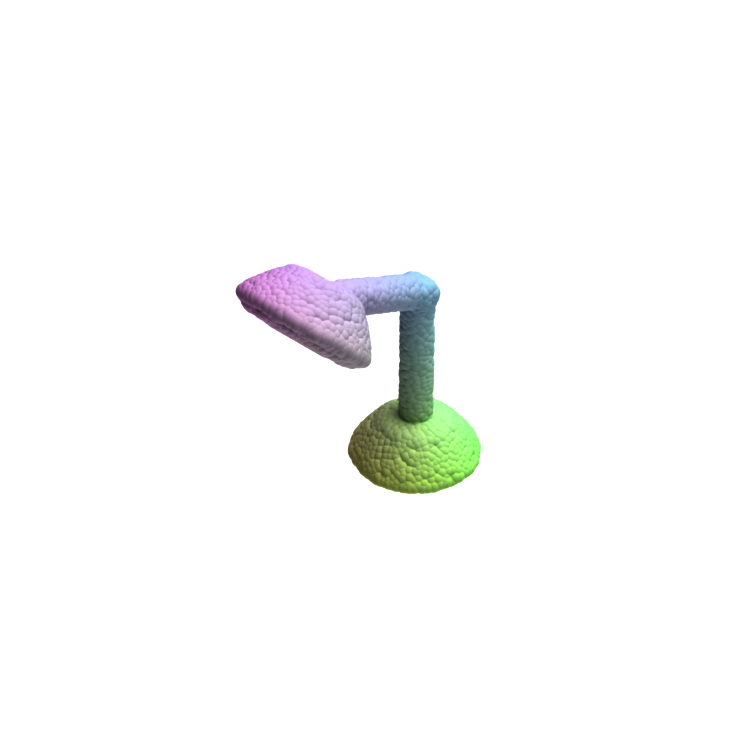}}{}\hfill
\jsubfig{\includegraphics[height=1.75cm, trim={7.0cm 7.0cm 7.0cm 7.0cm}, clip]{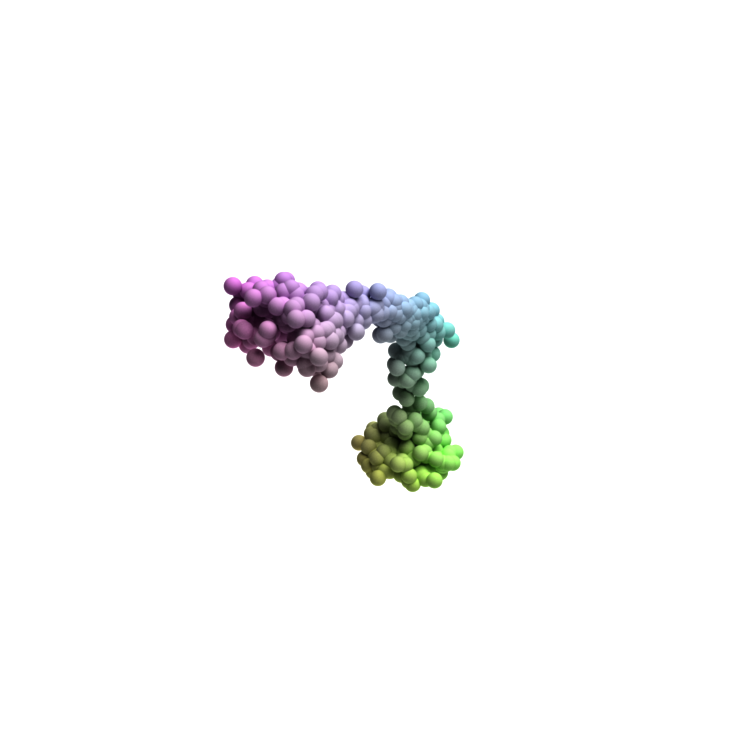}}{} \hfill
\jsubfig{\includegraphics[height=1.75cm, trim={4.0cm 4.0cm 4.0cm 4.0cm}, clip]{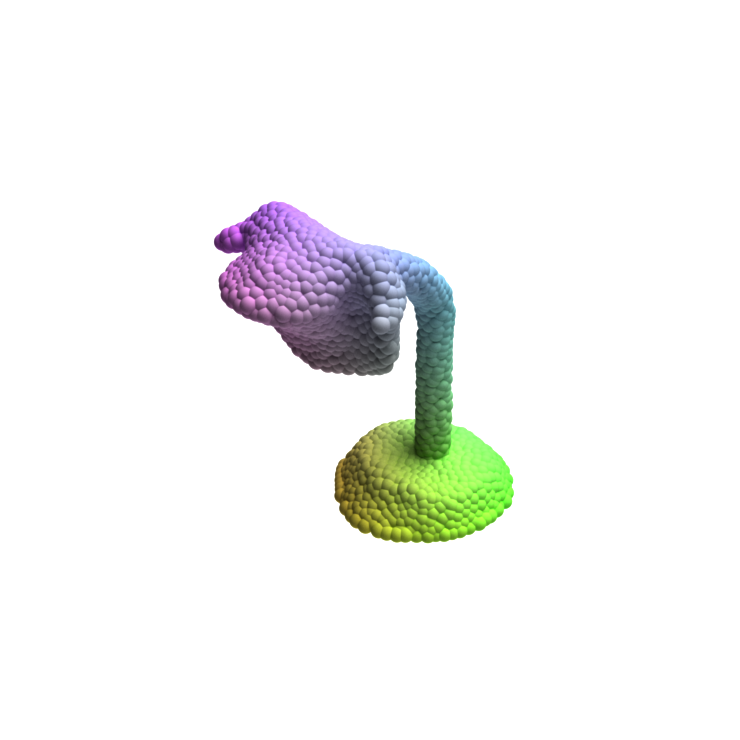}}{}\hfill
\jsubfig{\includegraphics[height=1.75cm, trim={7.0cm 7.0cm 7.0cm 7.0cm}, clip]{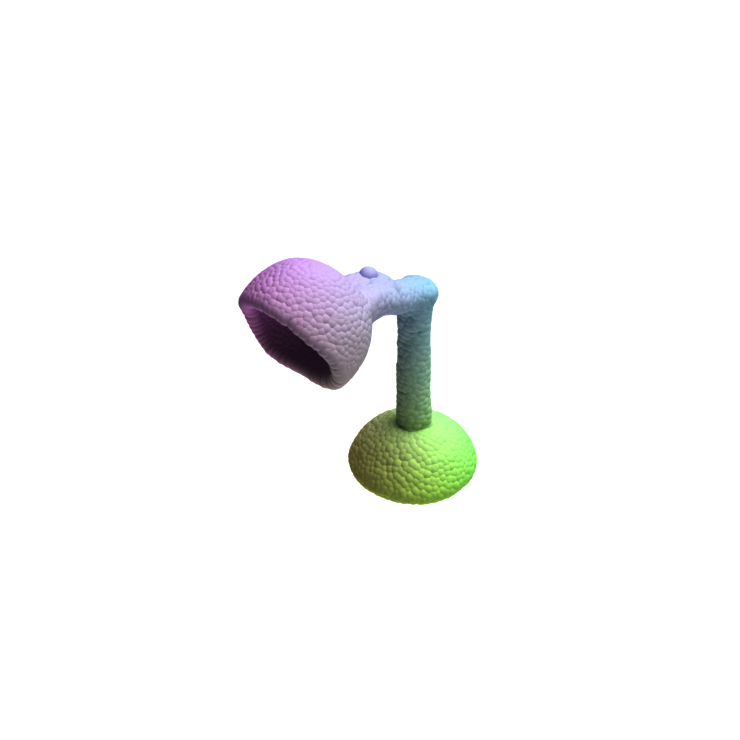}}{} \vspace{2.0pt}

\rotatebox{90}{\hspace{0.3cm}\footnotesize{\emph{No supports}}}
\rotatebox{90}{\hspace{0.2cm}\footnotesize{\emph{between legs}}}
\jsubfig{\includegraphics[height=1.75cm, trim={4.0cm 4.0cm 4.0cm 4.0cm}, clip]{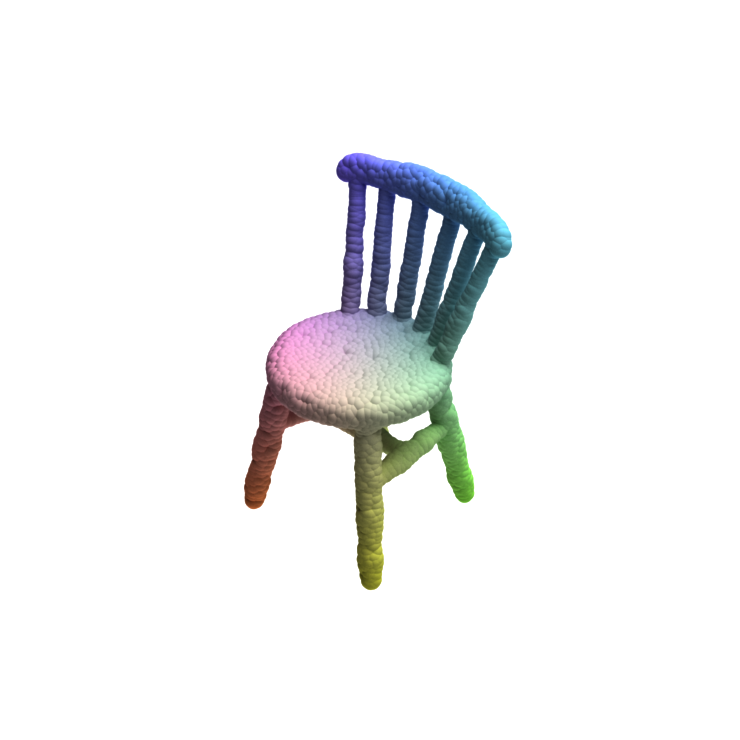}}{}\hfill
\jsubfig{\includegraphics[height=1.75cm, trim={4.0cm 4.0cm 4.0cm 4.0cm}, clip]{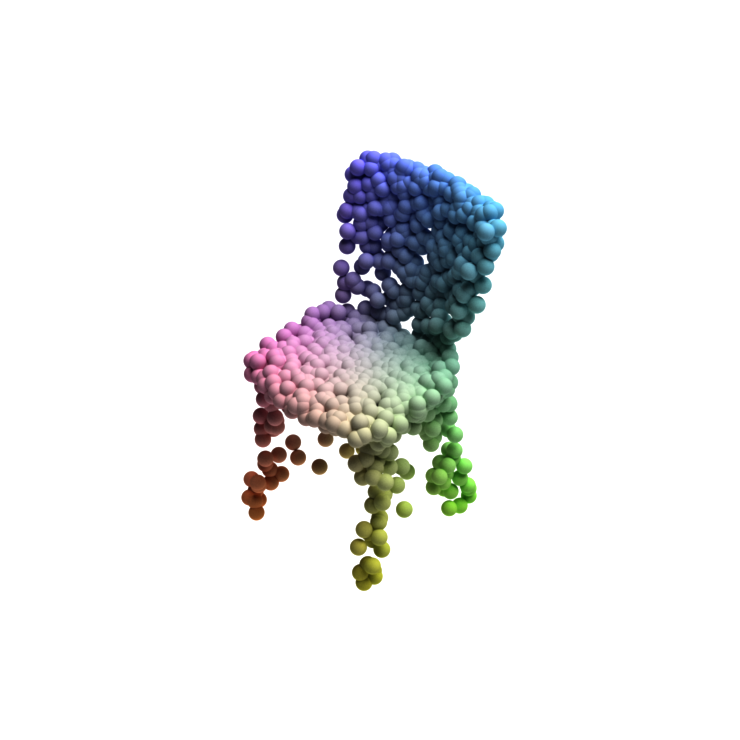}}{} \hfill
\jsubfig{\includegraphics[height=1.75cm, trim={2.5cm 2.5cm 2.5cm 2.5cm}, clip]{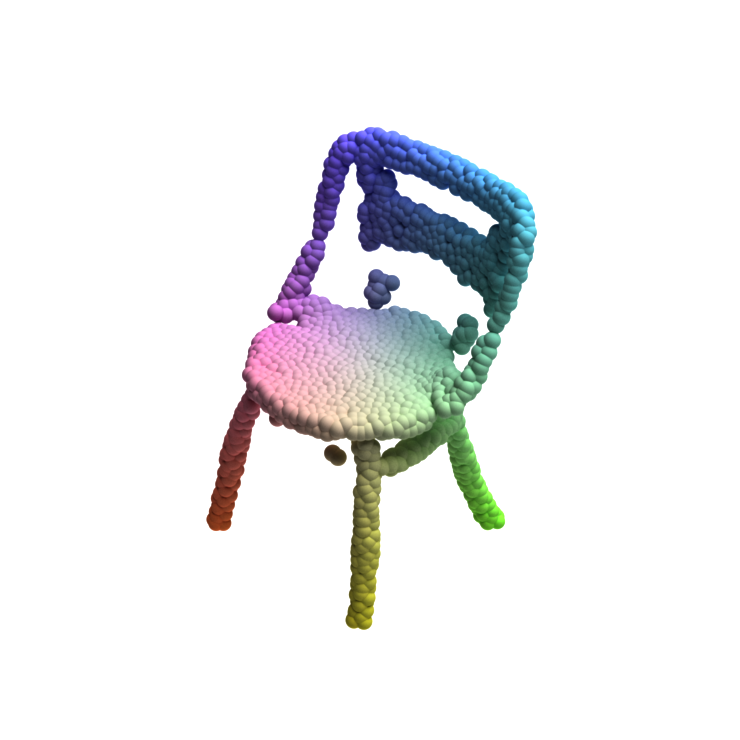}}{}\hfill
\jsubfig{\includegraphics[height=1.75cm, trim={4.0cm 4.0cm 4.0cm 4.0cm}, clip]{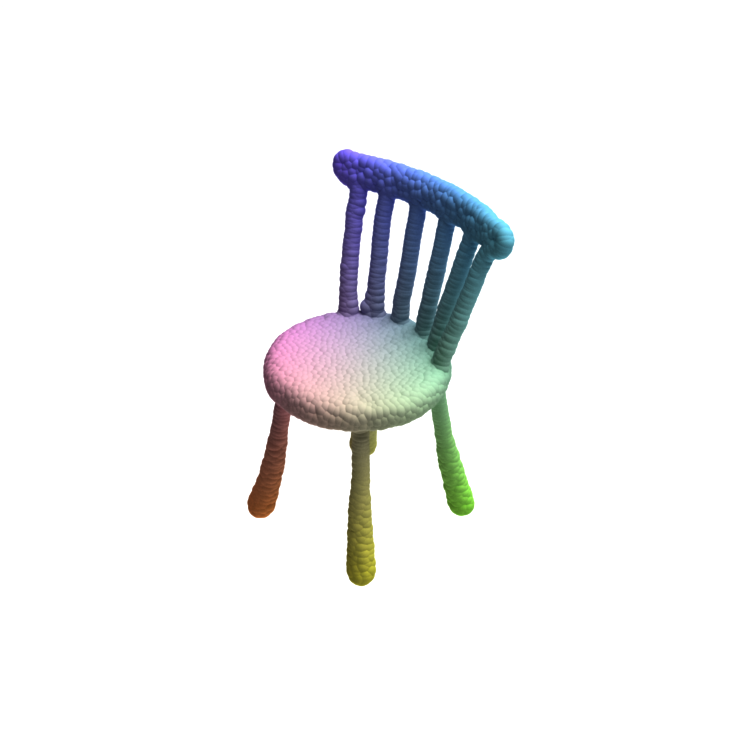}}{} \vspace{2.0pt}

\rotatebox{90}{\hspace{0.1cm}\footnotesize{\emph{There is a gap}}}
\rotatebox{90}{\hspace{0.3cm}\footnotesize{\emph{in the back}}}
\jsubfig{\includegraphics[height=1.75cm, trim={4.0cm 4.0cm 4.0cm 4.0cm}, clip]{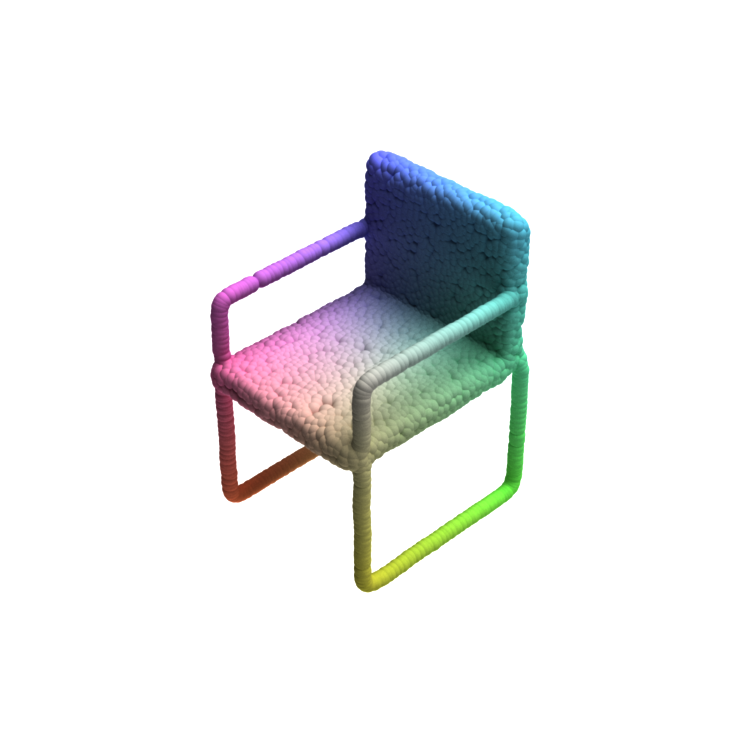}}{}\hfill
\jsubfig{\includegraphics[height=1.75cm, trim={4.0cm 4.0cm 4.0cm 4.0cm}, clip]{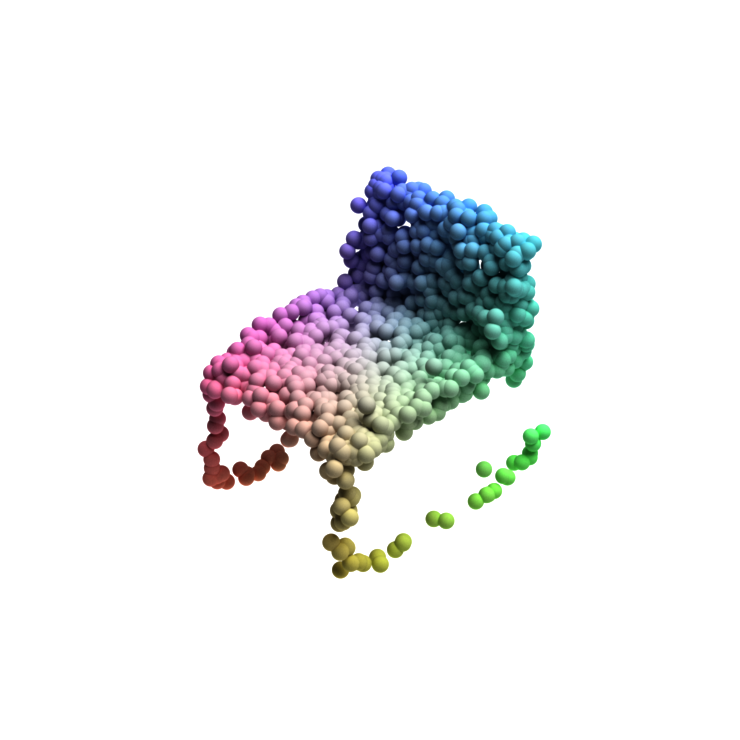}}{} \hfill
\jsubfig{\includegraphics[height=1.75cm, trim={2.5cm 2.5cm 2.5cm 2.5cm}, clip]{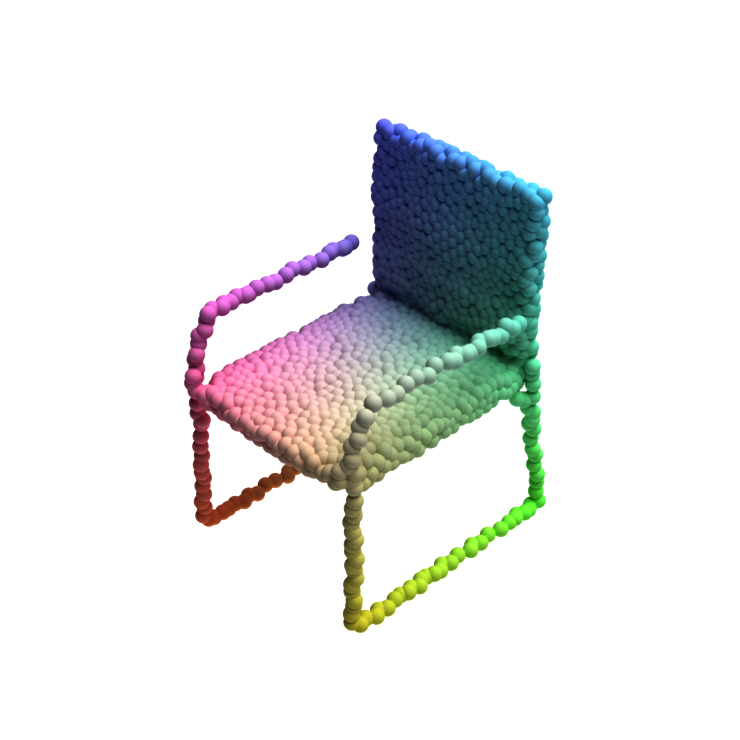}}{}\hfill
\jsubfig{\includegraphics[height=1.75cm, trim={4.0cm 4.0cm 4.0cm 4.0cm}, clip]{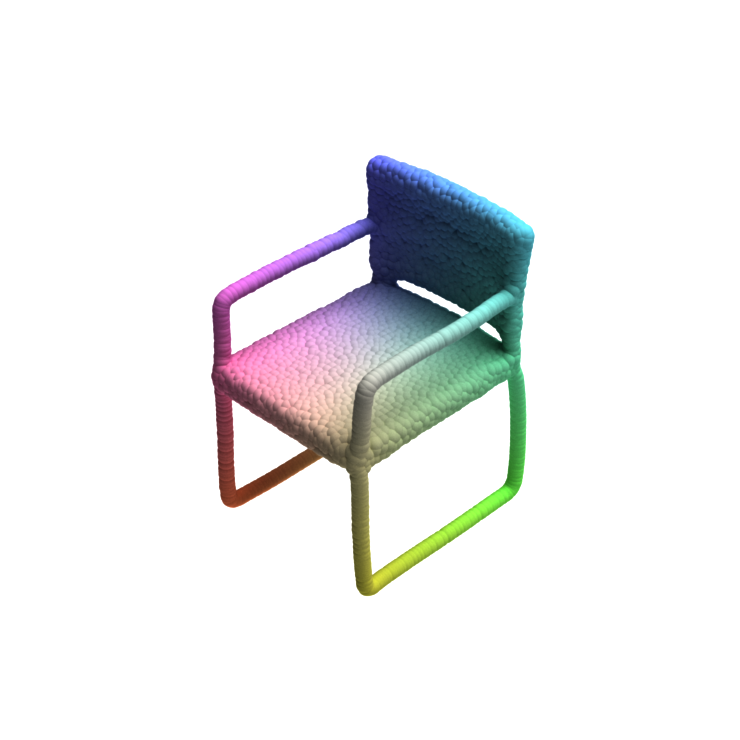}}{} \vspace{2.0pt}

\rotatebox{90}{\hspace{0.3cm}\footnotesize{\emph{The top has}}}
\rotatebox{90}{\hspace{0.3cm}\footnotesize{\emph{round edges}}}
\jsubfig{\includegraphics[height=1.75cm, trim={4.0cm 4.0cm 4.0cm 4.0cm}, clip]{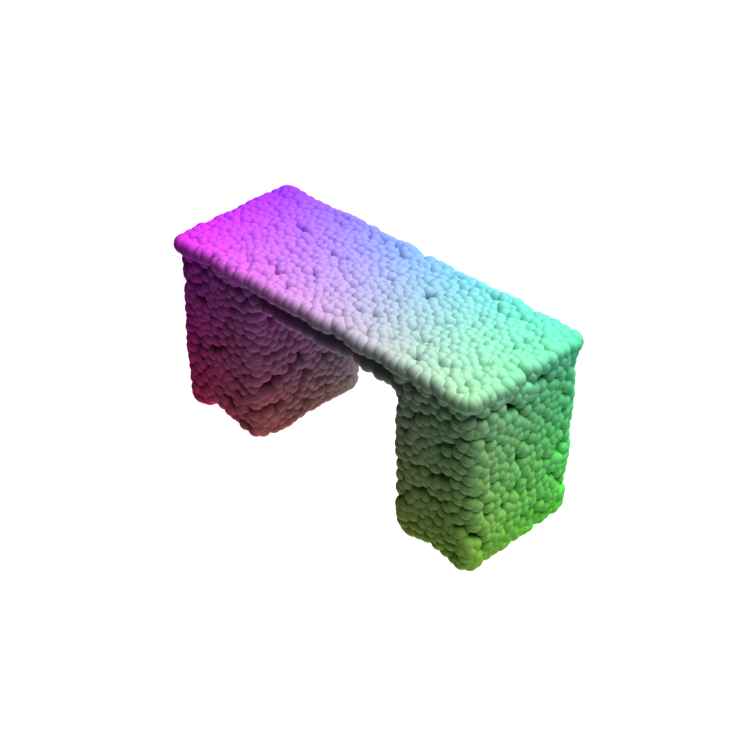}}{}\hfill
\jsubfig{\includegraphics[height=1.75cm, trim={4.0cm 4.0cm 4.0cm 4.0cm}, clip]{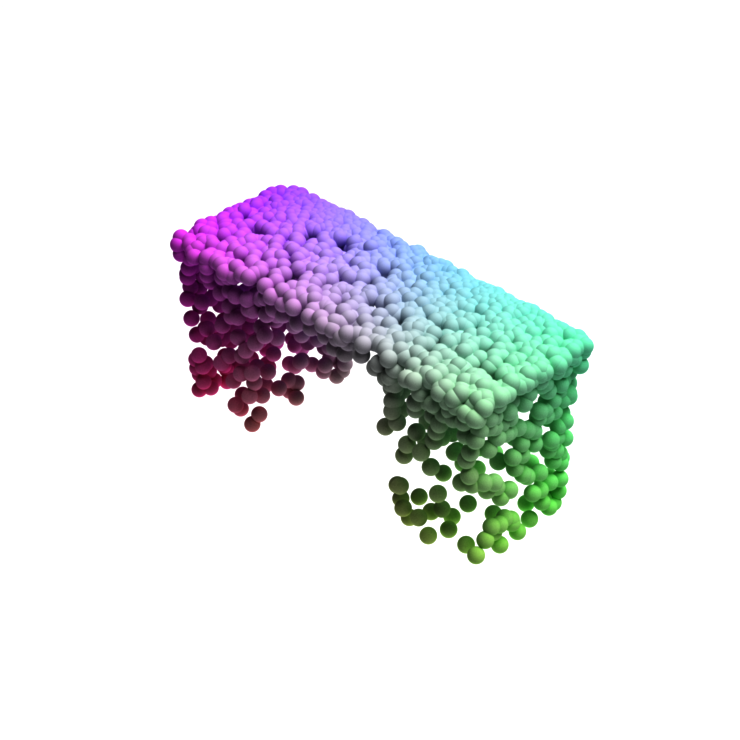}}{} \hfill
\jsubfig{\includegraphics[height=1.75cm, trim={4.0cm 4.0cm 4.0cm 4.0cm}, clip]{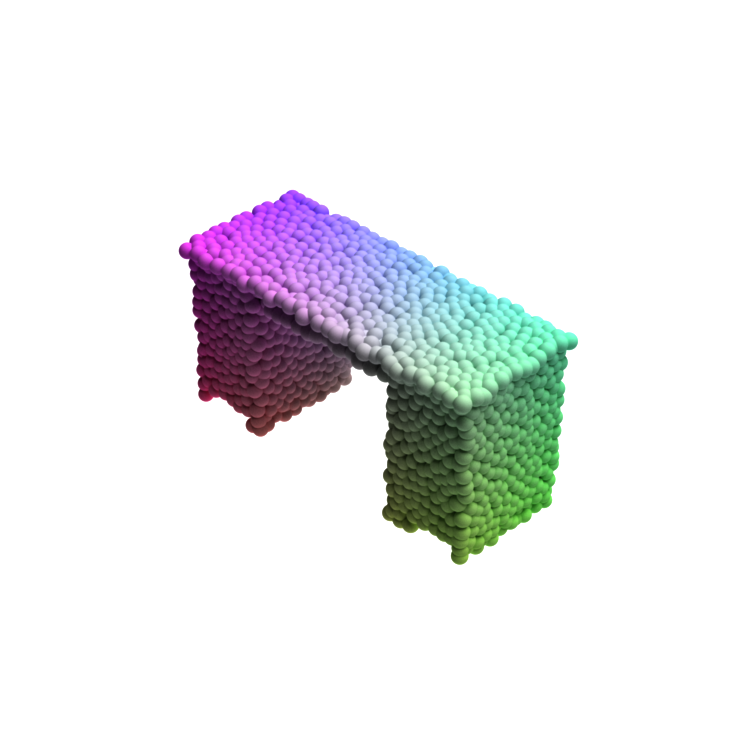}}{}\hfill
\jsubfig{\includegraphics[height=1.75cm, trim={4.0cm 4.0cm 4.0cm 4.0cm}, clip]{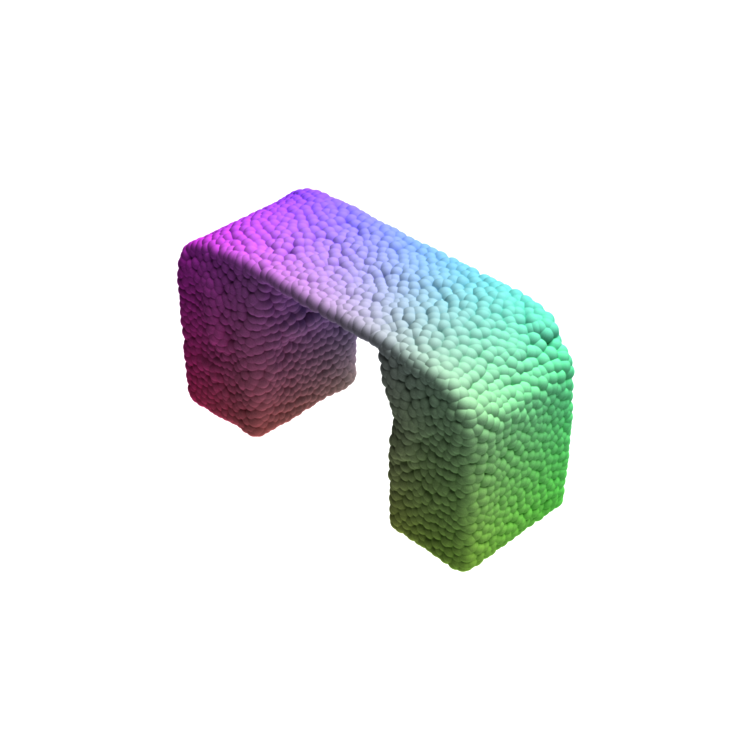}}{} \vspace{2.0pt}

\rotatebox{90}{\hspace{0.1cm}\footnotesize{\emph{The target has}}}
\rotatebox{90}{\hspace{0.1cm}\footnotesize{\emph{a smaller skirt}}}
\jsubfig{\includegraphics[height=1.75cm, trim={4.0cm 4.0cm 4.0cm 4.0cm}, clip]{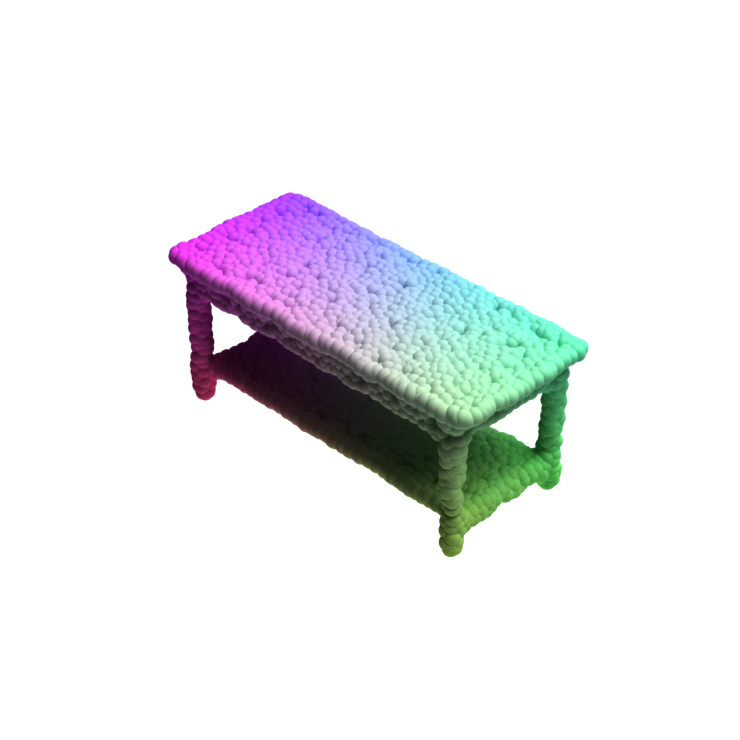}}{\footnotesize {Input}}\hfill
\jsubfig{\includegraphics[height=1.75cm, trim={4.0cm 4.0cm 4.0cm 4.0cm}, clip]{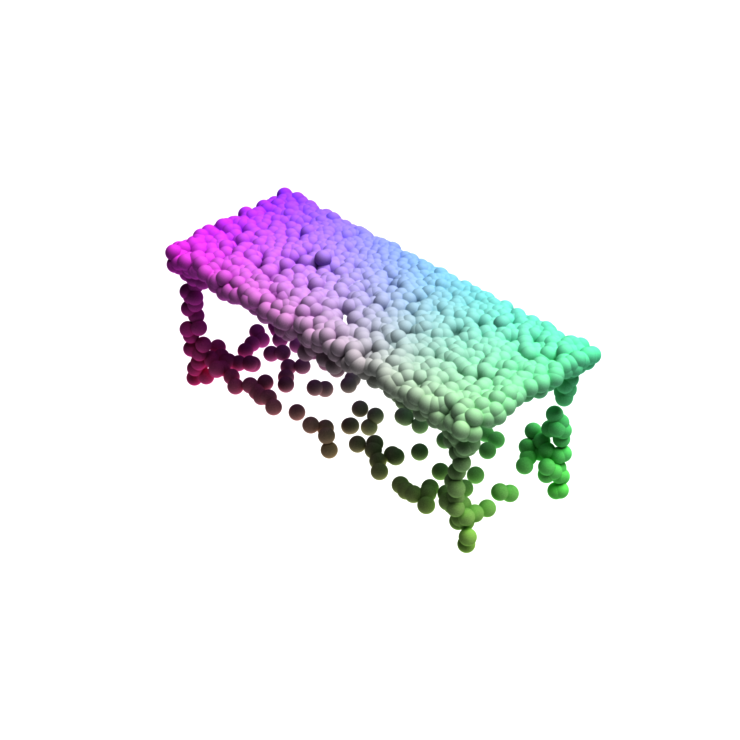}}{\footnotesize {Changeit3D}}\hfill
\jsubfig{\includegraphics[height=1.75cm, trim={4.0cm 4.0cm 4.0cm 4.0cm}, clip]{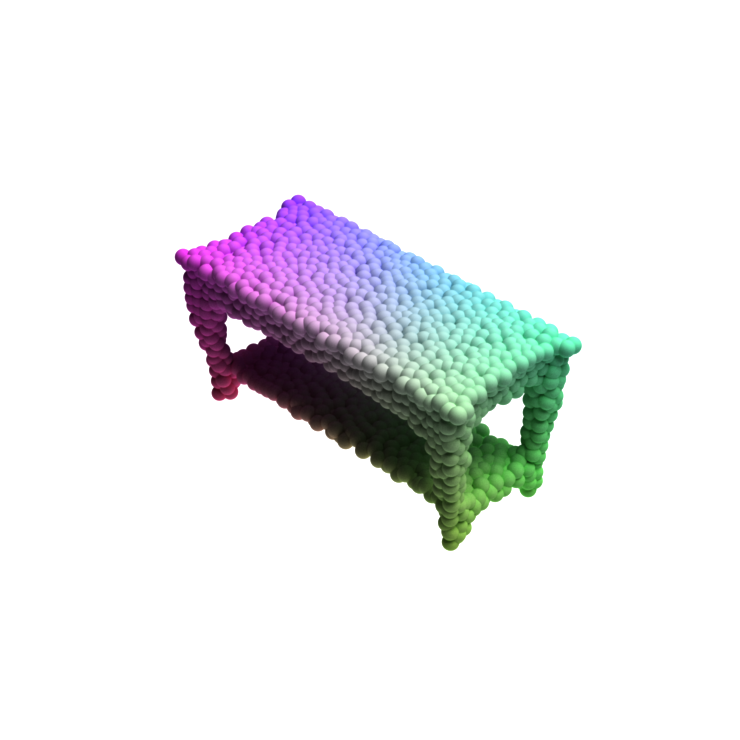}}{\footnotesize {Spice-E}}\hfill
\jsubfig{\includegraphics[height=1.75cm, trim={4.0cm 4.0cm 4.0cm 4.0cm}, clip]{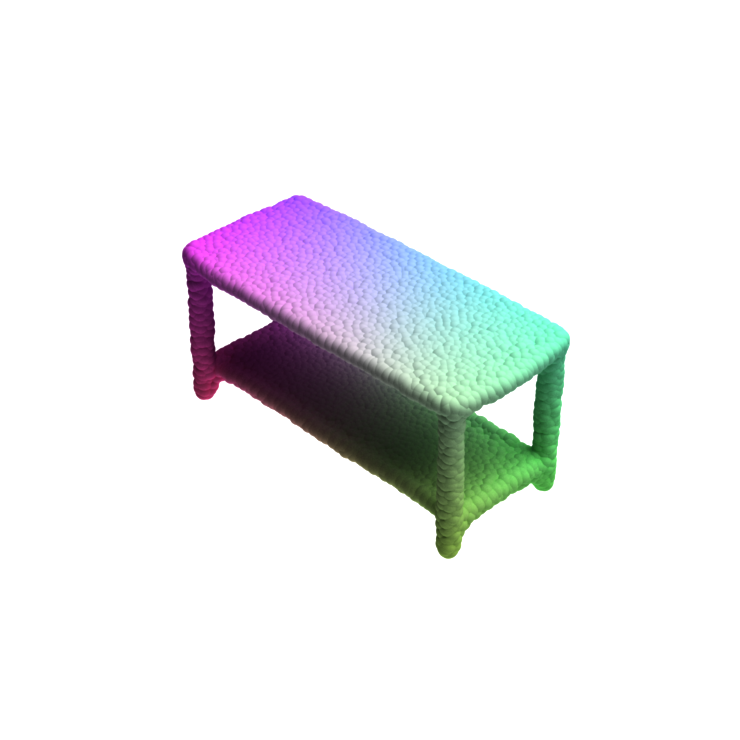}}{\footnotesize {Ours}}

\vspace{-5pt} 
\caption{
\textbf{Qualitative comparison}. We compare our method's outputs to those of ChangeIt3D \cite{achlioptas2023shapetalk} and Spice-E \cite{sella2024spice}. As illustrated above, our method outperforms these baselines in terms of edit fidelity, identity preservation and overall visual quality. Outputs for all methods are presented in their default resolution (2048 for ChangeIt3D, 4096 for Spice-E and our method).
}

\vspace{-0.8em}

\label{fig:comparisons_fig}
\end{figure}

\subsection{Quantitative Evaluation}
\label{sec:quantitative}

We perform a quantitative analysis over the ShapeTalk and l-ShapeTalk test splits, comparing performance against ChangeIt3D~\cite{achlioptas2022changeit3d} and Spice-E~\cite{sella2024spice}. To the best of our knowledge, these are the only methods that explicitly address the task of text-guided fine-grained shape editing which have publicly-available code and pretrained models. Note that their models are also trained on the ShapeTalk dataset.  To ensure a fair comparison, Spice-E's and our method's result point clouds were downsampled to ChangeIt3D's 2048 point output resolution prior to metric calculation.
Results are reported in Table \ref{tab:metrics}. 
As illustrated in the table, our method outperforms prior work across all metrics. We achieve significant improvements particularly in metrics that quantify shape quality (FPD and CD) and identity preservation (GD and l-GD). This can be attributed to the localized nature of our approach, which enables producing results that better resemble the input. Regarding edit fidelity, we wish to emphasize that prior work achieve a \emph{negative} average CLIP direction similarity score over l-Shapetalk while our method yields a positive value, illustrating that our shapes are edited in a manner that faithfully adheres to the target prompt according to the CLIP embeddings.

\begin{table}[t]
    \centering
    \resizebox{0.85\linewidth}{!}{ %
    \begin{tabular}{l c c c}
        \toprule
        Method & Changeit3D & Spice-E & Ours \\
        \midrule
        Preference \% & 0.09 & 0.16 & \textbf{0.75} \\
        \bottomrule
    \end{tabular}
    }
    \vspace{-6pt}
    \caption{\textbf{User Study.} The results indicate the percentage of the time for which each method was preferred. As illustrated above, the user study revealed a clear preference for our results.}
    \label{tab:user_study}
\end{table}

\begin{table*}[t]
\centering
\setlength{\tabcolsep}{2.5pt}
\def\arraystretch{1.0}
\begin{tabularx}{\textwidth}{lcccccccccccccc}
\toprule
 &           & \multicolumn{6}{c}{Shapetalk} &  & \multicolumn{6}{c}{l-Shapetalk} 
                          \\ 
\cmidrule(lr){3-8} \cmidrule(lr){10-15} 
Metric && $\text{CLIP}_{Sim}\uparrow$& $\text{CLIP}_{Dir}\downarrow$ & 
GD$\downarrow$ & CD$\downarrow$ & FPD$\downarrow$ & l-GD$\downarrow$ && $\text{CLIP}_{Sim}\uparrow$& $\text{CLIP}_{Dir}\downarrow$ & GD$\downarrow$ & CD$\downarrow$ & FPD$\downarrow$ & l-GD$\downarrow$ \\ \midrule
\inpaintingModel{} Only && 0.25&	1.01&	1.23&	0.09	&42.81&	0.51
 && 0.26&	1.01&	1.31&	0.12&	73.45&	0.39
\\  
$t_r = T$ && 0.25&	1.01 &	0.82&	0.10&	56.38&	0.14
 && 0.25&	1.01 &	0.92&	0.13	&102.52&	0.14
\\ 
Ours && \textbf{0.26}&	\textbf{0.99} & \textbf{0.34}	& \textbf{0.05}&	\textbf{33.64}	& \textbf{0.07}
 && \textbf{0.27}	& \textbf{0.99} &	\textbf{0.29}&	\textbf{0.04}	&\textbf{13.51}	&\textbf{0.05}
\\ 
\bottomrule
\end{tabularx}
\vspace{-8pt}
\caption{\textbf{Quantitative Comparison Breakdown}. We compare our outputs against \inpaintingModel{} (no inference time coordinate blending) and $t_r = T$ (inpainting at each inference step) across several metrics (Section \ref{sec:metrics}) over both the ShapeTalk and l-ShapeTalk test sets; additional details are provided in Section \ref{sec:ablations}. As can be observed from the table above, our inference-time coordinate blending mechanism significantly improves our model's performance, particularly in terms of identity preservation.}
\label{tab:metrics_ablations}
\end{table*}

\smallskip \noindent \textbf{User Study.}
To also evaluate to what extent our results are grounded in human perception,  we complemented our quantitative analysis with a perceptual user study. In this study, 60 participants were asked to answer 15 questions; two different variants were distributed, each containing random samples from the l-Shapetalk test set. In each question users were shown a text prompt, an input point cloud and editing results produced by ChangeIt3D, Spice-E and our method, arranged in random order. Users were instructed to \textit{``select the shape which best represents the input text while also maintaining the shape of the input point cloud as best as possible"}. Users were also instructed to select the shape that best matches the input if none of the outputs match the text prompt.
Results for this perceptual study are summarized in Table \ref{tab:user_study} and show a clear preference for our results. %

\smallskip \noindent We also provide a qualitative comparison in Figure \ref{fig:comparisons_fig}. These results further highlight our improved performance in terms of edit fidelity (\emph{e.g.}, second row from the bottom), structural fidelity (\emph{e.g.}, third row from the top, where our output better resembles a plausible chair), and identity preservation (\emph{e.g.}, first row, where only our output successfully preserves the lamp's structure outside the shade region). Additional qualitative comparisons against other paradigms enabling text-guided shape editing such as optimization based methods (Vox-E \cite{sella2023vox}, Fantasia3D \cite{chen2023fantasia3d}) and image editing based pipelines (InstructPix2Pix \cite{xu2023instructp2p} + One-2-3-45++ \cite{liu2024one}) are provided in the supplementary material.

\begin{figure} %

\rotatebox{90}{}
\rotatebox{90}{\hspace{0.1cm}\footnotesize{\emph{Thin backrest}}}
\hfill
\jsubfig{\includegraphics[height=1.7cm, trim={4.0cm 4.0cm 4.0cm 4.0cm}, clip]{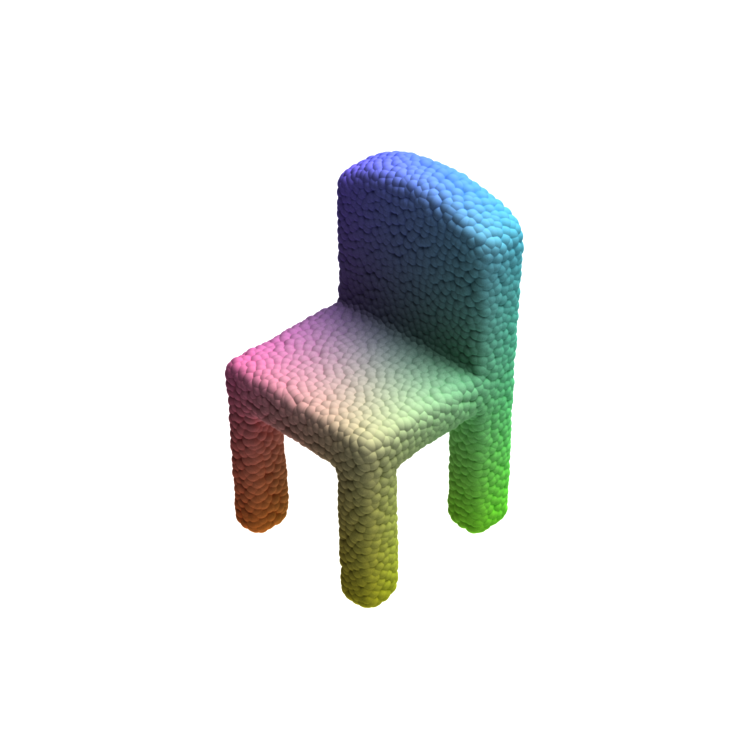}}{} \hfill
\jsubfig{\includegraphics[height=1.7cm, trim={4.0cm 4.0cm 4.0cm 4.0cm}, clip]{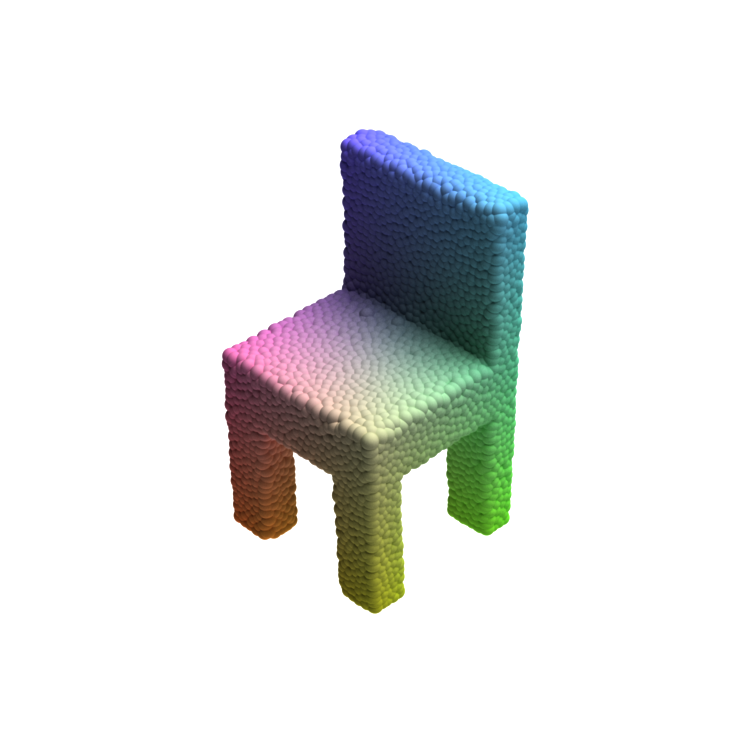}}{}\hfill
\jsubfig{\includegraphics[height=1.7cm, trim={4.0cm 4.0cm 4.0cm 4.0cm}, clip]{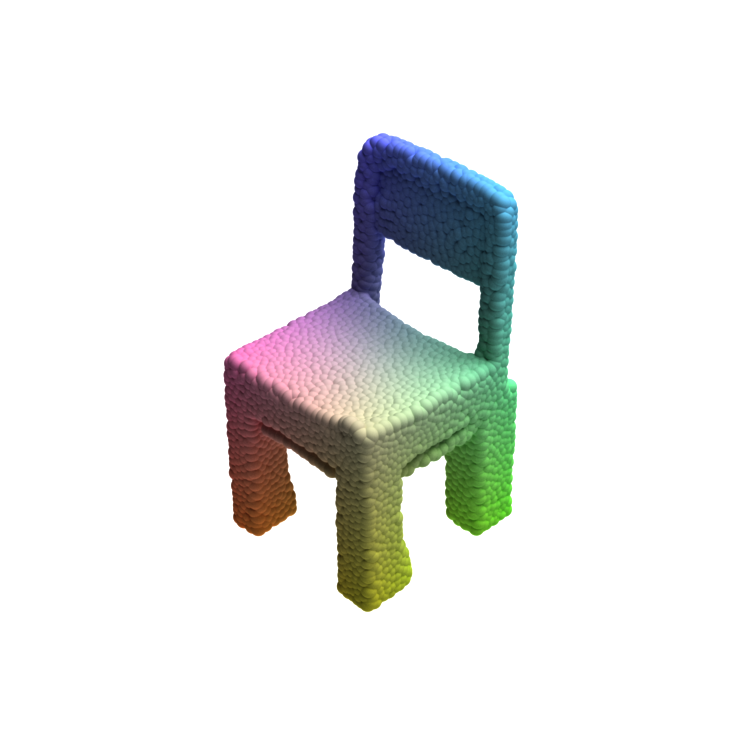}}{}\hfill
\jsubfig{\includegraphics[height=1.7cm, trim={4.0cm 4.0cm 4.0cm 4.0cm}, clip]{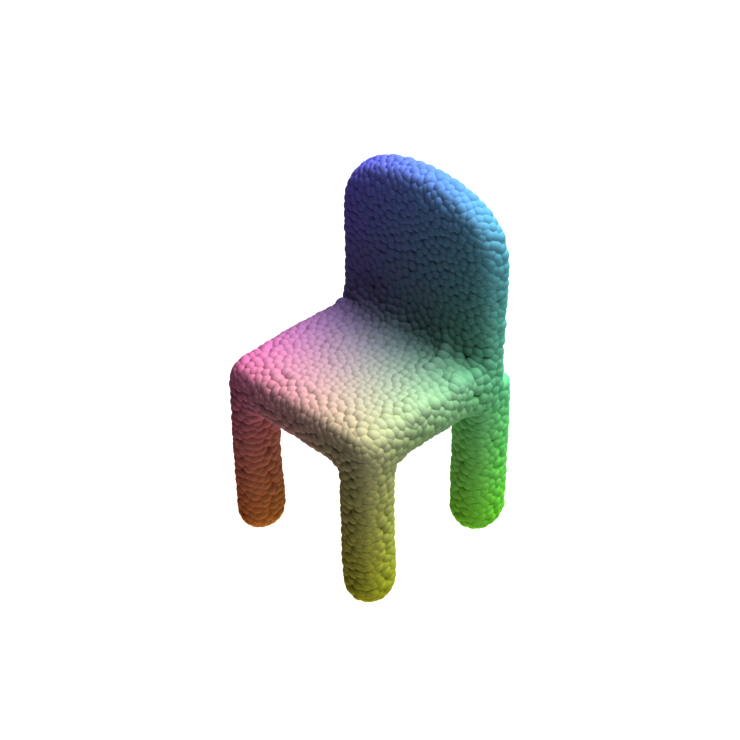}}{\footnotesize {}}
\\ \vspace{-7.0pt}

\rotatebox{90}{\hspace{0.5cm}\footnotesize{\emph{The legs}}}
\rotatebox{90}{\hspace{0.3cm}\footnotesize{\emph{are straight}}}
\hfill
\jsubfig{\includegraphics[height=1.7cm, trim={4.0cm 4.0cm 4.0cm 4.0cm}, clip]{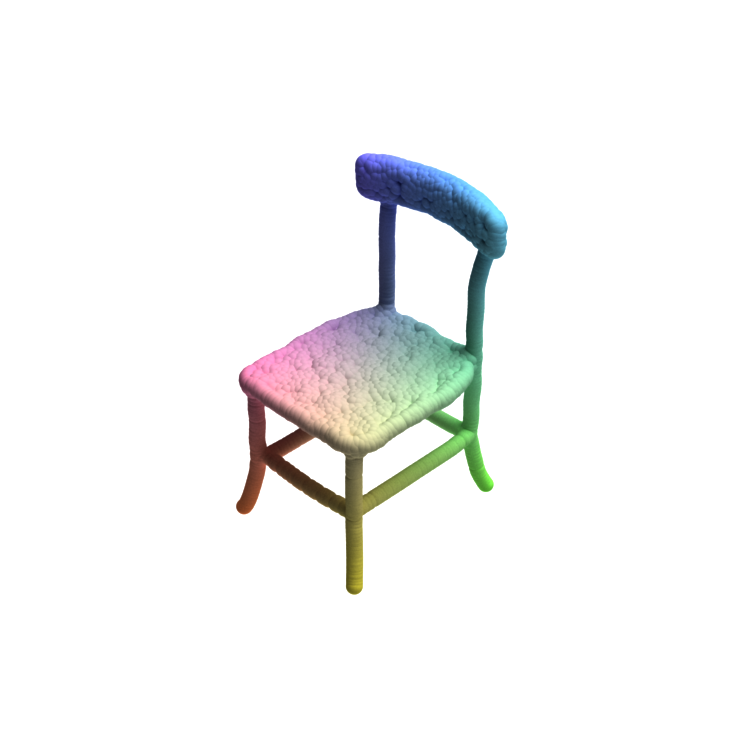}}{} \hfill
\jsubfig{\includegraphics[height=1.7cm, trim={4.0cm 4.0cm 4.0cm 4.0cm}, clip]{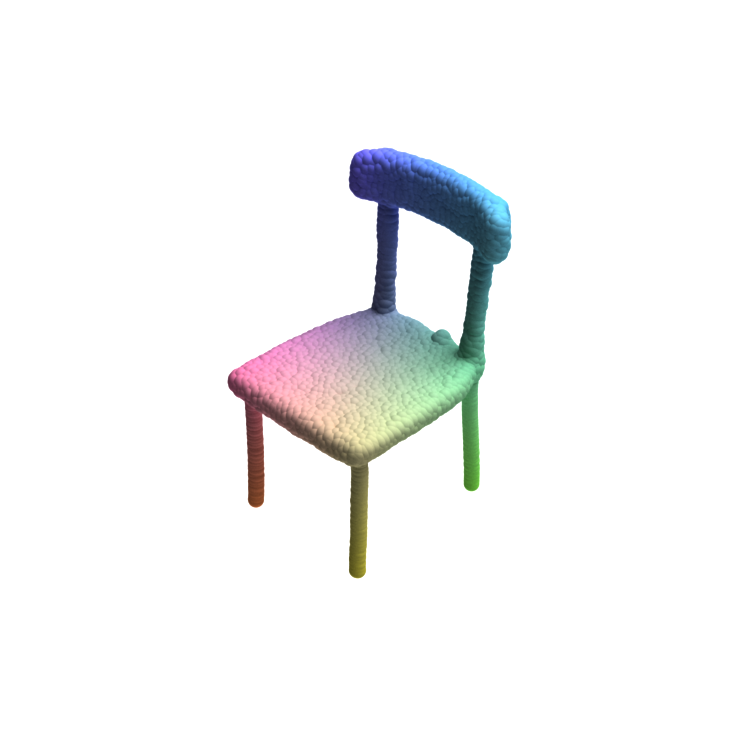}}{}\hfill
\jsubfig{\includegraphics[height=1.7cm, trim={4.0cm 4.0cm 4.0cm 4.0cm}, clip]{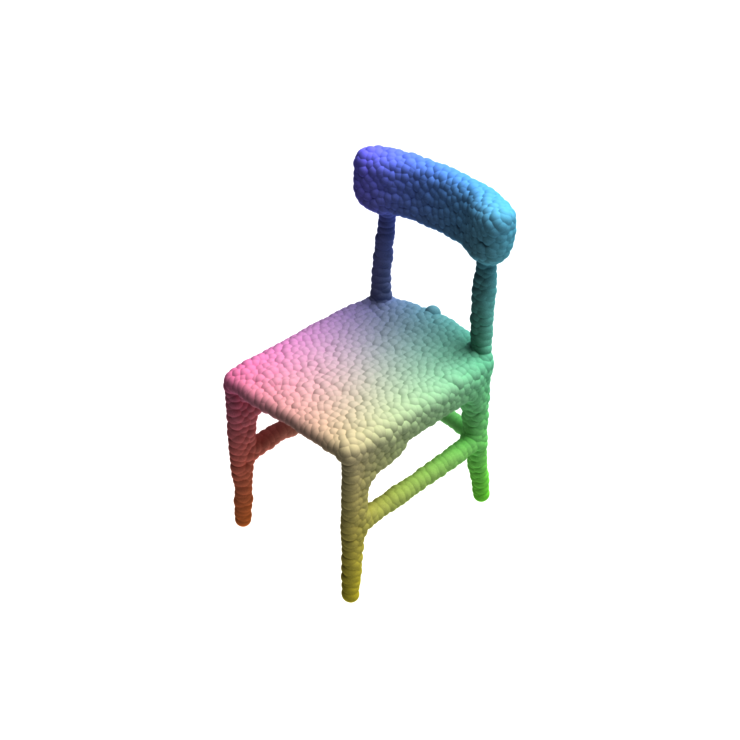}}{}\hfill
\jsubfig{\includegraphics[height=1.7cm, trim={4.0cm 4.0cm 4.0cm 4.0cm}, clip]{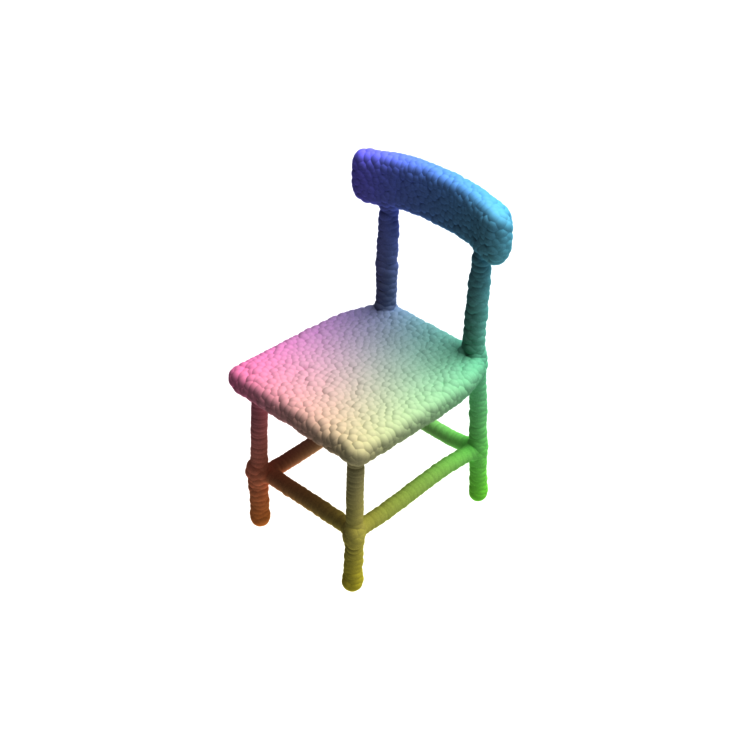}}{\footnotesize {}}
\\ \vspace{-12.0pt}

\rotatebox{90}{}
\rotatebox{90}{\hspace{0.2cm}\footnotesize{\emph{Smaller top}}}
\hfill
\jsubfig{\includegraphics[height=1.7cm, trim={4.0cm 4.0cm 4.0cm 4.0cm}, clip]{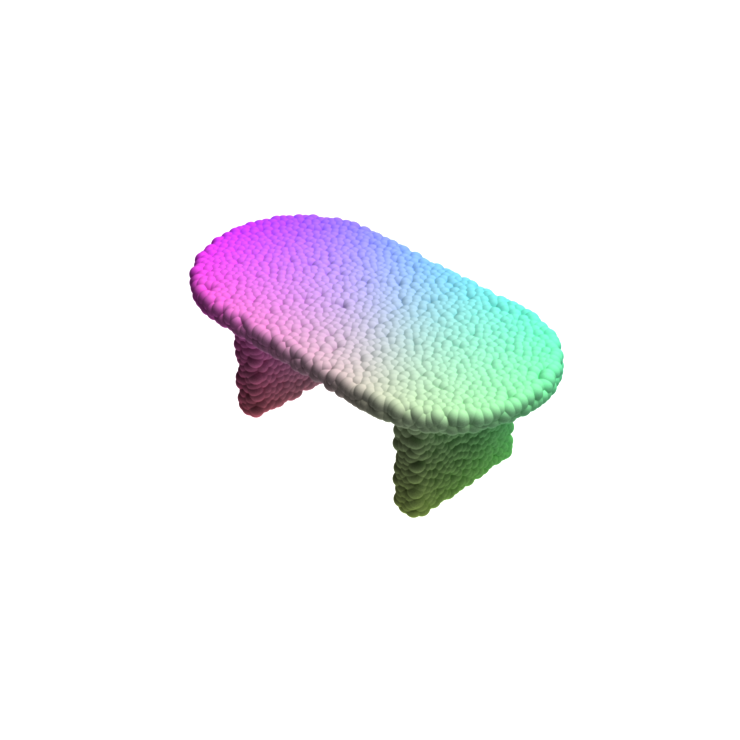}}{} \hfill
\jsubfig{\includegraphics[height=1.7cm, trim={4.0cm 4.0cm 4.0cm 4.0cm}, clip]{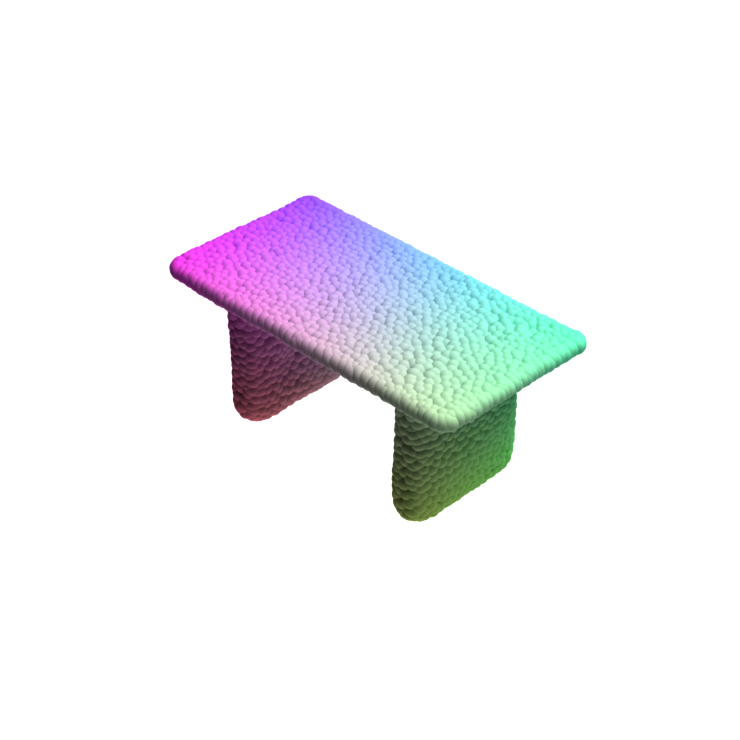}}{}\hfill
\jsubfig{\includegraphics[height=1.7cm, trim={4.0cm 4.0cm 4.0cm 4.0cm}, clip]{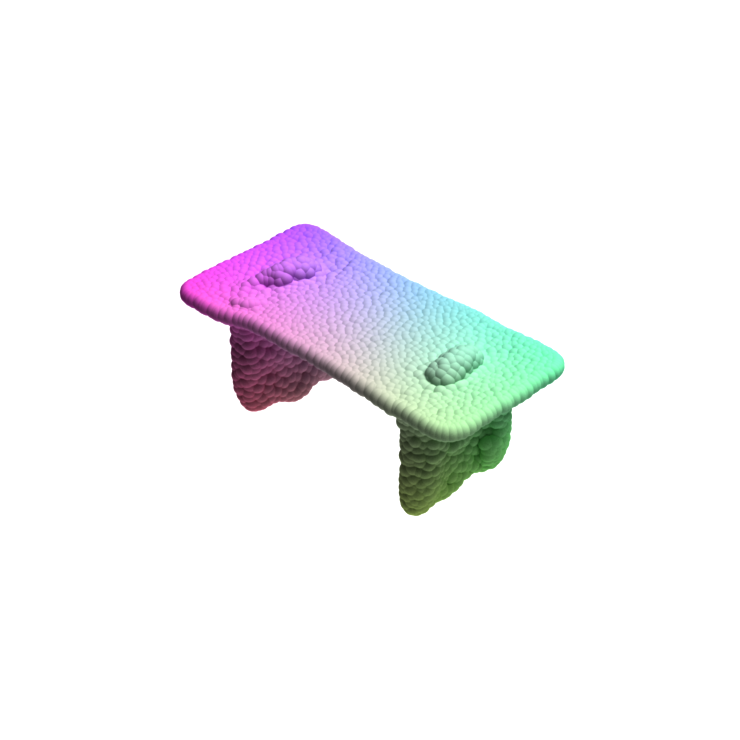}}{}\hfill
\jsubfig{\includegraphics[height=1.7cm, trim={4.0cm 4.0cm 4.0cm 4.0cm}, clip]{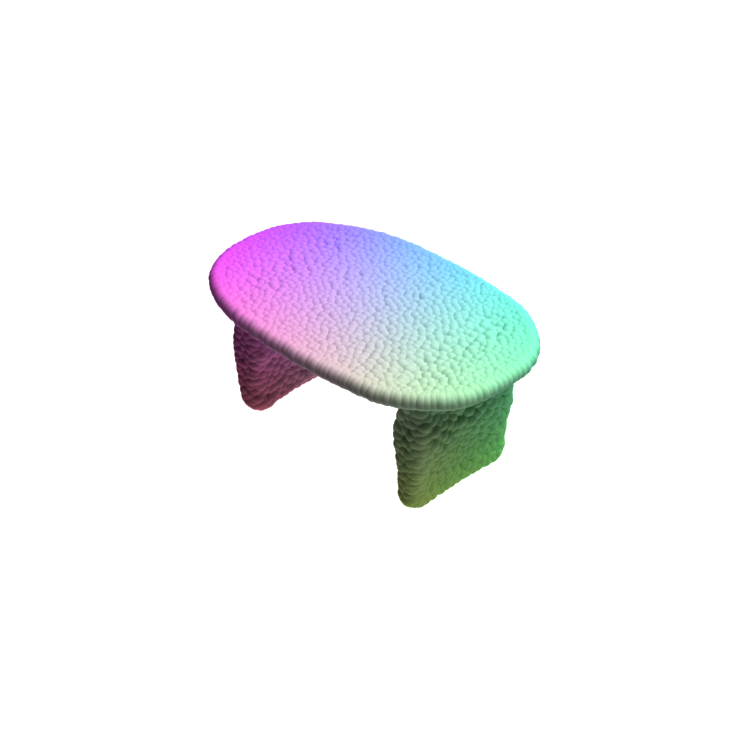}}{\footnotesize {}}
\\ \vspace{-10.0pt}

\rotatebox{90}{\hspace{0.1cm}\footnotesize{\emph{The table has}}}
\rotatebox{90}{\hspace{0.4cm}\footnotesize{\emph{thick legs}}}
\hfill
\jsubfig{\includegraphics[height=1.7cm, trim={4.0cm 4.0cm 4.0cm 4.0cm}, clip]{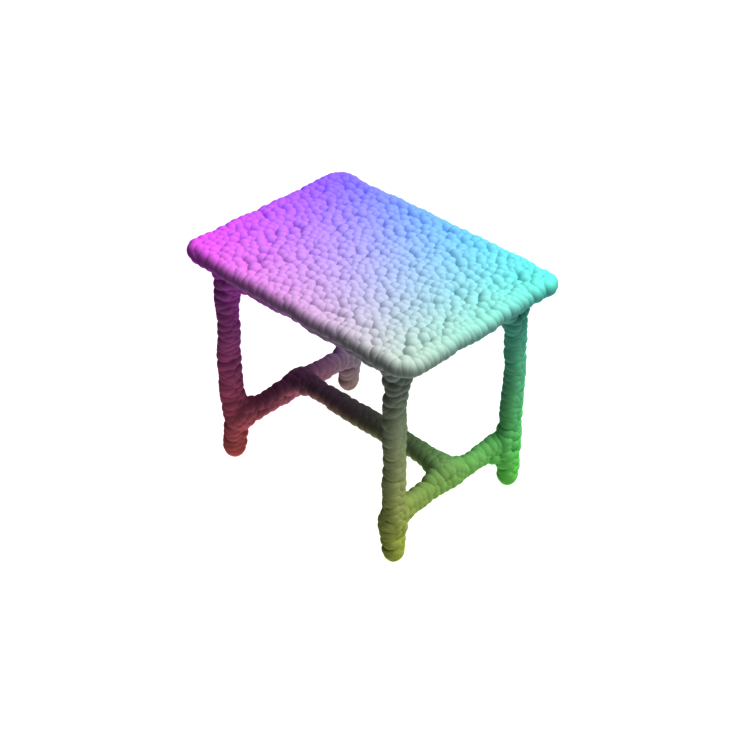}}{} \hfill
\jsubfig{\includegraphics[height=1.7cm, trim={4.0cm 4.0cm 4.0cm 4.0cm}, clip]{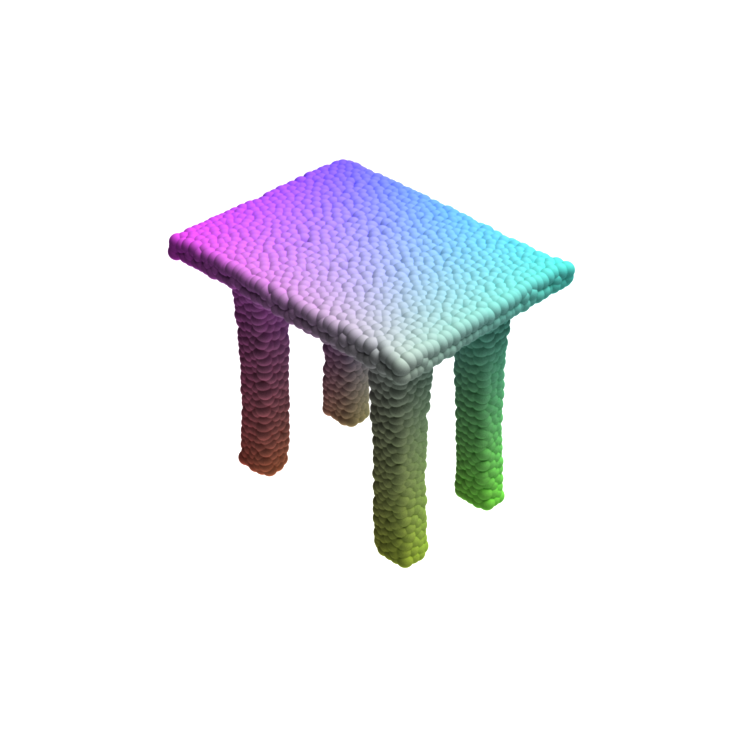}}{}\hfill
\jsubfig{\includegraphics[height=1.7cm, trim={4.0cm 4.0cm 4.0cm 4.0cm}, clip]{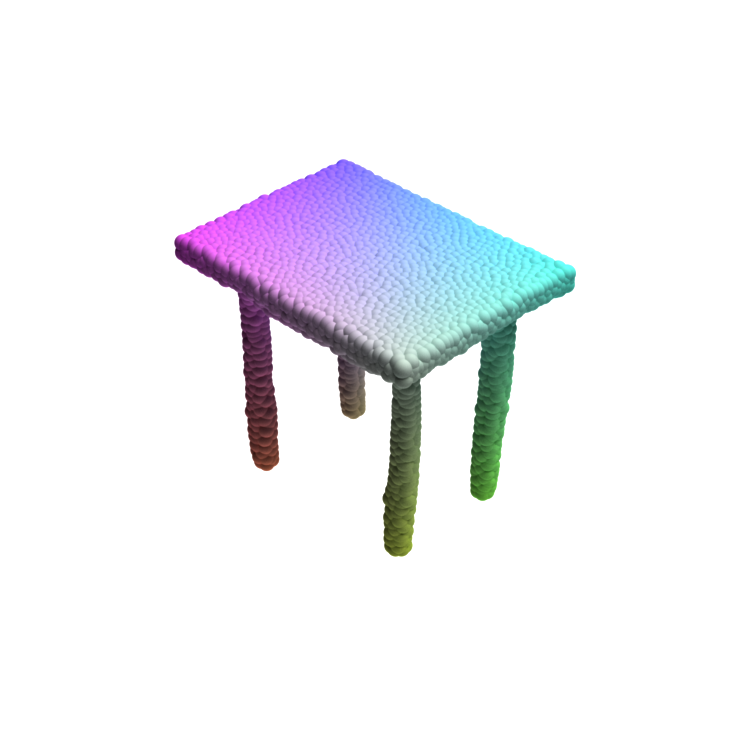}}{}\hfill
\jsubfig{\includegraphics[height=1.7cm, trim={4.0cm 4.0cm 4.0cm 4.0cm}, clip]{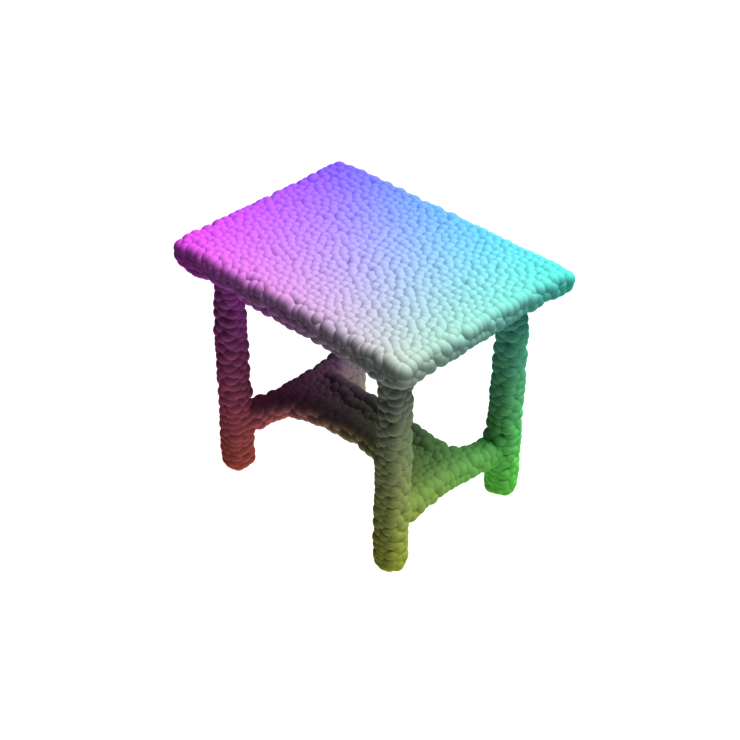}}{\footnotesize {}}
\\ \vspace{-10pt}

\rotatebox{90}{}
\rotatebox{90}{\footnotesize{\emph{Closed shade}}}
\hfill
\jsubfig{\includegraphics[height=1.7cm, trim={4.0cm 4.0cm 4.0cm 4.0cm}, clip]{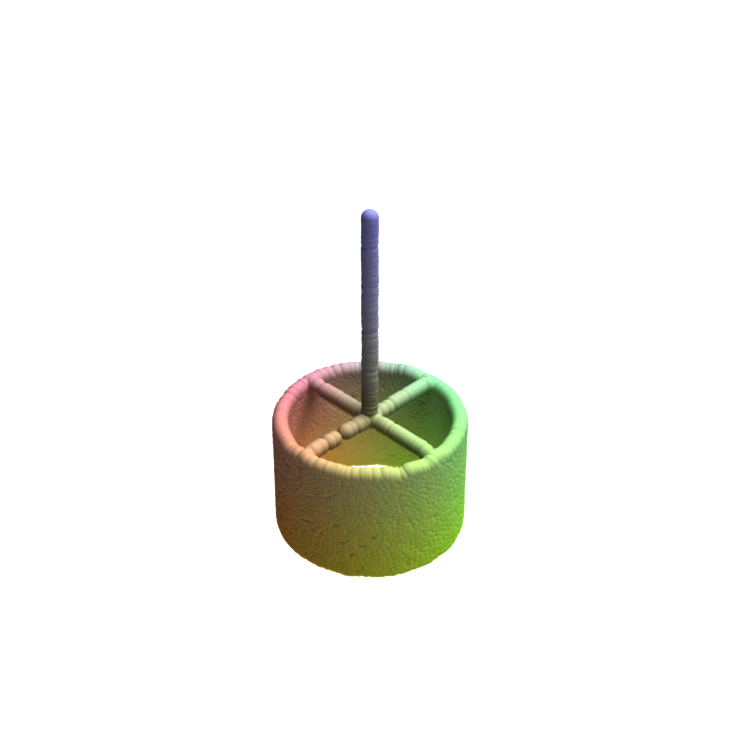}}{} \hfill
\jsubfig{\includegraphics[height=1.7cm, trim={4.0cm 4.0cm 4.0cm 4.0cm}, clip]{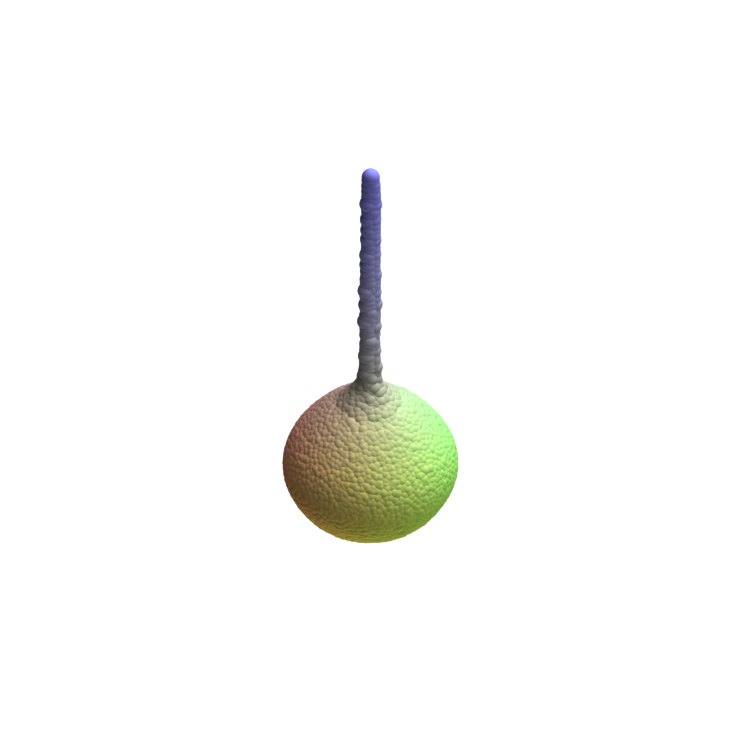}}{}\hfill
\jsubfig{\includegraphics[height=1.7cm, trim={4.0cm 4.0cm 4.0cm 4.0cm}, clip]{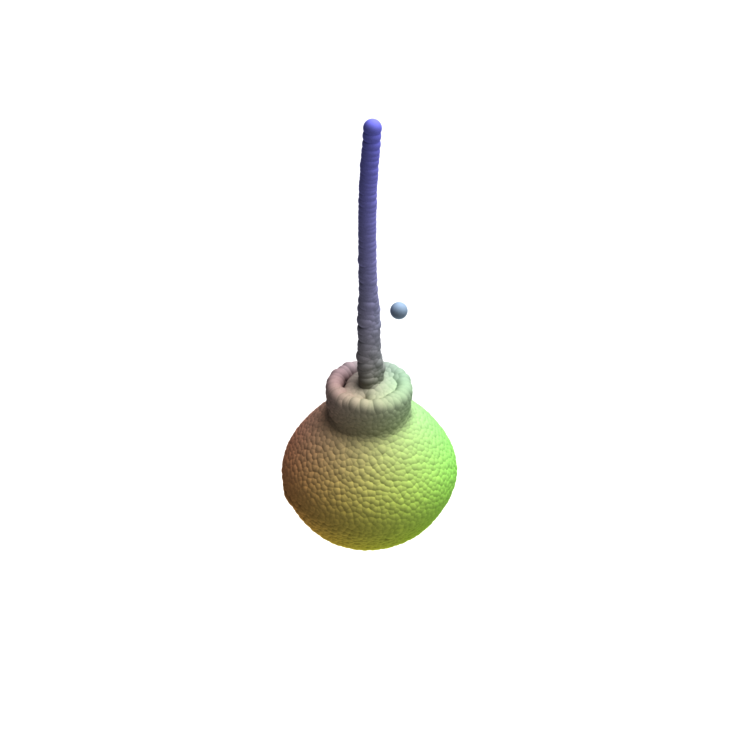}}{}\hfill
\jsubfig{\includegraphics[height=1.7cm, trim={4.0cm 4.0cm 4.0cm 4.0cm}, clip]{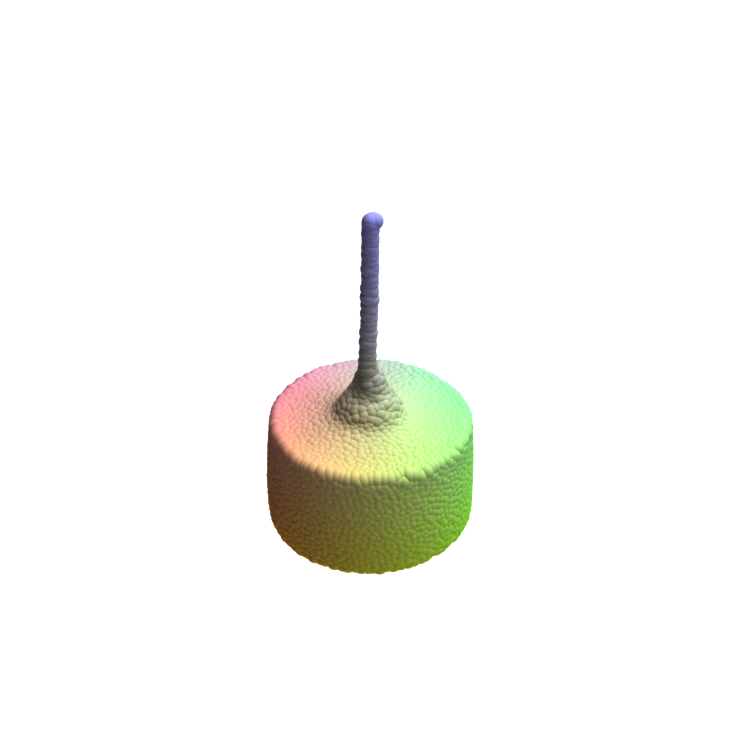}}{\footnotesize {}}
\\ \vspace{-10pt}

\rotatebox{90}{\hspace{0.4cm}\footnotesize{\emph{The base}}}
\rotatebox{90}{\hspace{0.3cm}\footnotesize{\emph{is smaller}}}
\hfill
\jsubfig{\includegraphics[height=1.7cm, trim={4.0cm 4.0cm 4.0cm 4.0cm}, clip]{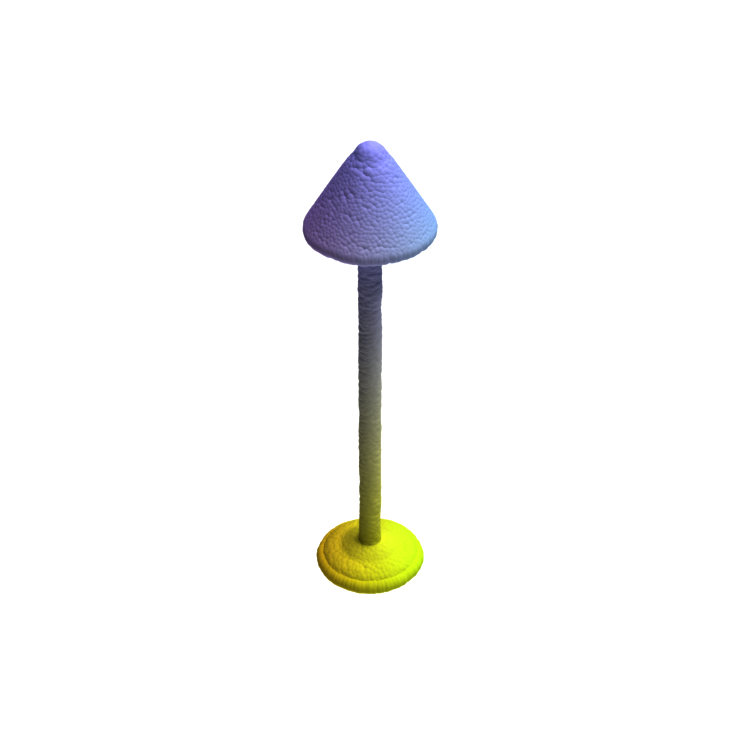}}{\footnotesize {Input}} \hfill
\jsubfig{\includegraphics[height=1.7cm, trim={4.0cm 4.0cm 4.0cm 4.0cm}, clip]{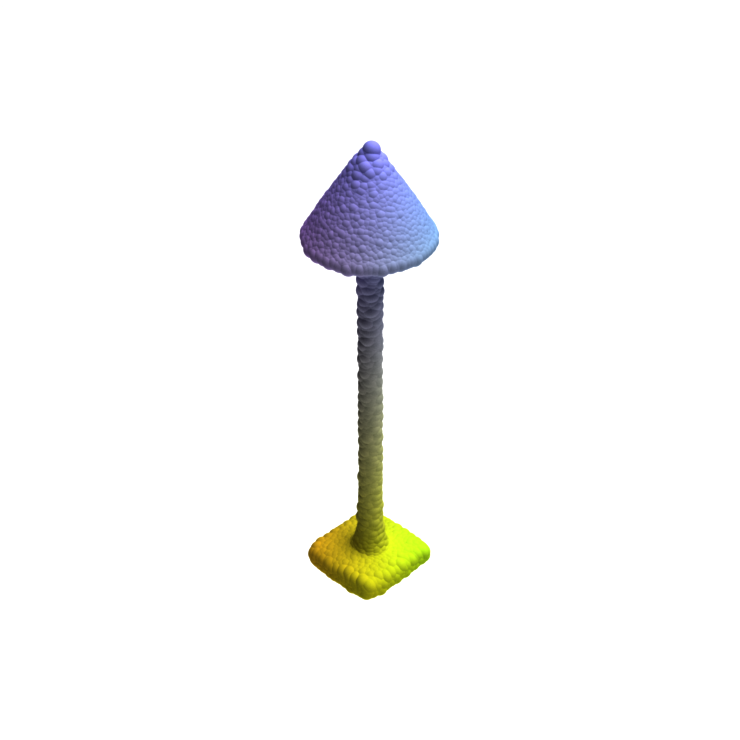}}{\footnotesize {\inpaintingModel{} only}}\hfill
\jsubfig{\includegraphics[height=1.7cm, trim={4.0cm 4.0cm 4.0cm 4.0cm}, clip]{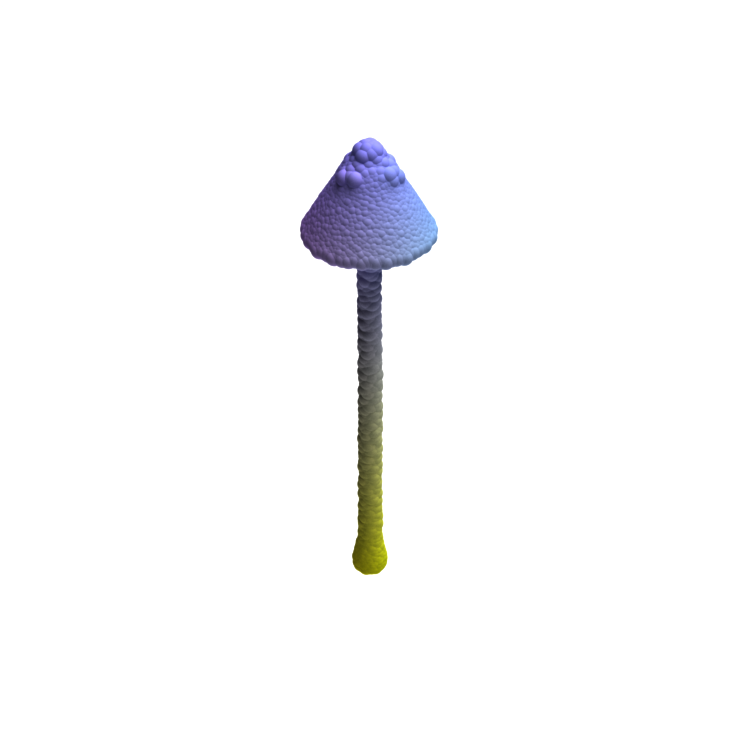}}{\footnotesize { $t_r=T$}}\hfill
\jsubfig{\includegraphics[height=1.7cm, trim={4.0cm 4.0cm 4.0cm 4.0cm}, clip]{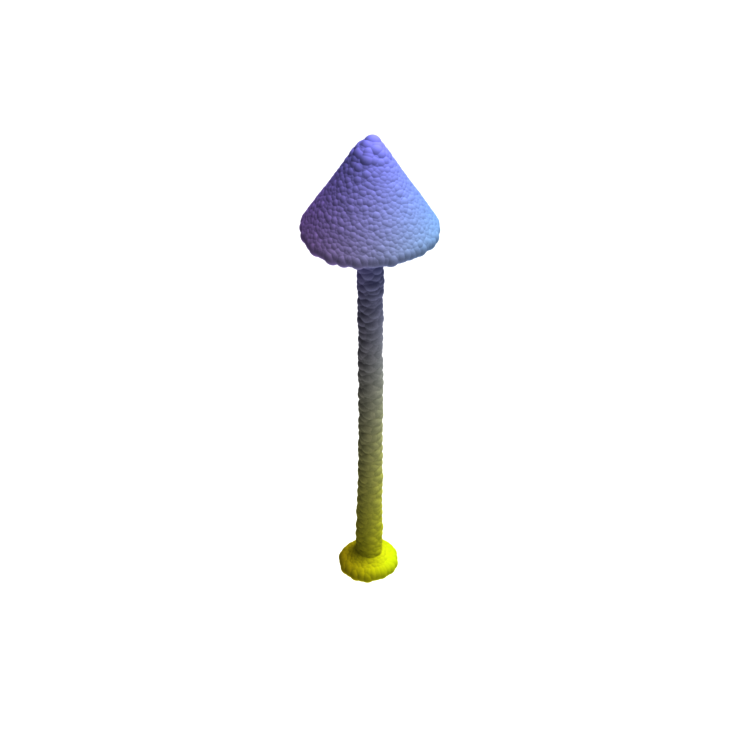}}{\footnotesize {Ours}}
\vspace{-5pt} 
\caption{\textbf{Qualitative ablation results}, obtained for samples from the ShapeTalk dataset. As detailed in Section \ref{sec:ablations}, we compare our method to \inpaintingModel{} (i.e. removing our inference-time coordinate blending algorithm) and our method when $t_r$ is set to $T$ (i.e. inpainting is performed at every inference step).
}

\label{fig:ablations}
\end{figure}

\label{sec:comparisons}

\subsection{Ablation Study}
\label{sec:ablations}
We provide an ablation study, quantifying to what extent our coordinate blending inference-time mechanism effects  performance. Specifically, we experiment with two variants: (i) performing coordinate blending at every step of the inference (setting $t_r = T$), and (ii) ablating this inference algorithm altogether (denoted as \inpaintingModel{} only) %

Results are reported in Table \ref{tab:metrics_ablations}; several examples are illustrated in Figure \ref{fig:ablations}. 
As illustrated in the table, our proposed coordinate blending mechanism is particularly important in terms of identity preservation, yielding significantly lower GD and l-GD values, in comparison to the \inpaintingModel{} Only experiment. This is also evident in the qualitative results which highlight that an inpainting only model cannot produce an edited part that aligns with the text prompt while also resembling the original part (e.g. second and fourth rows). Results also show that performing coordinate blending at each step ($t_r = T$) diminishes the method's ability to preserve identity within the edited region, yielding higher GD values. This is somewhat unsurprising as our method is essentially blind to the original part when $t_r$ is set to $T$. We also observe setting $t_r = T$ also hurts structural fidelity, yielding the highest FPD and CD values. This can also be seen in the third and fifth rows from the top of Figure \ref{fig:ablations}, in which $t_r = T$ produces noticeable artifacts. We hypothesize that this deterioration in performance is the result of a significantly altered inference process in relation to the original process for which \inpaintingModel{} was trained. %

\section{Conclusion}
We introduced BlendedPC, a new approach for performing fine-grained, localized editing of point clouds using textual instructions. Technically, we formulated localized semantic editing as an inpainting problem, and tackled it by leveraging a powerful diffusion model coupled with a novel inference-time algorithm. %
Our technique enables localized editing that is both aligned with the target text and remains faithful to the original point cloud at a level unmatched by concurrent work, bringing us a step closer to the goal of democratizing 3D content editing and empowering users to shape 3D worlds with greater ease and creativity.

{
    \small
    \bibliographystyle{ieeenat_fullname_iccv}
    \bibliography{main}
}

\maketitlesupplementary

\medskip
\medskip
We refer readers to the interactive visualizations at \href{index.html}{index.html}. In this document, we provide  implementation details (Section \ref{sec:details}), additional experiments and comparisons (Section \ref{sec:experiments}) and 
a discussion of limitations (Section \ref{sec:limitations}).

\section{Implementation Details}
\label{sec:details}
\subsection{Training}
\label{sec:inpaint_e_supp}

We trained our model for each category using object-specific ShapeTalk and l-ShapeTalk subsets. To evaluate generalization, we also trained a unified model across all three categories (Chair, Table, and Lamp). For all models, we used a batch size of 6 and a learning rate of $11 \times 10^{-4}$. The number of epochs, iterations, and training hours for each category are detailed in Table \ref{tab:training_times}. To encourage the network to rely more on structural guidance rather than text, we dropped the text guidance with a probability of 0.5, replacing the textual prompt with an empty string. As explained in Section 3.2.2 of the main paper, we also replaced the conditional point cloud input with the target point cloud with a probability of 0.1 to support our Inversion-Free Coordinate Blending mechanism. All training was conducted on a single RTX A5000 GPU (24GB VRAM) for each model.

\medskip \noindent \textbf{Data.} To generate the partial point clouds used as guidance during training and to construct the l-ShapeTalk dataset, we used the baseline Llama 3 \cite{dubey2024llama} model provided by \href{https://github.com/unslothai/unsloth}{unsloth}. Specifically, we instructed Llama 3 with the following prompt for chairs:

\begin{verbatim}
### Instruction: Return a single word 
using only one of the following options. 
Options: back, leg, arm, seat, unknown. 
### Input: What part of a chair does the 
next utterance describe? 
If none of the parts are described in 
the utterance return unknown. Utterance: 
it has larger height. 
### Response: unknown (EOT-token)
\end{verbatim}

We adjusted the prompt for each category, instructing the language model to extract appropriate part names specific to that category. To enhance accuracy, we fine-tuned the model for this task using 200 manually labeled examples. 
Samples where the model returned \textit{"unknown"} were excluded from the training set, leading to the creation of the l-ShapeTalk dataset.

For evaluation, we applied the same method but removed the option to return \textit{"unknown"}, requiring the language model to extract the part name it deemed most suitable to convey the text prompt.

The part name in each l-ShapeTalk sample was provided as input to a segmentation model to extract binary masks. For segmentation, we used the PyTorch implementation of PointNet by \href{https://github.com/axinc-ai/ailia-models/tree/master/point_segmentation/pointnet_pytorch}{ailia-models}. These binary masks were inverted and then multiplied element-wise with the target point cloud to generate the guidance partial point cloud.

\begin{table}[t]
    \centering
    \resizebox{0.85\linewidth}{!}{ %
    \begin{tabular}{l c c c}
        \toprule
        Model & Iterations & Epochs & Training Time (hours) \\
        \midrule
        Chair & 13.5 M & 250 & 119 \\
        Table & 12.1 M & 250 & 107 \\
        Lamp & 8.2 M & 250 & 74  \\
        Airplane & 0.9 M & 250 & 52  \\
        Guitar & 0.4 M & 250 & 31  \\
        Knife & 0.3 M & 250 & 25  \\
        Cap & 0.2 M & 250 & 18  \\
        Skateboard & 0.1 M & 250 & 16  \\
        Unified & 13.5 M & 100 & 98  \\
        \bottomrule
    \end{tabular}
    }
    \vspace{-2pt}
    \caption{\textbf{Training time.} We report the number of iterations, epochs and the overall training time for each category.}
    \label{tab:training_times}
\end{table}

\subsection{Evaluation}

ShapeTalk prompts often describe relationships between objects (e.g., \textit{"it has a taller backrest"}), which makes them unsuitable for evaluating individual instances. To address this, we use Llama 3 to translate ShapeTalk prompts into more \textit{descriptive} prompts, such as converting \textit{"it has a taller backrest"} to \textit{"a chair with a tall backrest"}.

We use two CLIP-based \cite{radford2021learning} metrics, $\text{CLIP}_{Sim}$ and $\text{CLIP}_{Dir}$, to evaluate edit fidelity. For the CLIP image encoder, we rendered a single image for each point cloud from a consistent viewpoint.

$\text{CLIP}_{Sim}$ evaluates the similarity between the final output and its textual description. We encoded the rendered image of the output point cloud and the descriptive prompt using their respective CLIP encoders and calculated the cosine similarity between the resulting encodings. 

$\text{CLIP}_{Dir}$ assesses the semantic direction in CLIP space, capturing the relationship between both the guidance and output shapes. We encoded a rendered image of the input $E_I(I_{in})$ and output point clouds $E_I(I_{out})$, along with the descriptive prompt $E_T(T_{edit})$ and a simple prompt describing the general object shape (e.g., \textit{"a chair"}) $E_T(T_{general})$. The differences between the two text encodings ($\Delta T = E_T(T_{edit}) - E_T(T_{general})$) and the two image encodings ($\Delta I = E_I(I_{out}) -E_I(I_{in})$) were calculated, and their cosine similarity, subtracted from $1$,  was used as the the metric:
\begin{equation}
    \text{CLIP}_{Dir} = 1 - \frac{\Delta I \Delta T}{|\Delta I| | \Delta T |}
\end{equation}

All experiments were conducted using the \textit{"openai/clip-vit-base-patch32"} model, accessed through the \href{https://huggingface.co/docs/transformers/en/model_doc/clip}{Hugging Face} library.

l-GD measures the geometric distance between the input and output shapes in the ``non-edit'' region. Contrary to other non CLIP based metrics we opted to implement the calculation of this metric ourselves. The reasoning behind this is that the ChangeIt3D implementation produced l-GD values that are higher than their matching GD scores, which we felt was unintuitive. These higher values could possibly be caused by noisy segmentation, faulty part name extraction or a varying method of normalization in comparison to GD. In our implementation we use an LLM (llama3) to extract part names from the edit prompts and use an off-the-shelf segmentation method (PointNet) to extract part segmentations before calculating the chamfer distance. %
To ensure that our implementation is consistent with the trend produced by the ChangeIt3D script, we compared these two implementations and present the results in Table \ref{tab:lgd}. These results show that while our implementation produces lower values across the board (which are lower then their matching GD scores, presented in the main paper) the overall trend is consistant, and our method acheives the lowest l-GD scores in both implementations.

To calculate the rest of the metrics used in the quantitative analysis we used the \href{https://github.com/optas/changeit3d/blob/main/changeit3d/scripts/evaluate_change_it_3d.py}{evaluate\_change\_it\_3d.py} script available in \href{https://github.com/optas/changeit3d/tree/main}{ChangeIt3D's github repo}.

When calculating all quantitative metrics we ensured a fair comparison by downsampling all outputs for all methods down to a 2048 point resolution.

\begin{table}[t]
    \centering
    \resizebox{0.98\linewidth}{!}{ %
    \begin{tabular}{l c c c c c}
        \toprule
        & \multicolumn{2}{c}{Shapetalk} & & \multicolumn{2}{c}{l-Shapetalk} \\
        l-GD implementation & ChangeIt3D script & Ours && ChangeIt3D script & Ours \\
        \midrule
        ChangeIt3D & 0.82 & 0.19 && 0.89 & 0.20 \\
        Spice-E & 0.94 & 0.31 && 0.98 & 0.39 \\
        Ours & \textbf{0.78} & \textbf{0.07} && \textbf{0.55} & \textbf{0.05} \\
        \bottomrule
    \end{tabular}
    }
    \vspace{-2pt}
    \caption{\textbf{Comparison between different l-GD calculation implementations.} We compare l-GD scores calculated by the ChangeIt3d script against l-GD scores calculated by our methodology, which involves extracting part names from the editing prompts using an LLM (llama3) and segmenting output and input shapes using PointNet. Notice that l-GD scores calculated by the ChangeIt3D script are higher than the GD values presented in the main paper (Table \ref{tab:metrics}) which is unintuitive. Also note that our method acheives the lowest l-GD scores in both methodologies.}
    \label{tab:lgd}
\end{table}

\medskip \noindent \textbf{User Study.} As detailed in Section 4.2, our user study consisted of two forms, A and B, each containing 15 questions and answered by 60 different users. In each question, users were shown a text prompt and an input point cloud, then asked to select one of the editing results generated by ChangeIt3D, Spice-E, and our method, displayed in random order. Both the prompts and input point clouds were randomly selected from the l-ShapeTalk test set. The users were instructed:

\textit{"Your task is to select the target object that best represents the prompt while maintaining the shape of the source object as closely as possible. In other words, choose the most suitable target object that effectively embodies the editing of the source object based on the prompt. IMPORTANT - If none of the target objects align with the prompt, select the target object that best preserves the shape of the source object."}

Similarly to how other quantitative metrics were calculated, users were shown downsampled versions of Spice-E's and our method's results such that all point clouds for all method's were of the same point resolution (2048).

\medskip \noindent \textbf{Baselines.} We used the official \href{https://docs.google.com/forms/d/e/1FAIpQLSdOouzvK0zmjvmBoiQhbfnhe1Kac72XNmHXzshn6_KUEjw8QQ/viewform}{Changeit3D  implementation} to access the ShapeTalk dataset and the pre-trained model weights. The script \href{https://github.com/optas/changeit3d/blob/main/changeit3d/scripts/evaluate_change_it_3d.py}{evaluate\_change\_it\_3d.py} was employed to reproduce their results. Similarly, we used the \href{https://github.com/TAU-VAILab/Spice-E?tab=readme-ov-file}{Spice-E implementation} for obtaining their result. The outputs were converted into point clouds using marching cubes and random point sampling.

\section{Additional Experiments}
\label{sec:experiments}

To better view 3D results, we recommended viewing the supplemental HTML page, which includs fly-through visualizations demonstrating the quality of our results from multiple views.

\subsection{Comparison to Semantic Editing Paradigms}
In this section, we complement our main paper's comparisons with a qualitative analysis of alternative approaches to text-guided semantic editing of 3D shapes.

\begin{figure*}
\centering

\rotatebox{90}{\footnotesize{\emph{Rounded back}}}
\jsubfig{\jsubfig{\jsubfig{\includegraphics[height=2.85cm, trim={0.0cm 0.0cm 0.0cm 0.0cm}, clip]{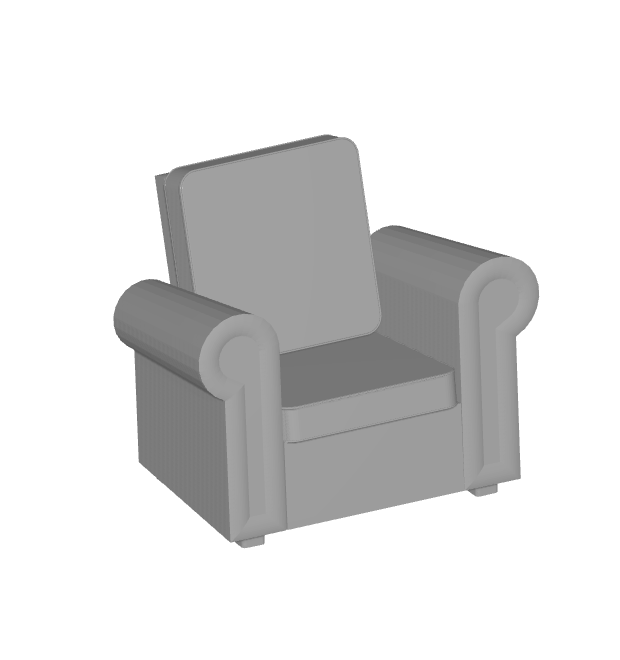}}{}
}{}

\jsubfig{
\jsubfig{\includegraphics[height=2.1cm, trim={0.0cm 0.0cm 0.0cm 1.0cm}, clip]{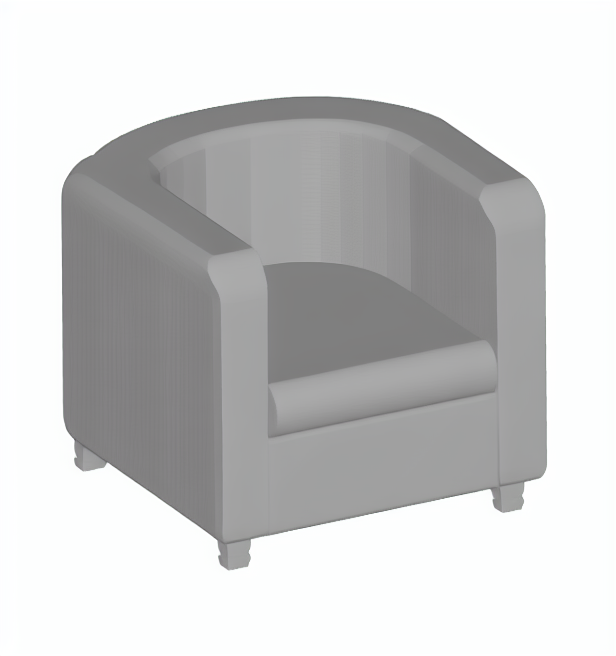}}{}
\jsubfig{\includegraphics[height=2.85cm, trim={15.0cm 0.0cm 15.0cm 0.0cm}, clip]{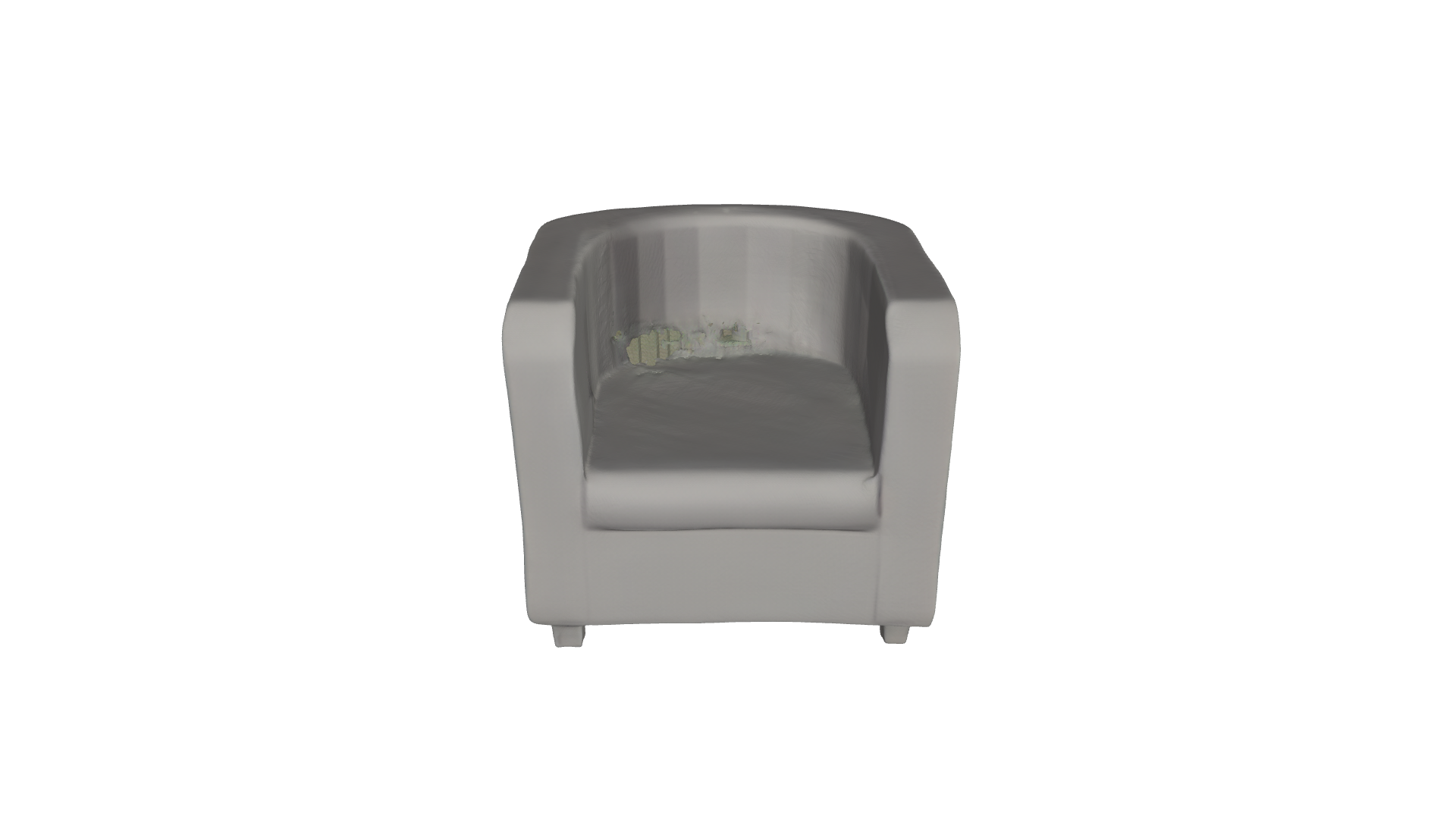}}{}
\jsubfig{\includegraphics[height=2.85cm, trim={15.0cm 0.0cm 15.0cm 0.0cm}, clip]{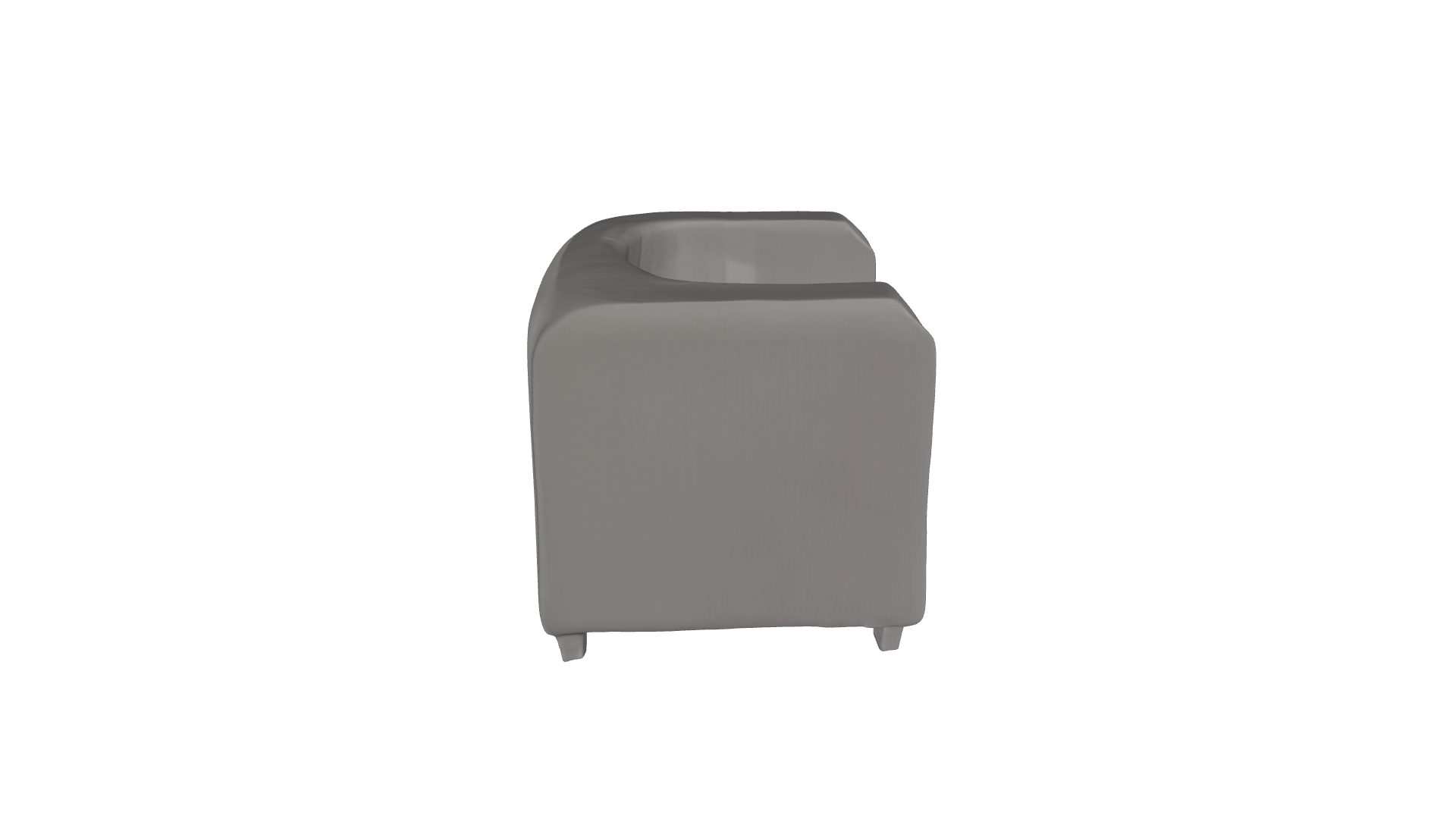}}{}}{} 

\jsubfig{\jsubfig{\includegraphics[height=2.85cm, trim={15.0cm 0.0cm 15.0cm 0.0cm}, clip]{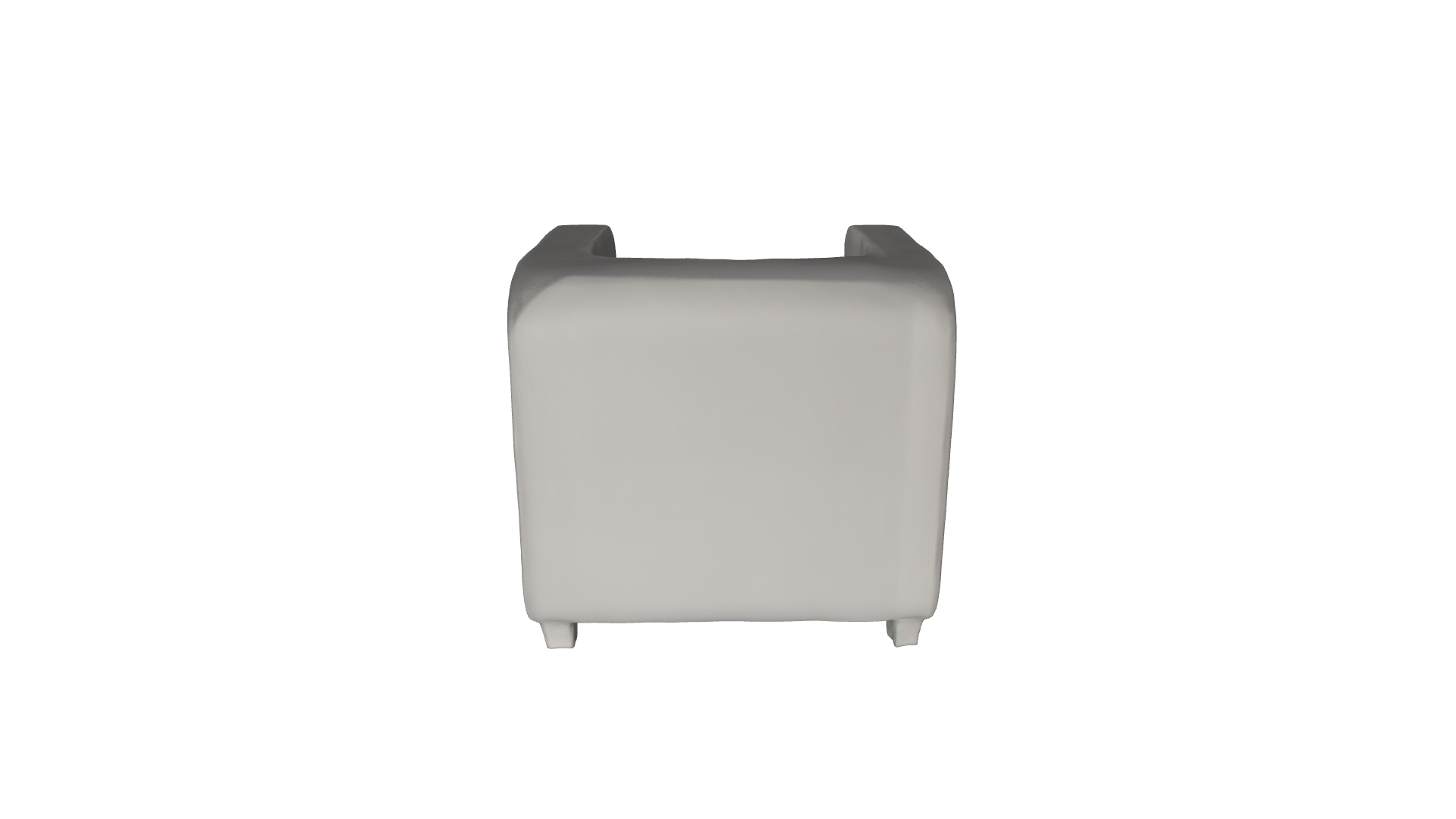}}{}
\jsubfig{\includegraphics[height=2.85cm, trim={15.0cm 0.0cm 15.0cm 0.0cm}, clip]{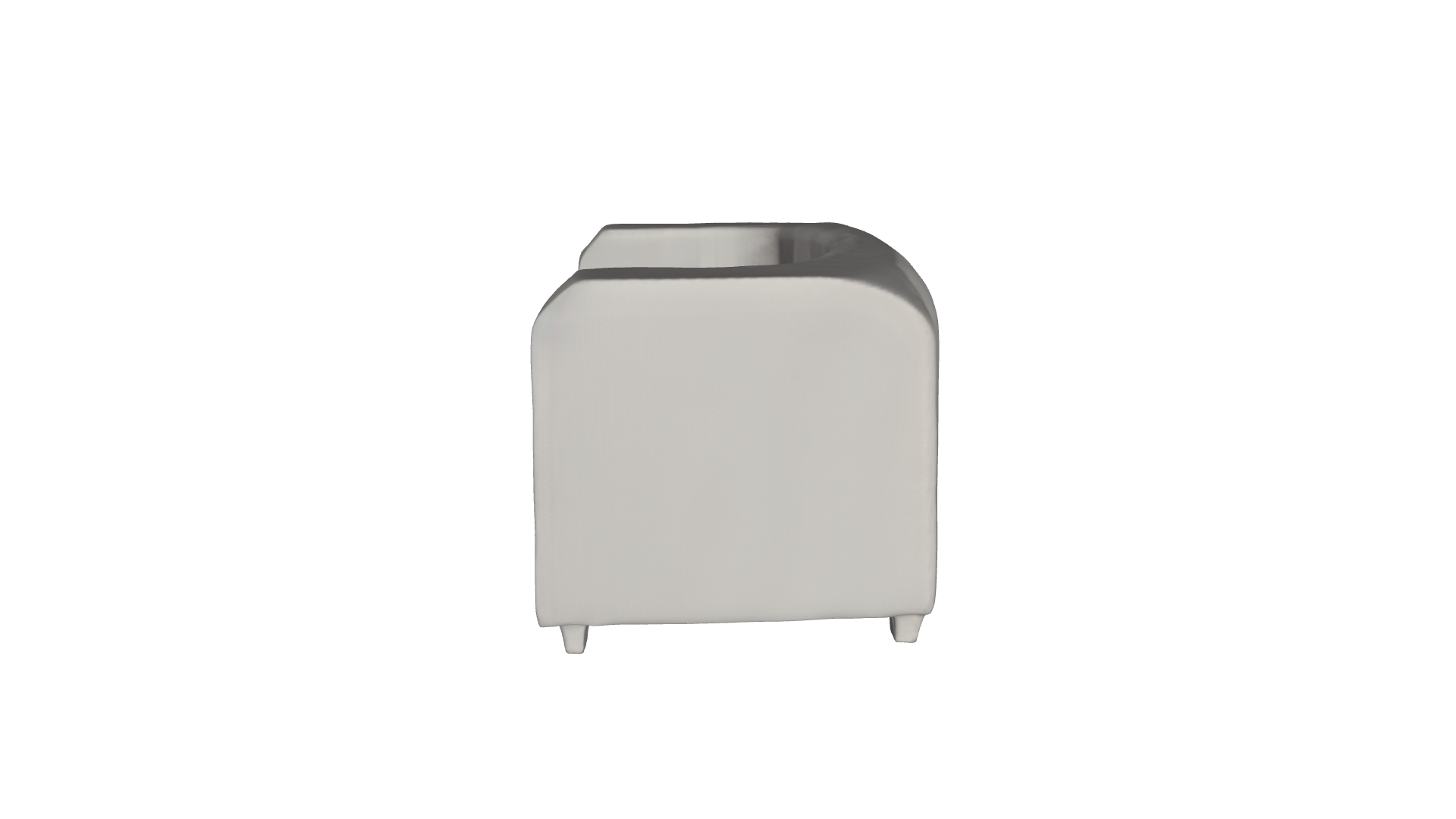}}{}}{}}{}

\rotatebox{90}{\footnotesize{\emph{Four legs}}}
\jsubfig{\jsubfig{\jsubfig{\includegraphics[height=2.5cm, trim={0.0cm 1.0cm 0.0cm 0.0cm}, clip]{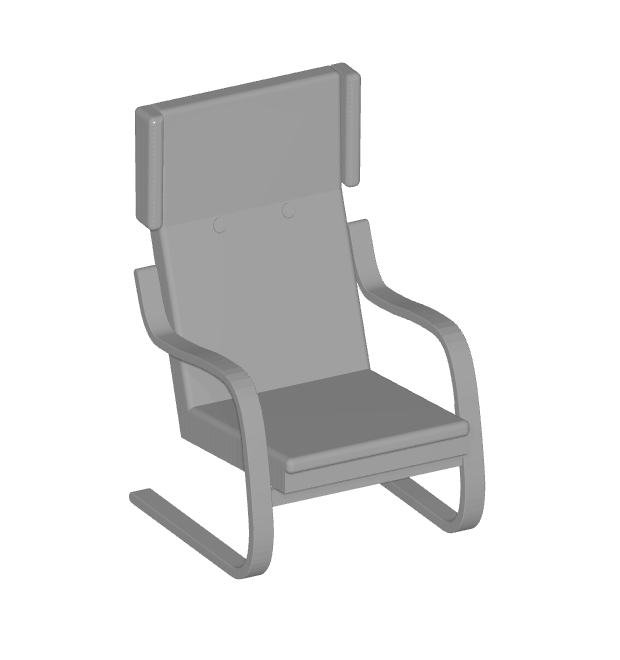}}{}
}{}

\jsubfig{
\jsubfig{\includegraphics[height=2.4cm, trim={0.0cm 0.0cm 0.0cm 0.0cm}, clip]{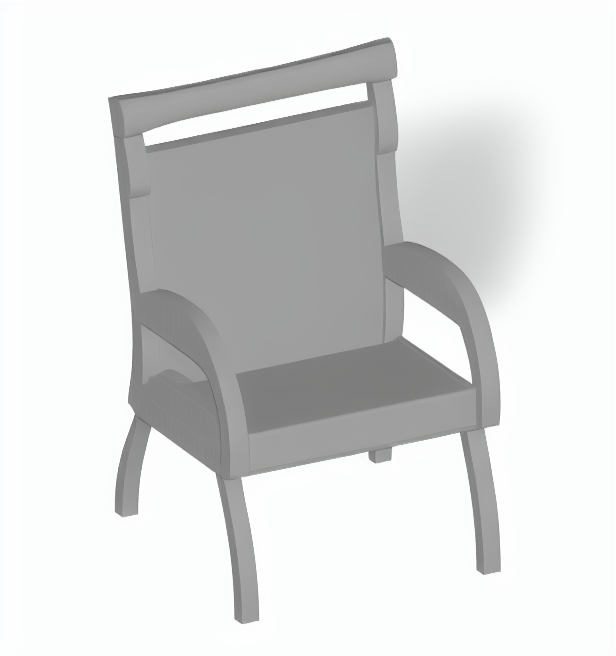}}{}
\jsubfig{\includegraphics[height=2.85cm, trim={15.0cm 0.0cm 15.0cm 0.0cm}, clip]{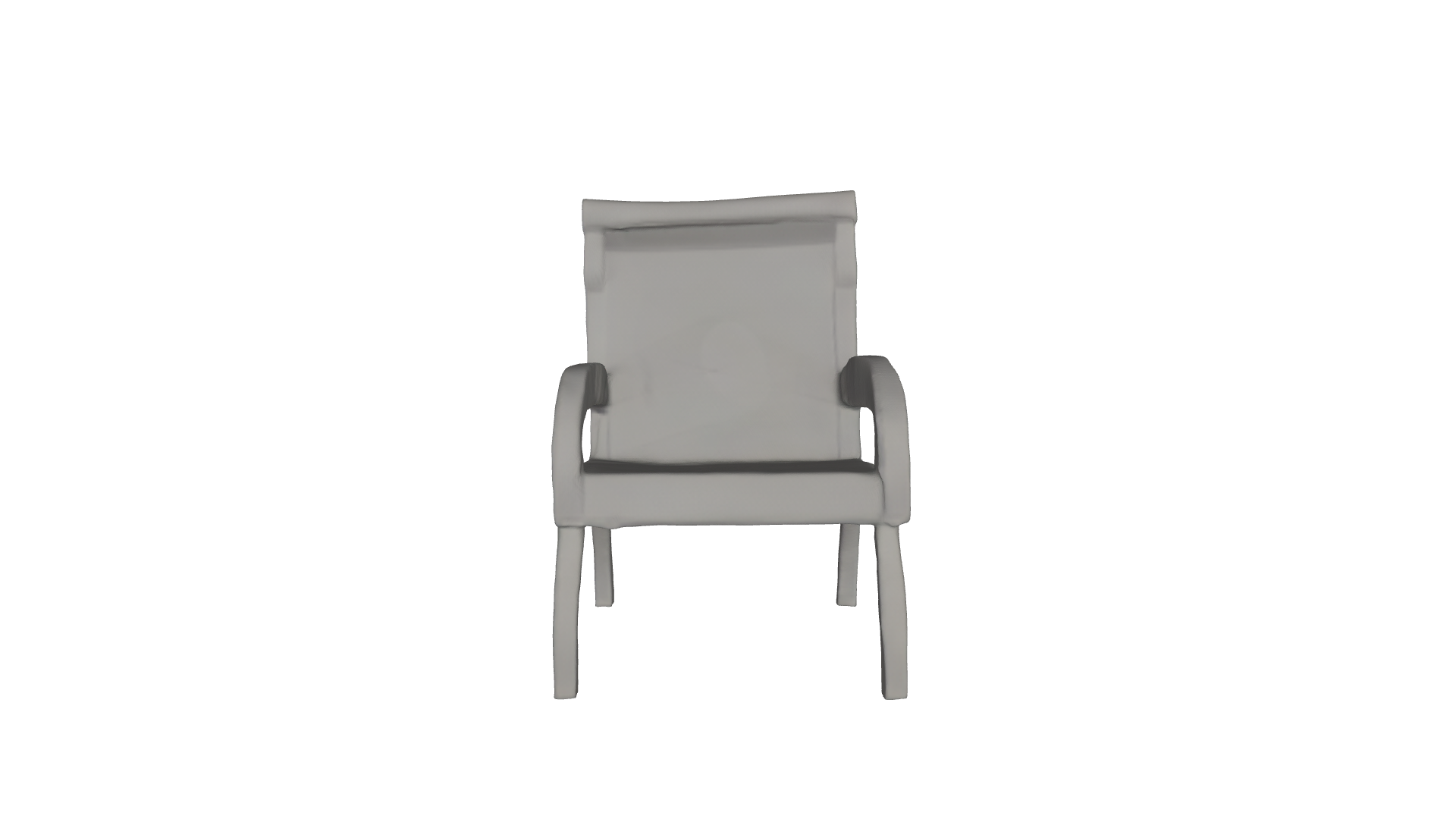}}{}
\jsubfig{\includegraphics[height=2.85cm, trim={15.0cm 0.0cm 15.0cm 0.0cm}, clip]{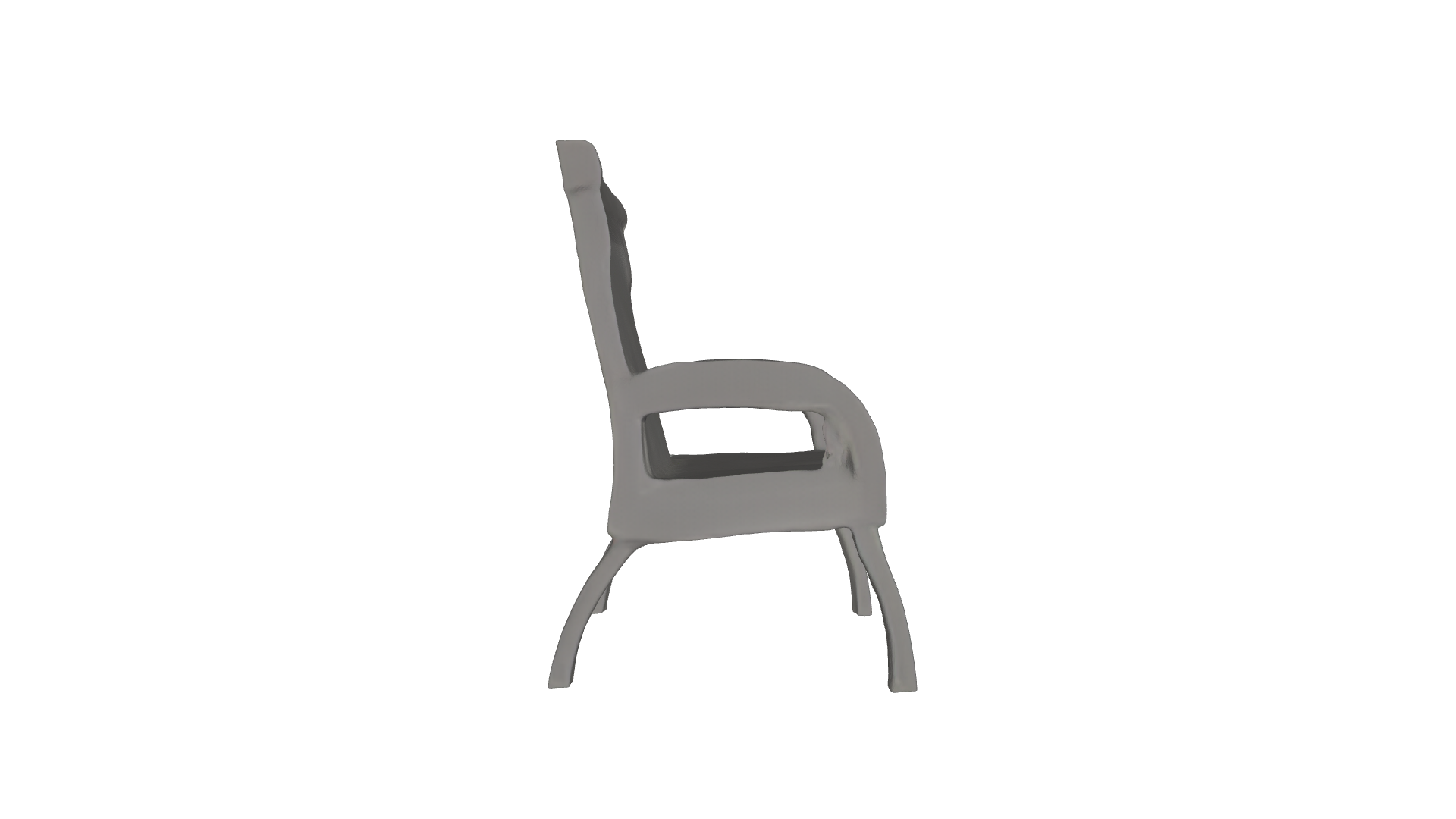}}{}}{} 

\jsubfig{\jsubfig{\includegraphics[height=2.85cm, trim={15.0cm 0.0cm 15.0cm 0.0cm}, clip]{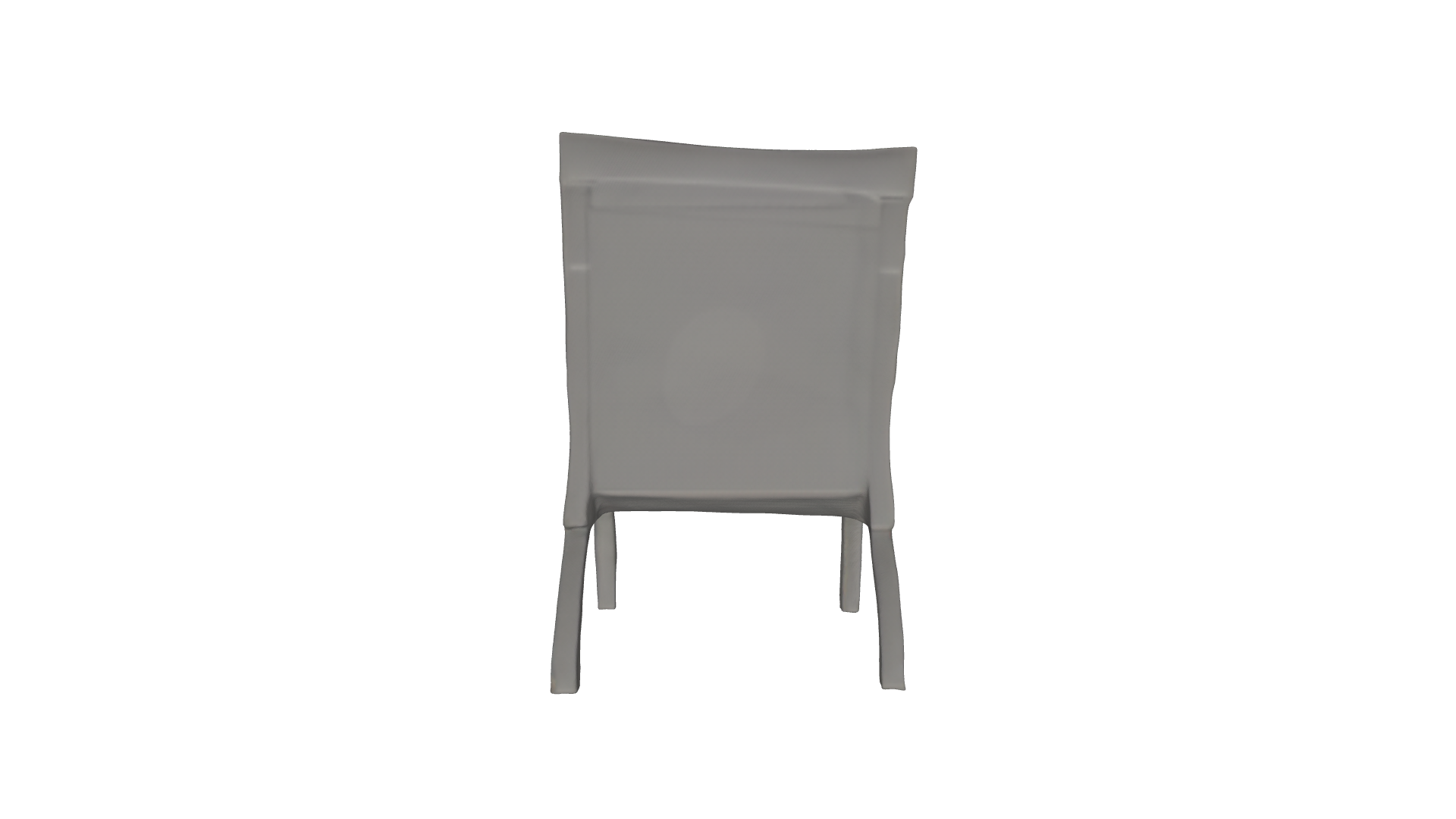}}{}
\jsubfig{\includegraphics[height=2.85cm, trim={15.0cm 0.0cm 15.0cm 0.0cm}, clip]{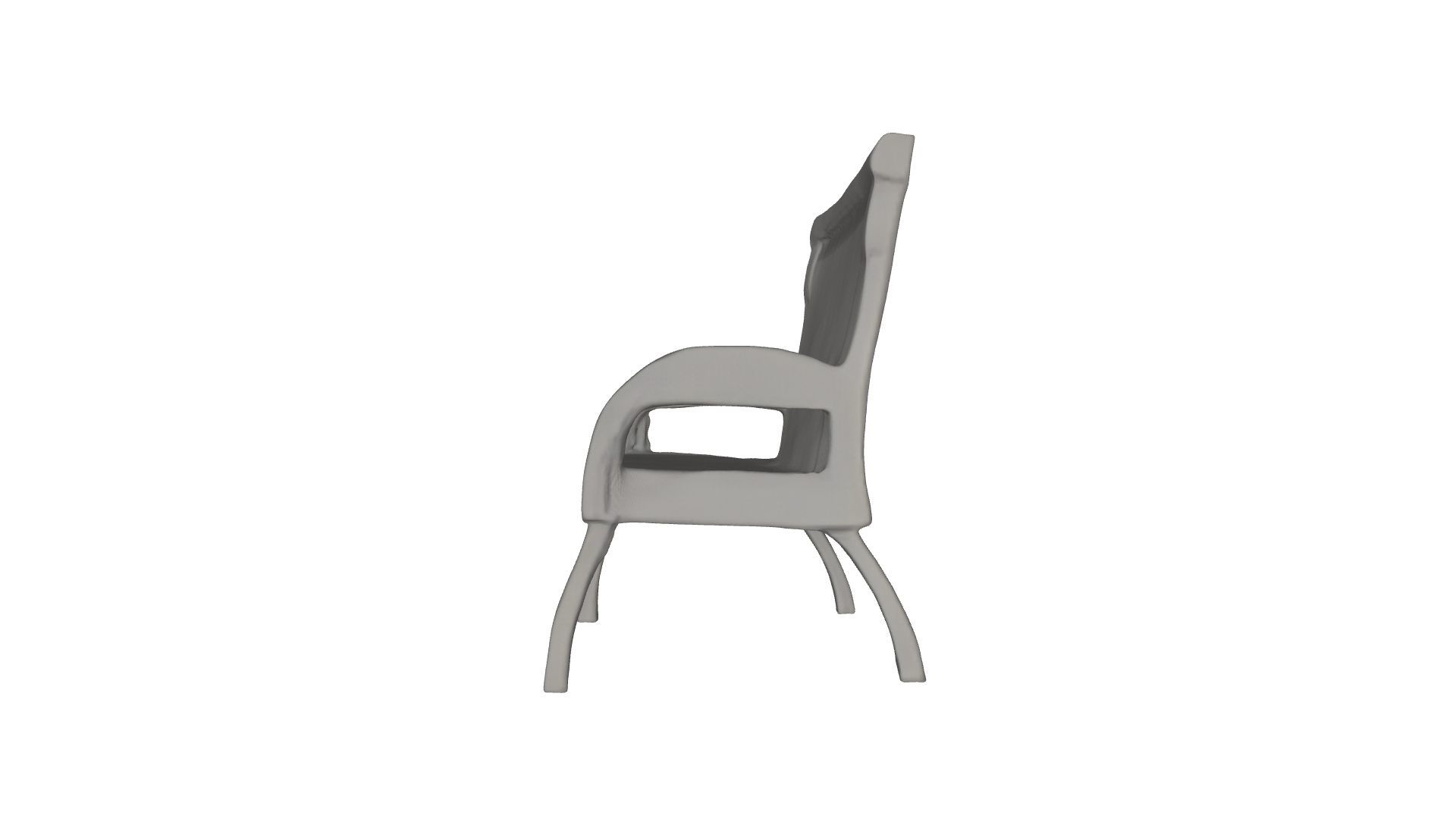}}{}}{}}{}

\rotatebox{90}{\footnotesize{\emph{Thinner legs}}}
\jsubfig{\jsubfig{\jsubfig{\includegraphics[height=2.85cm, trim={0.0cm 0.0cm 0.0cm 0.0cm}, clip]{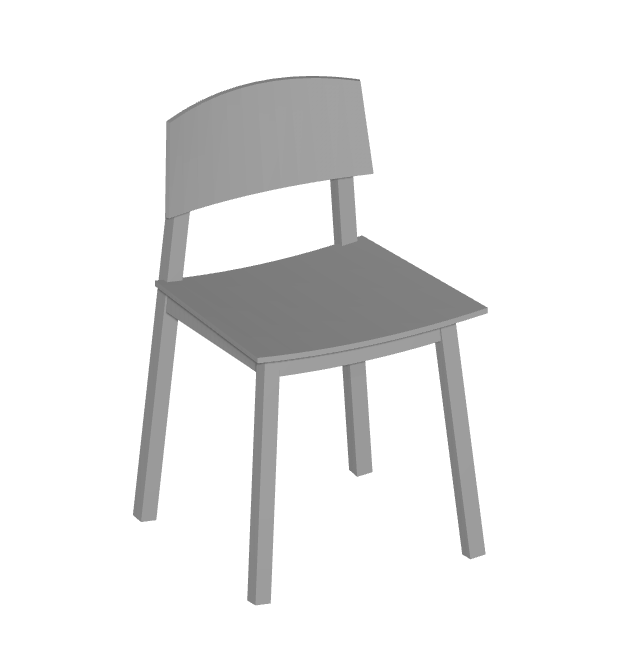}}{}
}{}
\jsubfig{
\jsubfig{\includegraphics[height=2.5cm, trim={0.0cm 0.0cm 0.0cm 0.0cm}, clip]{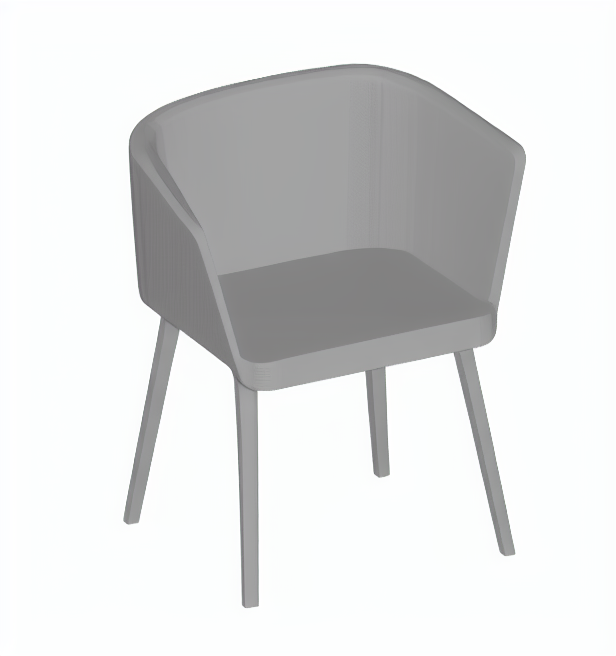}}{}
\jsubfig{\includegraphics[height=2.85cm, trim={15.0cm 0.0cm 15.0cm 0.0cm}, clip]{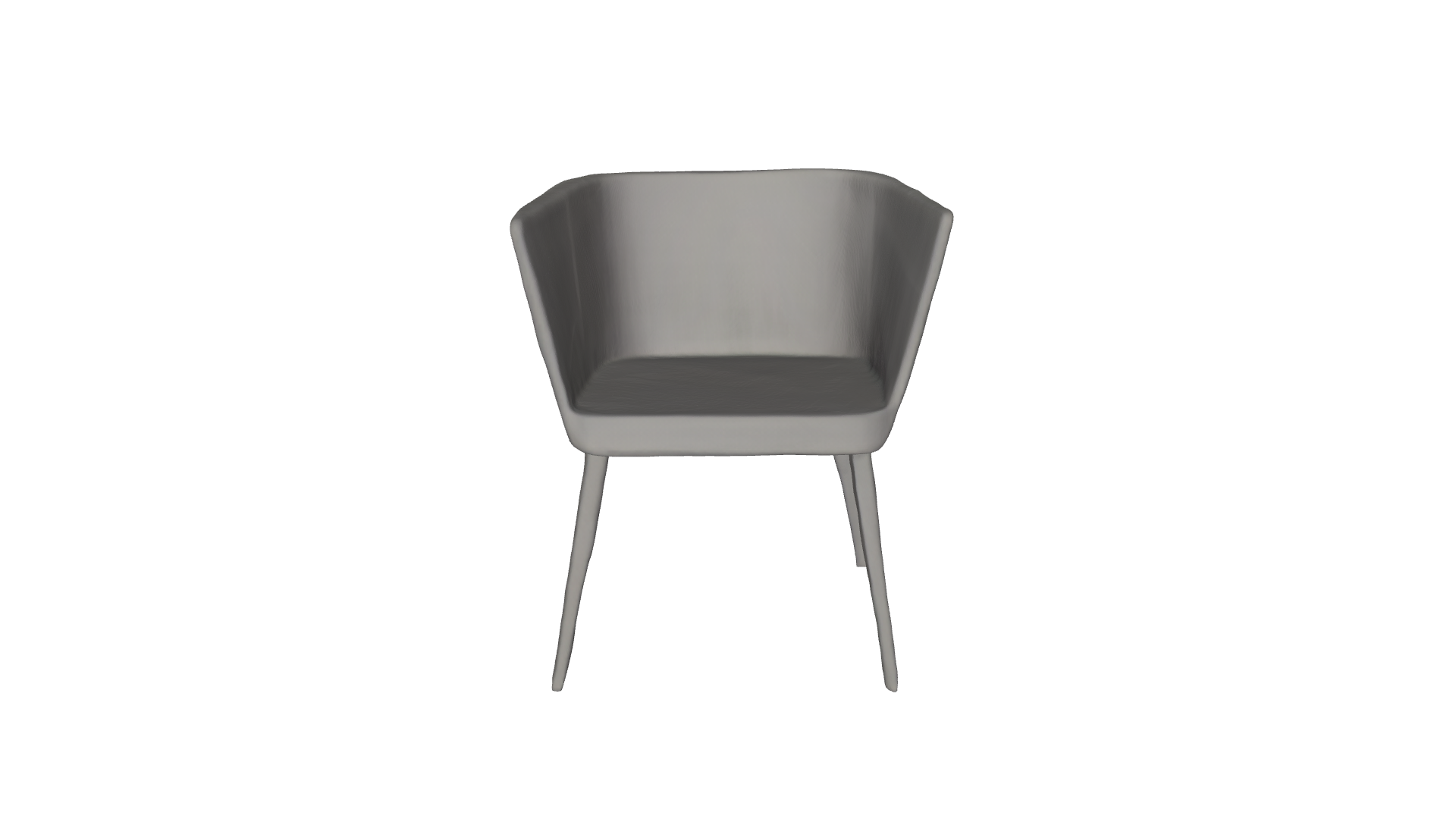}}{}
\jsubfig{\includegraphics[height=2.85cm, trim={15.0cm 0.0cm 15.0cm 0.0cm}, clip]{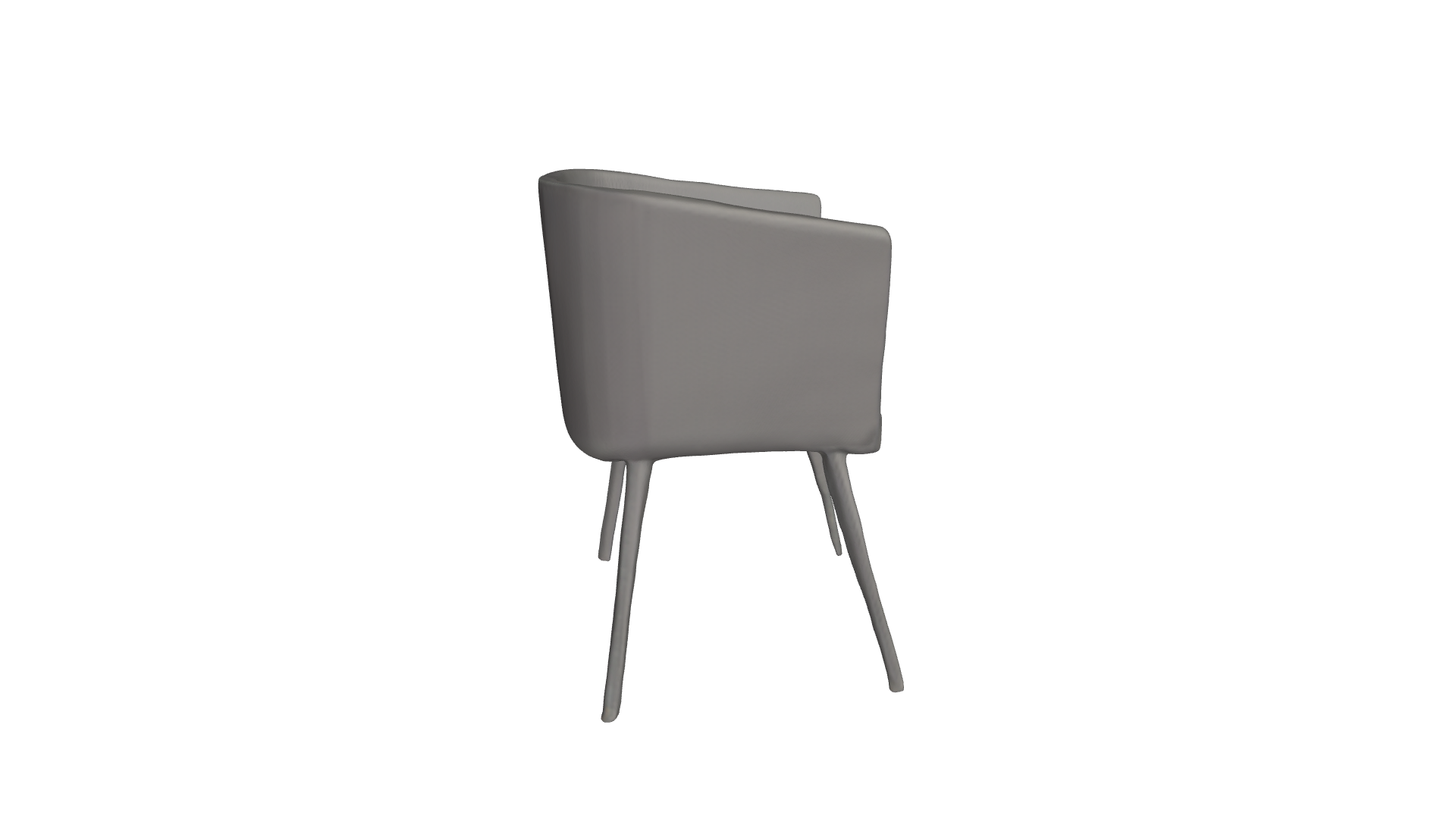}}{}}{} 

\jsubfig{\jsubfig{\includegraphics[height=2.85cm, trim={15.0cm 0.0cm 15.0cm 0.0cm}, clip]{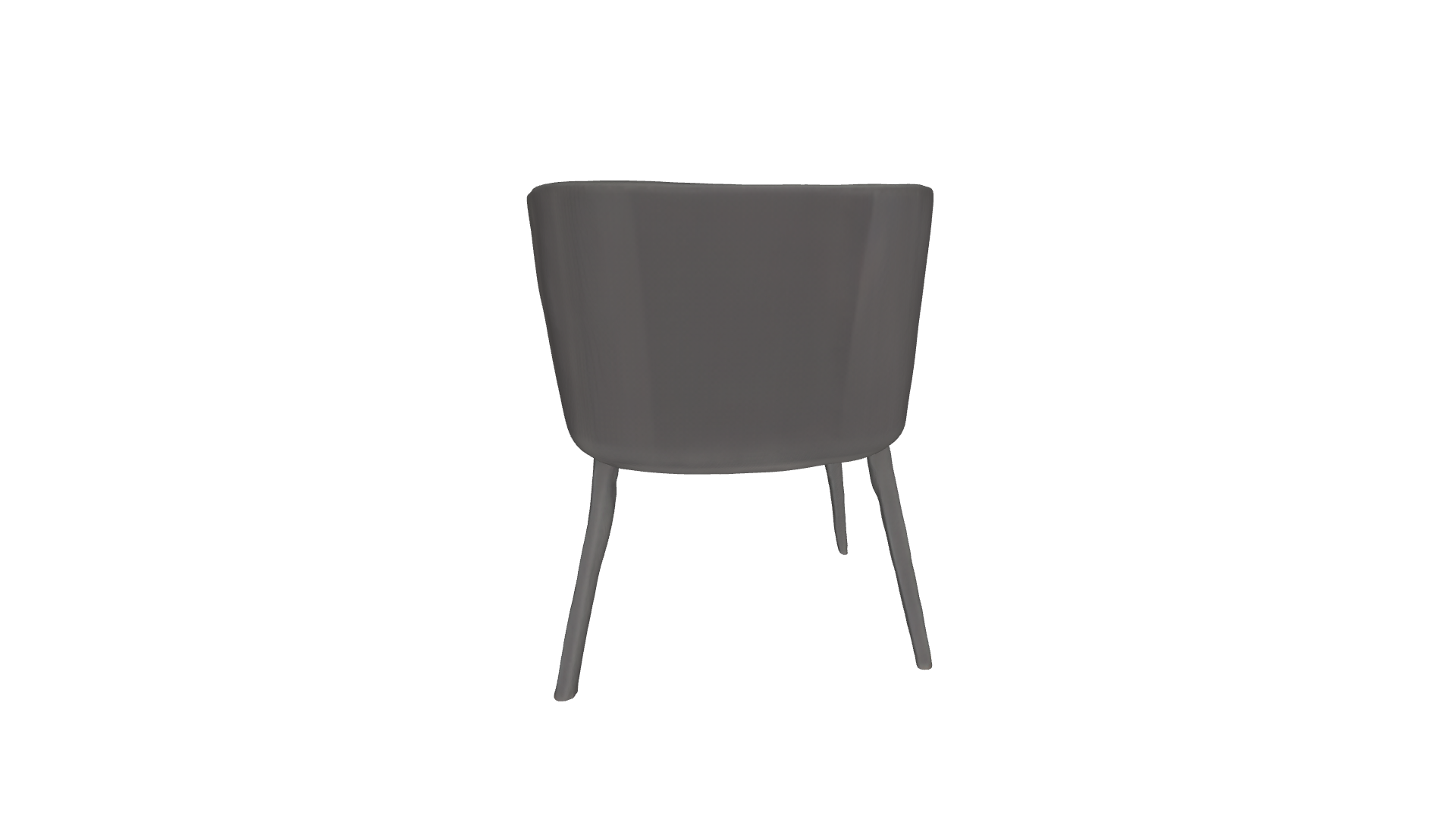}}{}

\jsubfig{\includegraphics[height=2.85cm, trim={15.0cm 0.0cm 15.0cm 0.0cm}, clip]{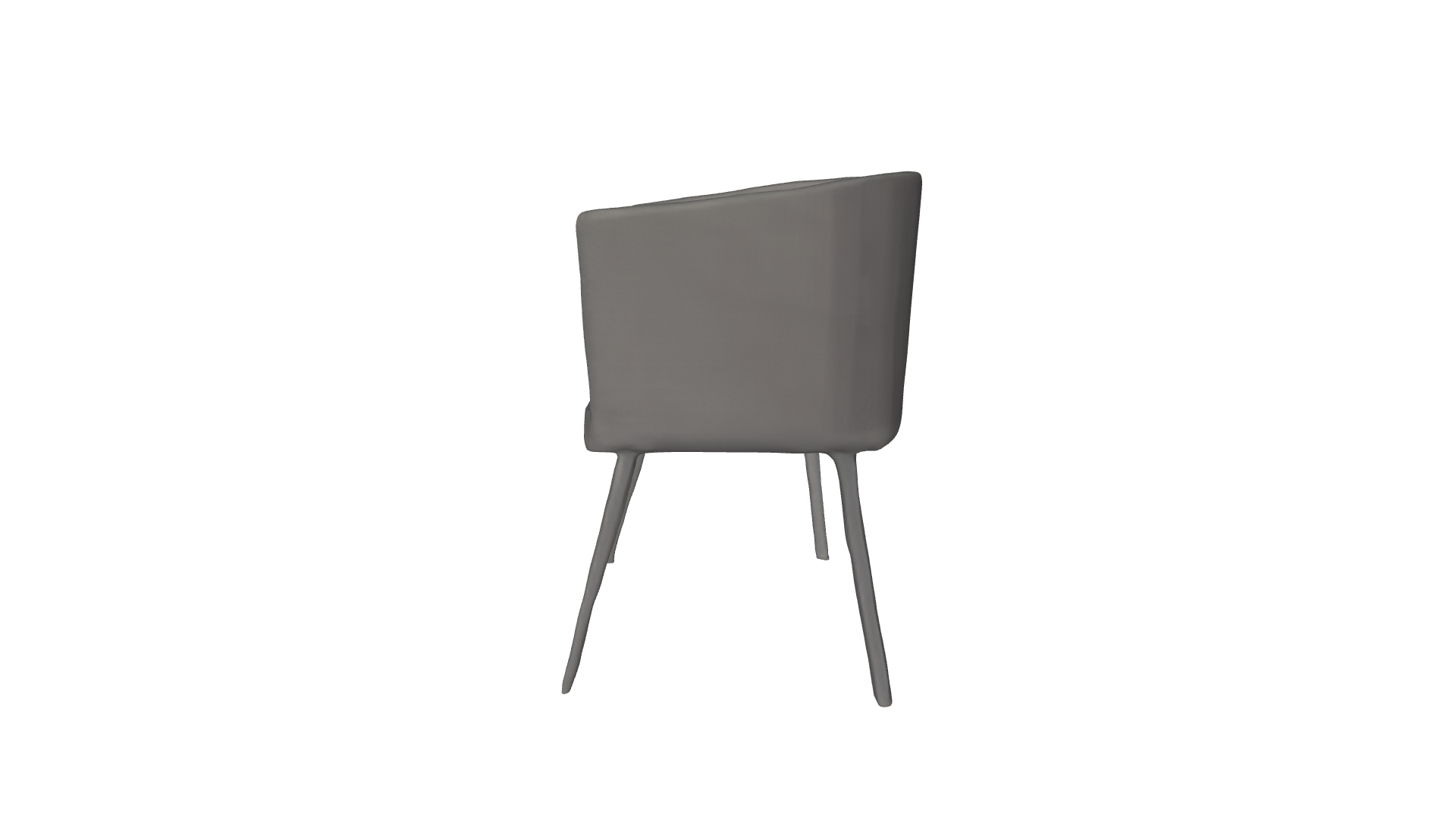}}{}}{}}{}

\rotatebox{90}{\footnotesize{\emph{Taller backrest}}}
\jsubfig{\jsubfig{\jsubfig{\includegraphics[height=2.85cm, trim={0.0cm 0.0cm 0.0cm 0.0cm}, clip]{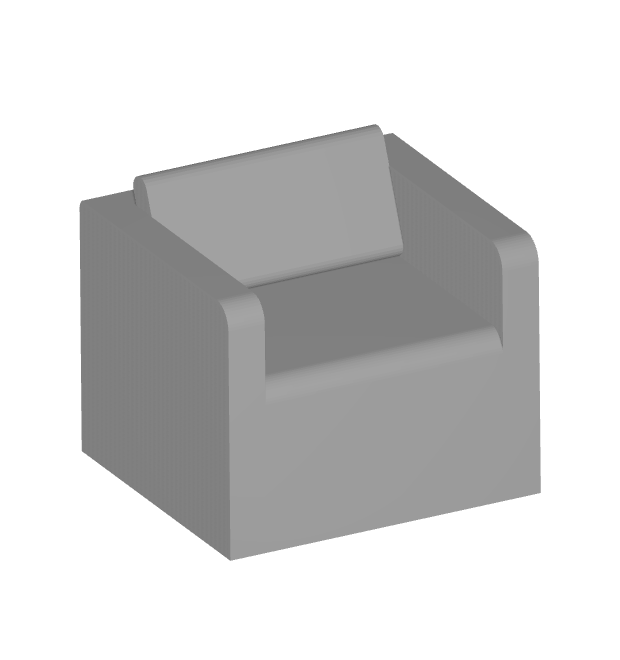}}{Guidance}}{}

\jsubfig{
\jsubfig{\includegraphics[height=2.1cm, trim={0.0cm 0.0cm 0.0cm 0.0cm}, clip]{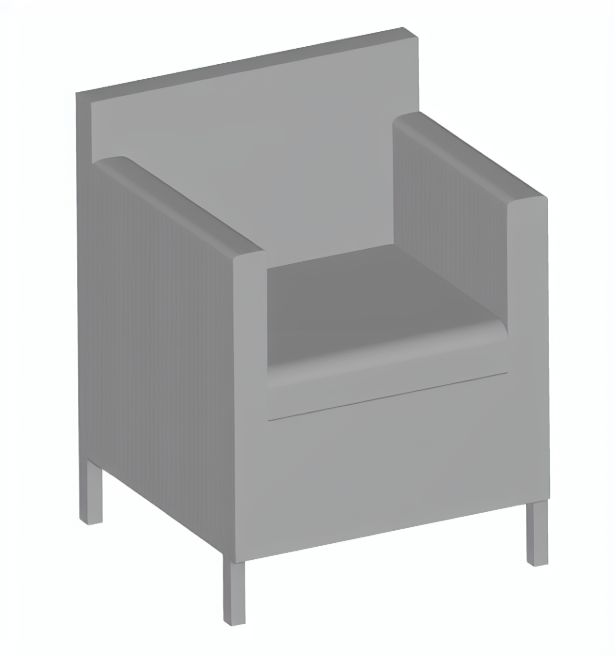}}{InstructPix2Pix output}
\jsubfig{\includegraphics[height=2.85cm, trim={15.0cm 0.0cm 15.0cm 0.0cm}, clip]{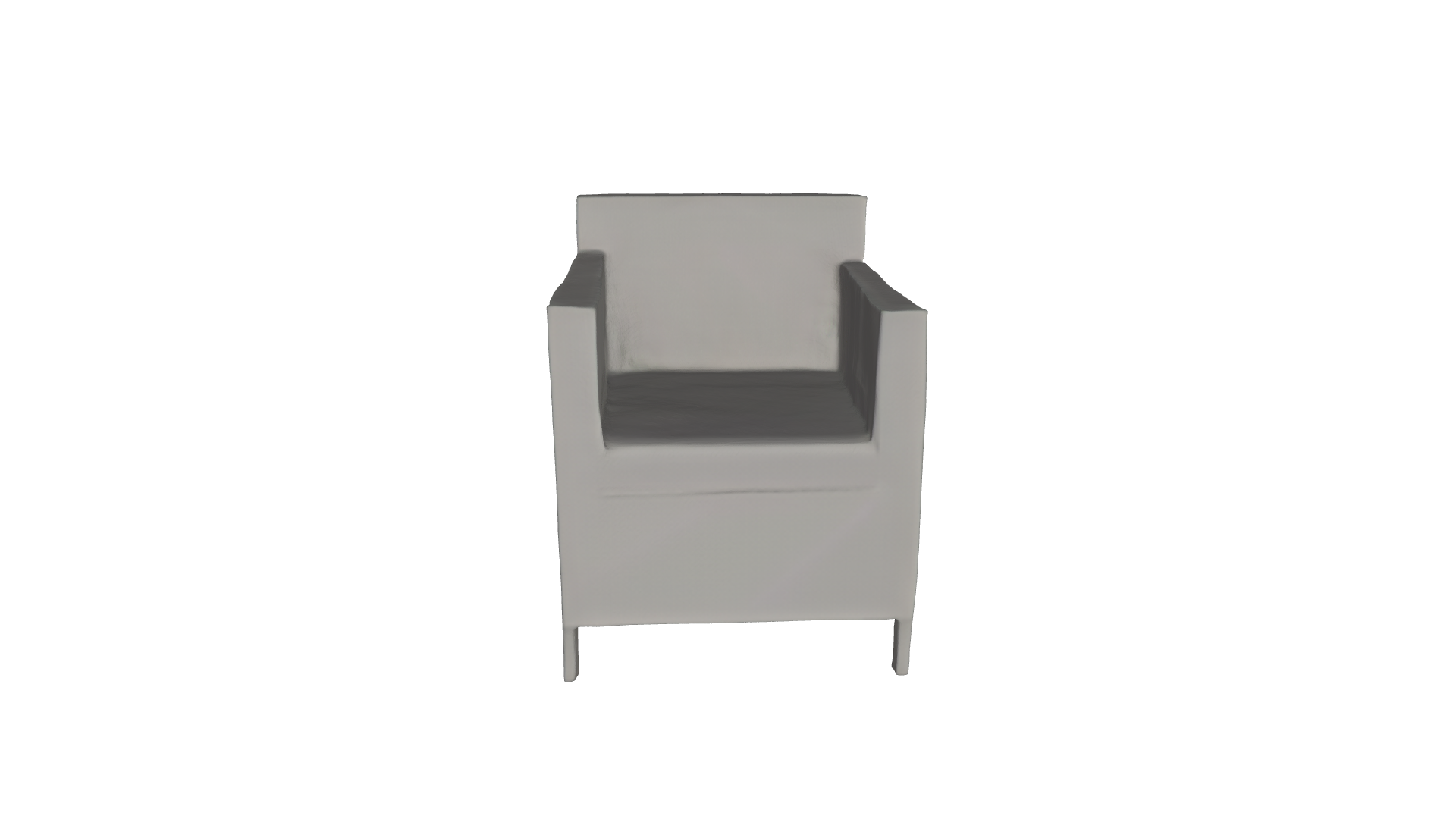}}{View 1}

\jsubfig{\includegraphics[height=2.85cm, trim={15.0cm 0.0cm 15.0cm 0.0cm}, clip]{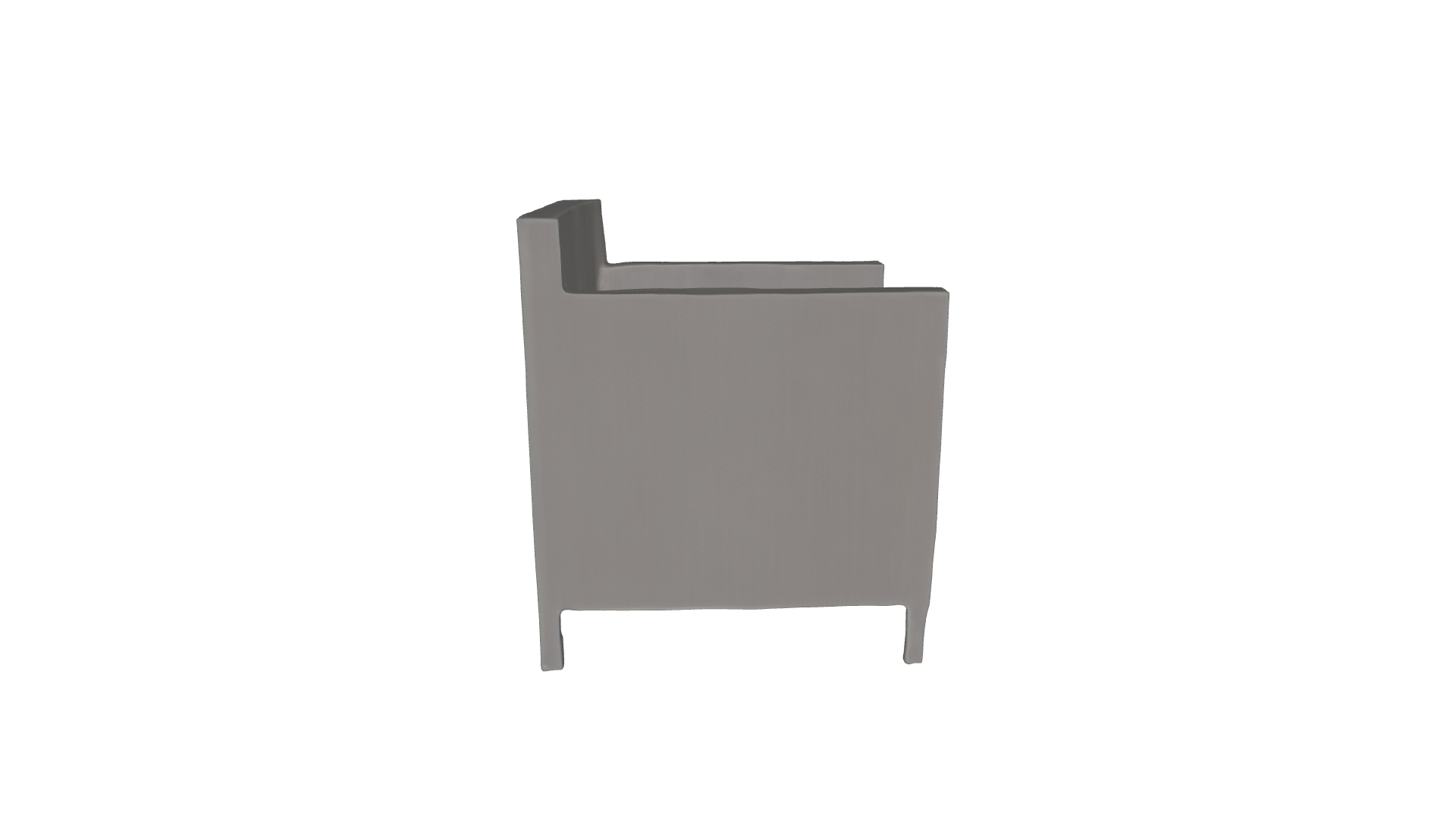}}{View 2}}{} 

\jsubfig{\jsubfig{\includegraphics[height=2.85cm, trim={15.0cm 0.0cm 15.0cm 0.0cm}, clip]{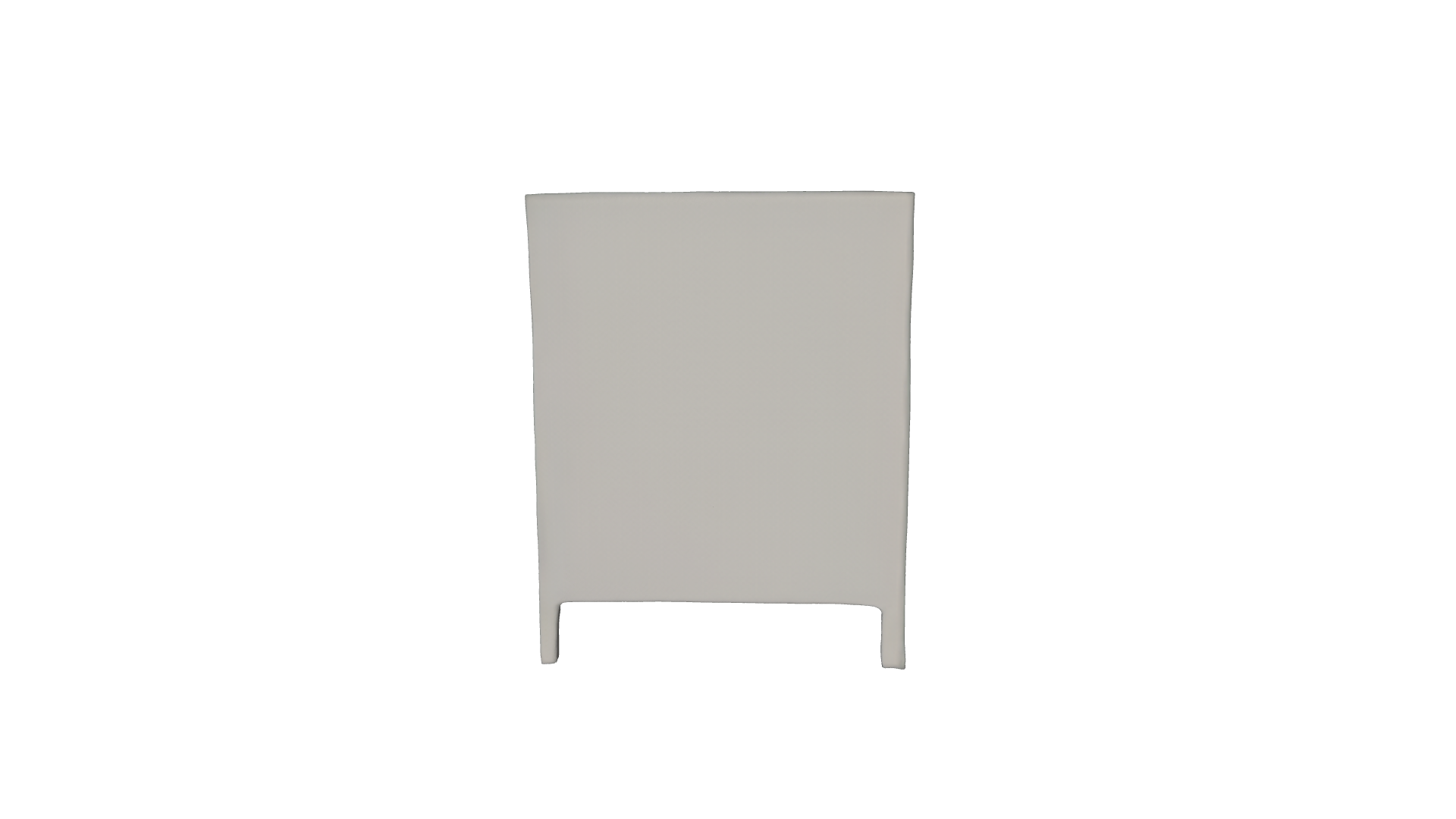}}{View 3}

\jsubfig{\includegraphics[height=2.85cm, trim={15.0cm 0.0cm 15.0cm 0.0cm}, clip]{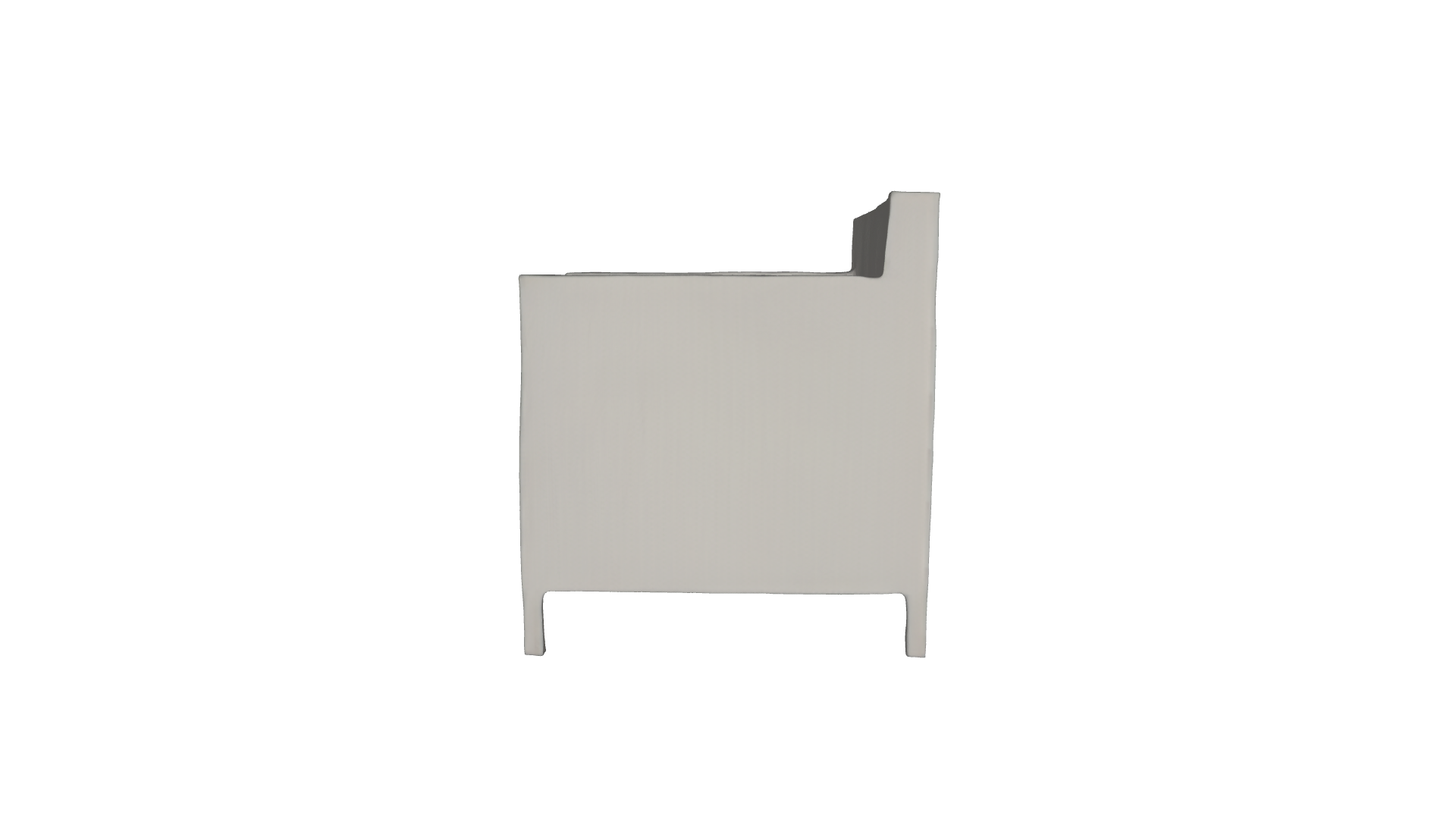}}{View 4}}{}}{}

\vspace{10pt} 
\caption{\textbf{Image Editing Followed by Single-View 3D Reconstruction.}
We first use InstructPix2Pix to edit the rendered image and then apply One-2-3-45++ for single-view reconstruction. In addition to InstructPix2Pix's challenges with precise localization, One-2-3-45++ introduces minor defects in the reconstructed shapes.}
\label{fig:instruct_one}
\vspace{30pt} 
\end{figure*}

\begin{figure*}

\rotatebox{90}{\hspace{0.3cm}\footnotesize{\emph{Rounded back}}}
\jsubfig{\jsubfig{\jsubfig{\includegraphics[height=2.125cm, trim={0.0cm 0.0cm 0.0cm 0.0cm}, clip]{images/baselines/chair_ShapeNet_c790496caf5f2904f14c803eab703899/pix2pix_source.png}}{}
\jsubfig{\includegraphics[height=2.125cm, trim={0.0cm 0.0cm 0.0cm 0.0cm}, clip]{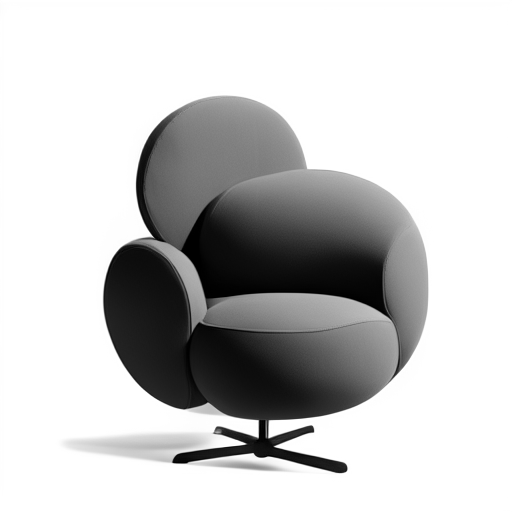}}{}
\jsubfig{\includegraphics[height=2.125cm, trim={0.0cm 0.0cm 0.0cm 0.0cm}, clip]{images/baselines/chair_ShapeNet_c790496caf5f2904f14c803eab703899/pix2pix_output.png}}{}}{}

\jsubfig{\jsubfig{\includegraphics[height=2.125cm, trim={15.0cm 0.0cm 15.0cm 0.0cm}, clip]{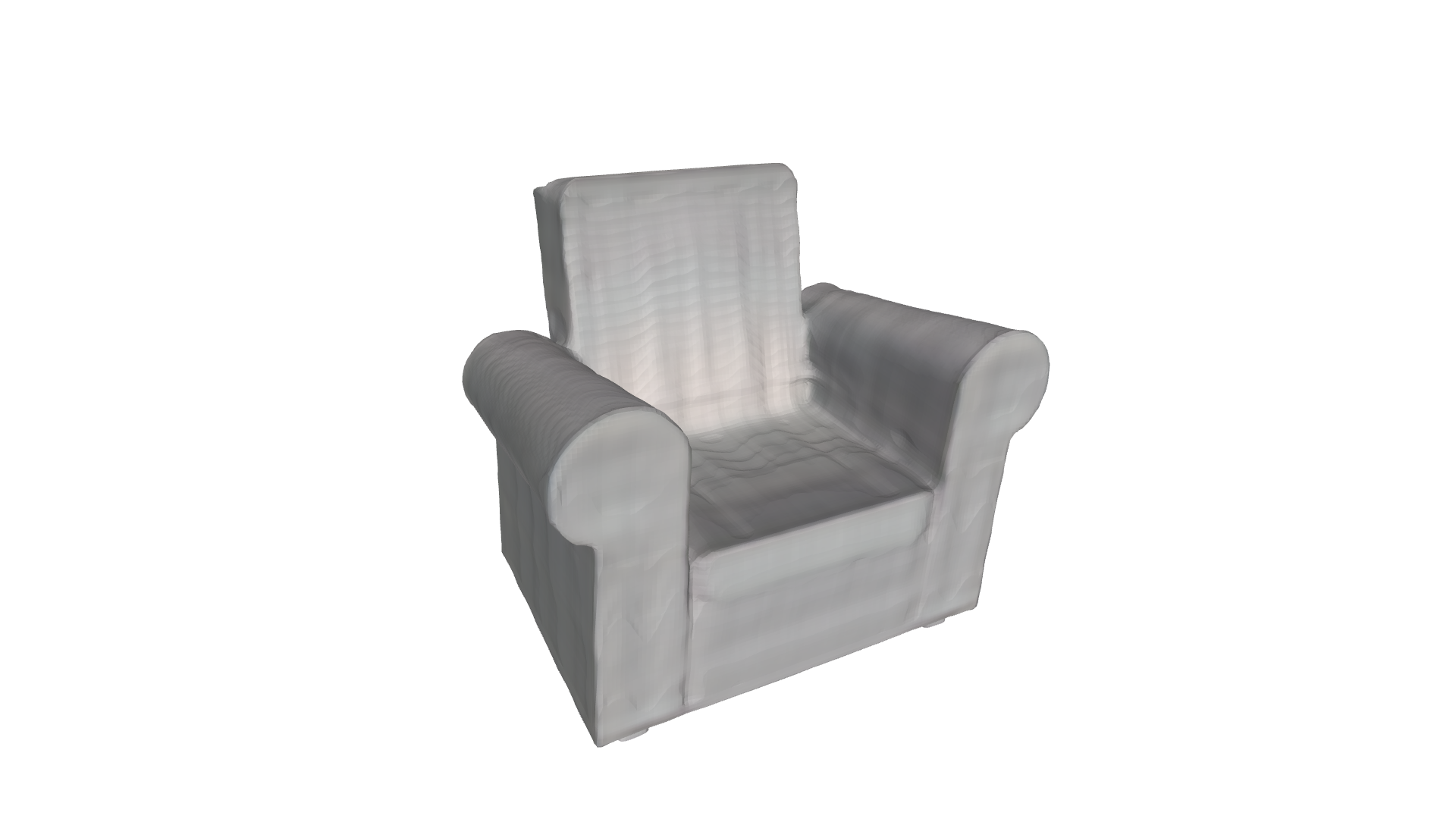}}{}
\jsubfig{\includegraphics[height=2.125cm, trim={3.5cm 2.5cm 3.5cm 4.5cm}, clip]{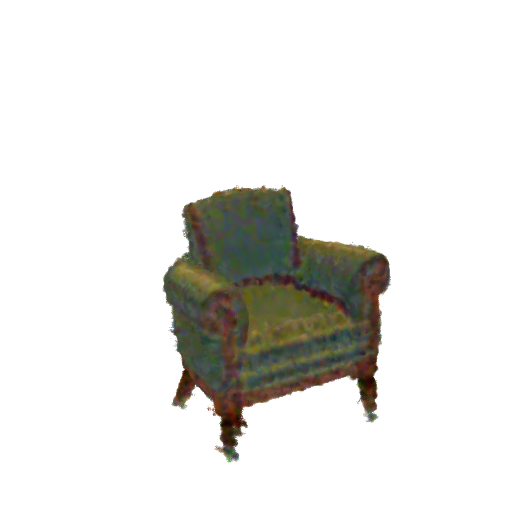}}{}
\jsubfig{\includegraphics[height=2.125cm, trim={15.0cm 0.0cm 15.0cm 0.0cm}, clip]{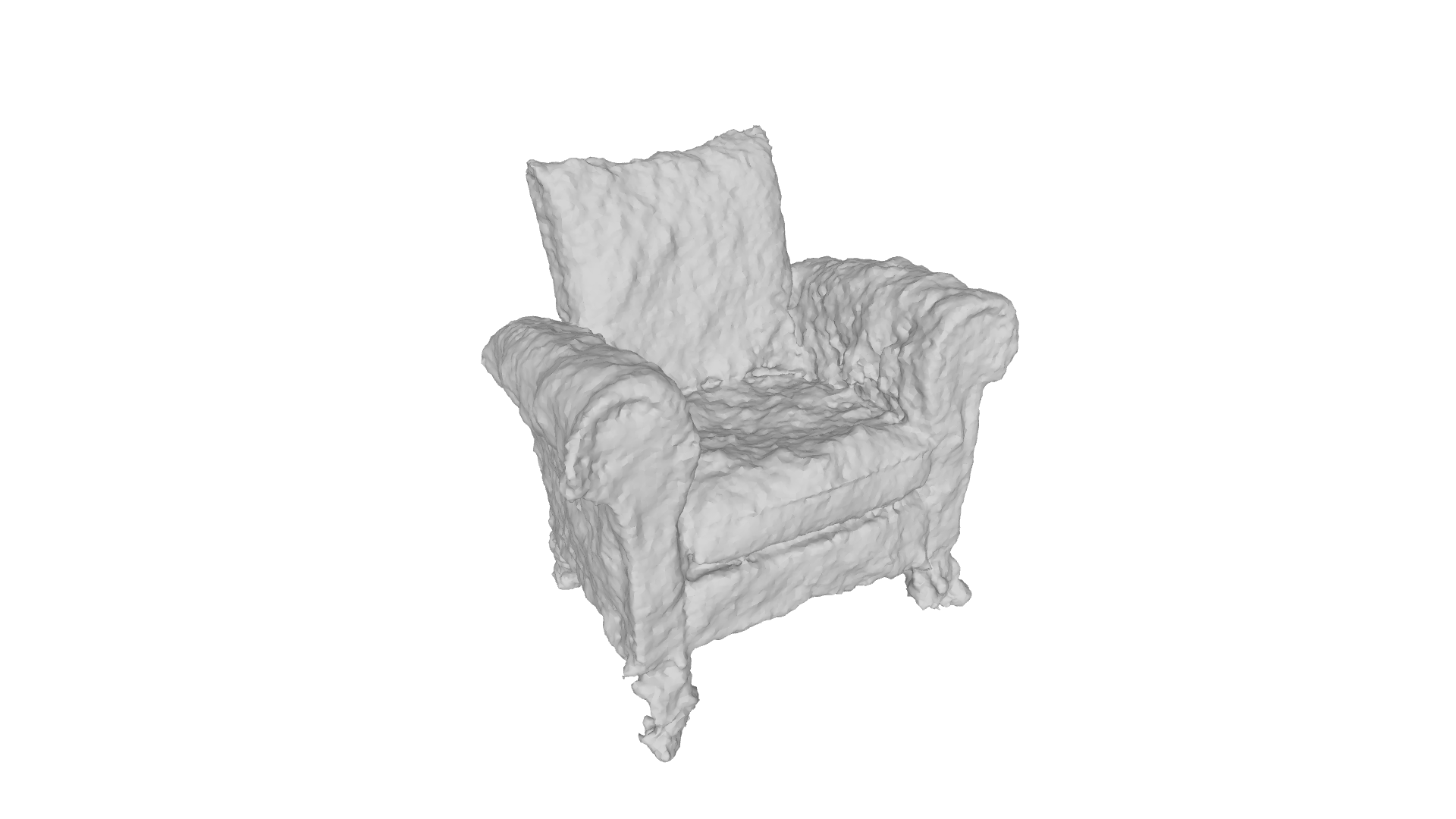}}{}}{} 

\jsubfig{\jsubfig{\includegraphics[height=2.125cm, trim={4.0cm 4.0cm 4.0cm 4.0cm}, clip]{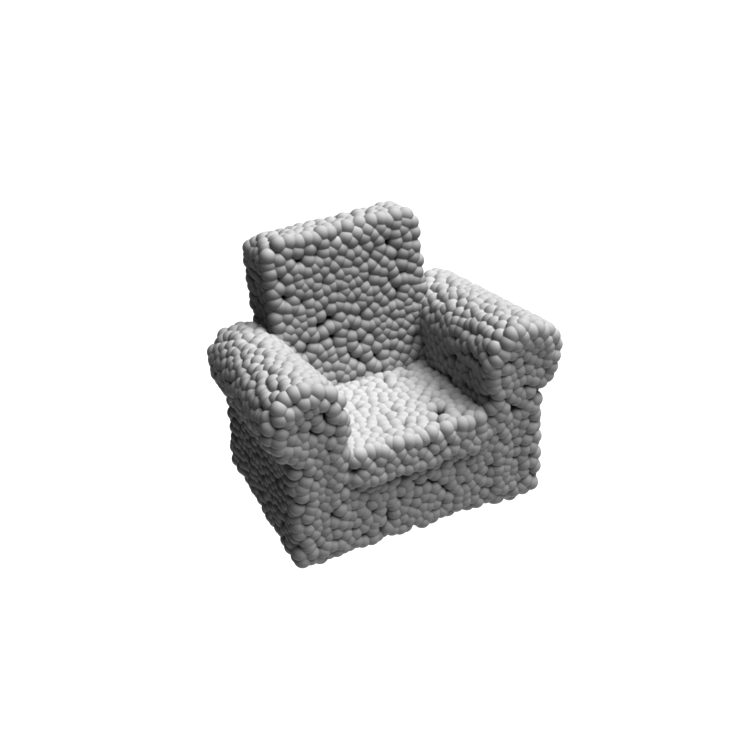}}{}
\jsubfig{\includegraphics[height=2.125cm, trim={4.0cm 4.0cm 4.0cm 4.0cm}, clip]{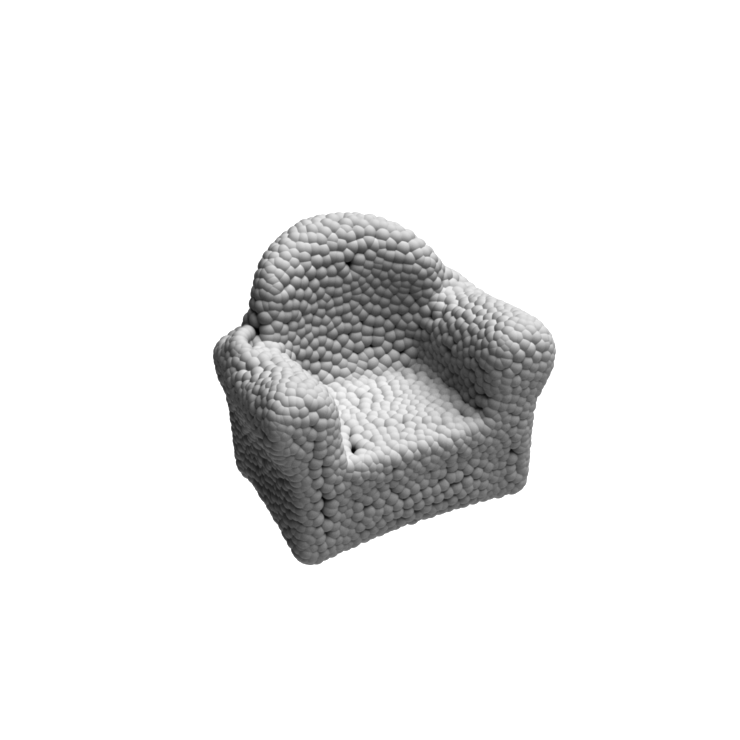}}{}}{}}{}

\rotatebox{90}{\hspace{0.8cm}\footnotesize{\emph{Four legs}}}
\jsubfig{\jsubfig{\jsubfig{\includegraphics[height=2.125cm, trim={0.0cm 0.0cm 0.0cm 0.0cm}, clip]{images/baselines/chair_ShapeNet_5a30a8edad307d8b04cb542e2c50eb4/pix2pix_source.png}}{}
\jsubfig{\includegraphics[height=2.125cm, trim={0.0cm 0.0cm 0.0cm 0.0cm}, clip]{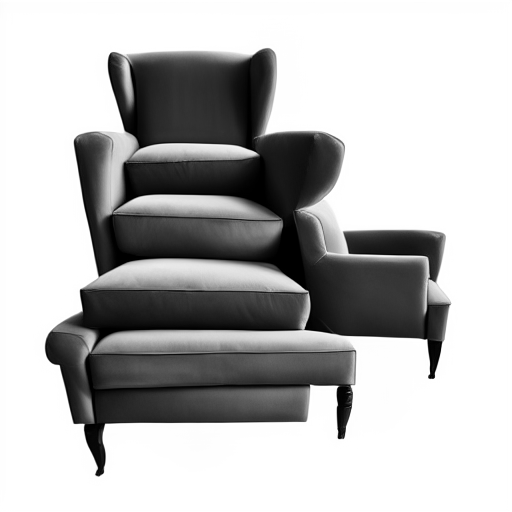}}{}
\jsubfig{\includegraphics[height=2.125cm, trim={0.0cm 0.0cm 0.0cm 0.0cm}, clip]{images/baselines/chair_ShapeNet_5a30a8edad307d8b04cb542e2c50eb4/pix2pix_output.png}}{}}{}

\jsubfig{\jsubfig{\includegraphics[height=2.125cm, trim={15.0cm 0.0cm 15.0cm 0.0cm}, clip]{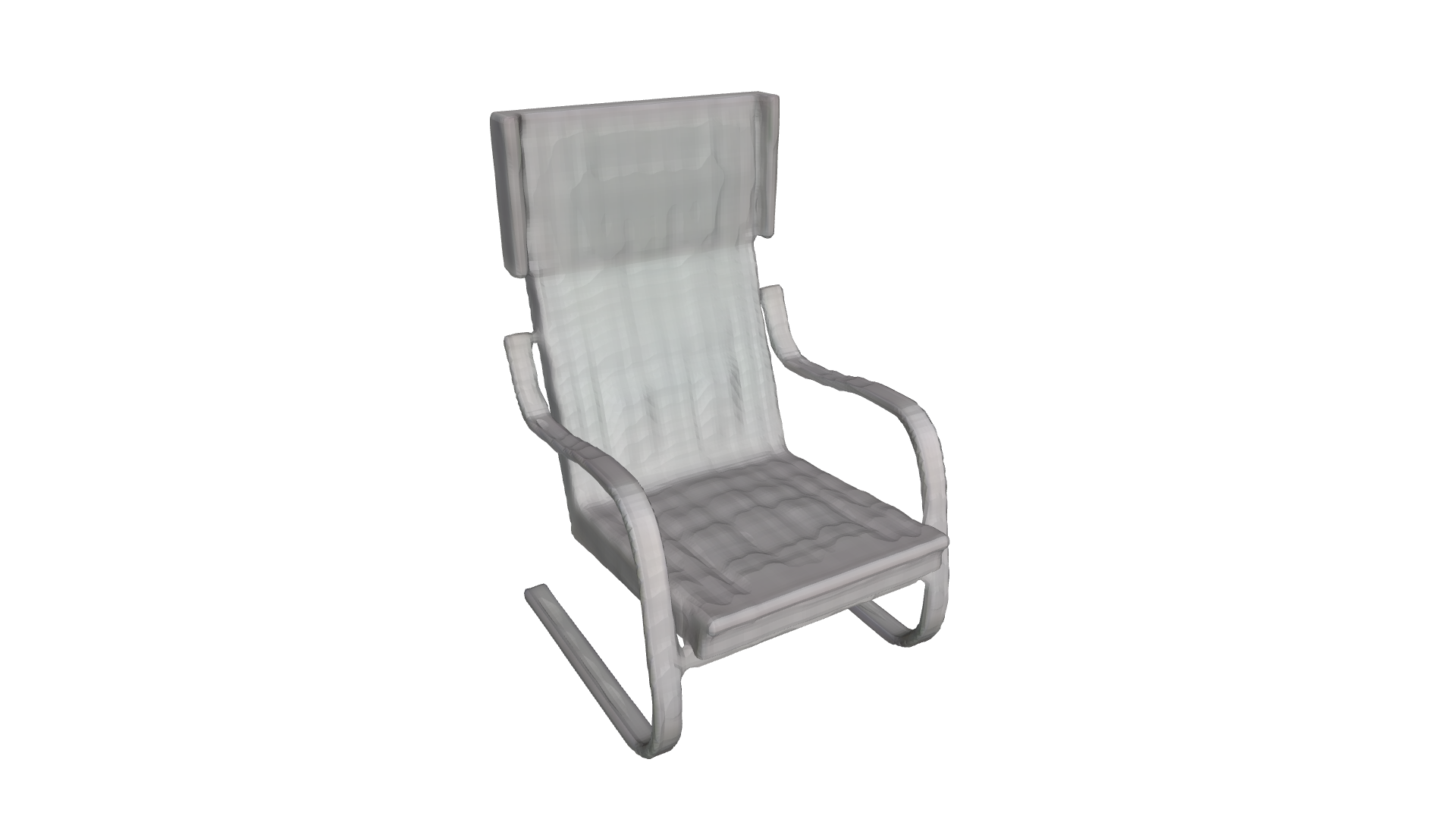}}{}
\jsubfig{\includegraphics[height=2.125cm, trim={3.5cm 2.0cm 3.5cm 5.0cm}, clip]{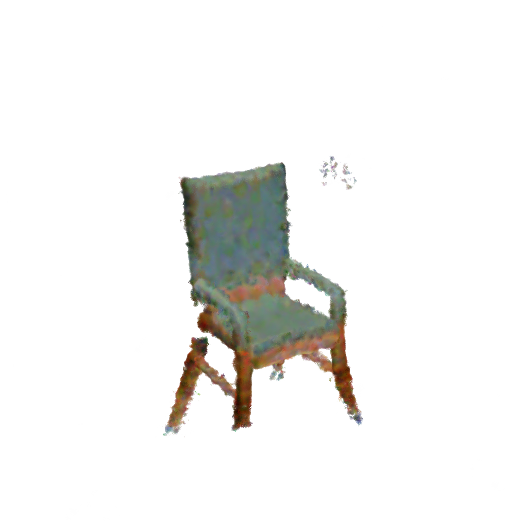}}{}
\jsubfig{\includegraphics[height=2.125cm, trim={15.0cm 0.0cm 15.0cm 0.0cm}, clip]{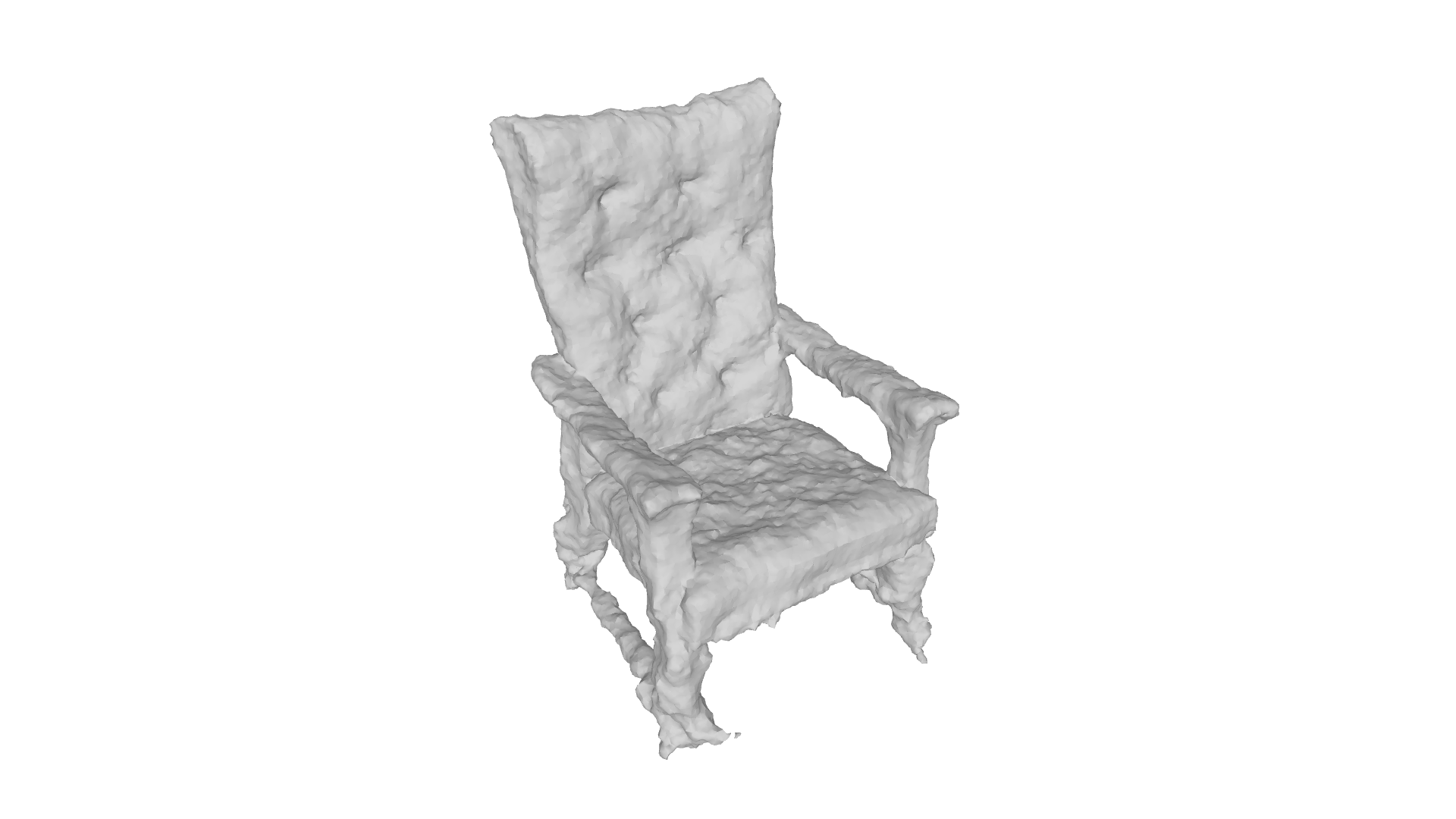}}{}}{} 

\jsubfig{\jsubfig{\includegraphics[height=2.125cm, trim={4.0cm 4.0cm 4.0cm 4.0cm}, clip]{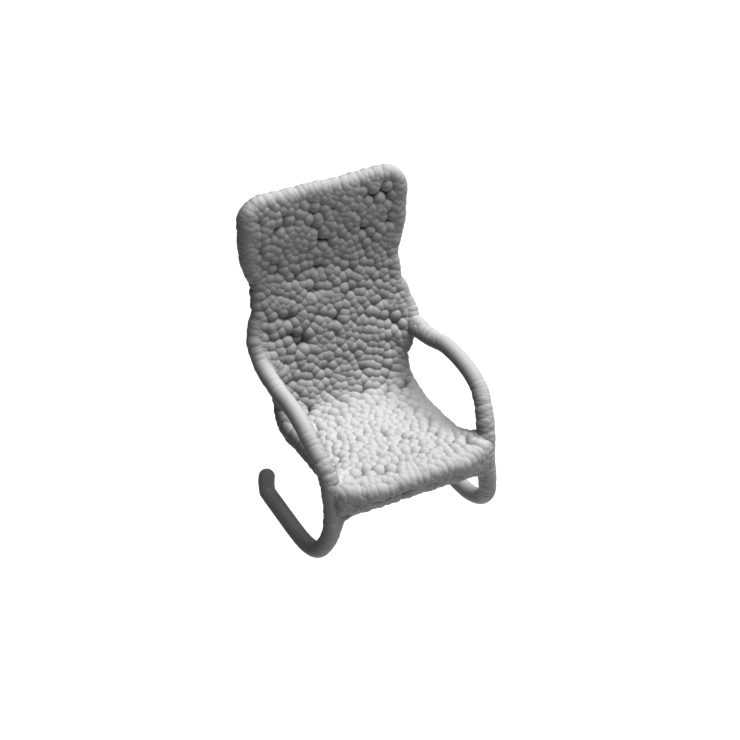}}{}
\jsubfig{\includegraphics[height=2.125cm, trim={4.0cm 4.0cm 4.0cm 4.0cm}, clip]{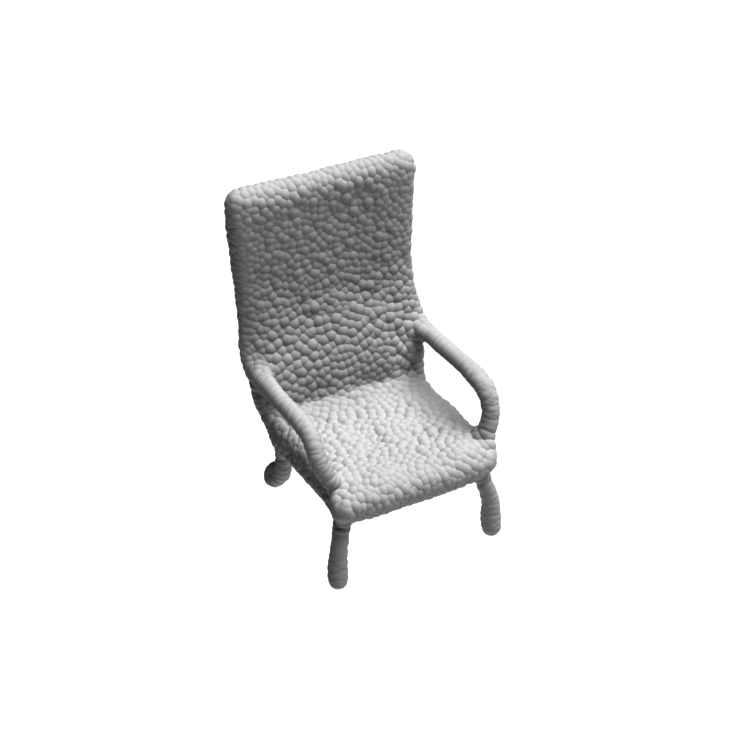}}{}}{}}{}

\rotatebox{90}{\hspace{0.7cm}\footnotesize{\emph{Thinner legs}}}
\jsubfig{\jsubfig{\jsubfig{\includegraphics[height=2.125cm, trim={0.0cm 0.0cm 0.0cm 0.0cm}, clip]{images/baselines/chair_ShapeNet_c45ff54d4192633684cd6dc1b226aa5b/pix2pix_source.png}}{}
\jsubfig{\includegraphics[height=2.125cm, trim={0.0cm 0.0cm 0.0cm 0.0cm}, clip]{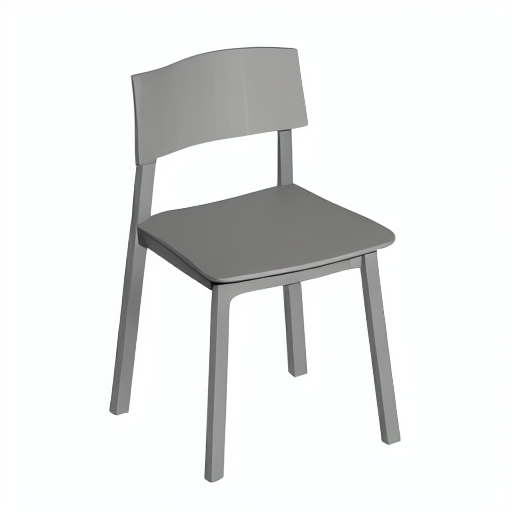}}{}
\jsubfig{\includegraphics[height=2.125cm, trim={0.0cm 0.0cm 0.0cm 0.0cm}, clip]{images/baselines/chair_ShapeNet_c45ff54d4192633684cd6dc1b226aa5b/pix2pix_output.png}}{}}{}

\jsubfig{\jsubfig{\includegraphics[height=2.125cm, trim={15.0cm 0.0cm 15.0cm 0.0cm}, clip]{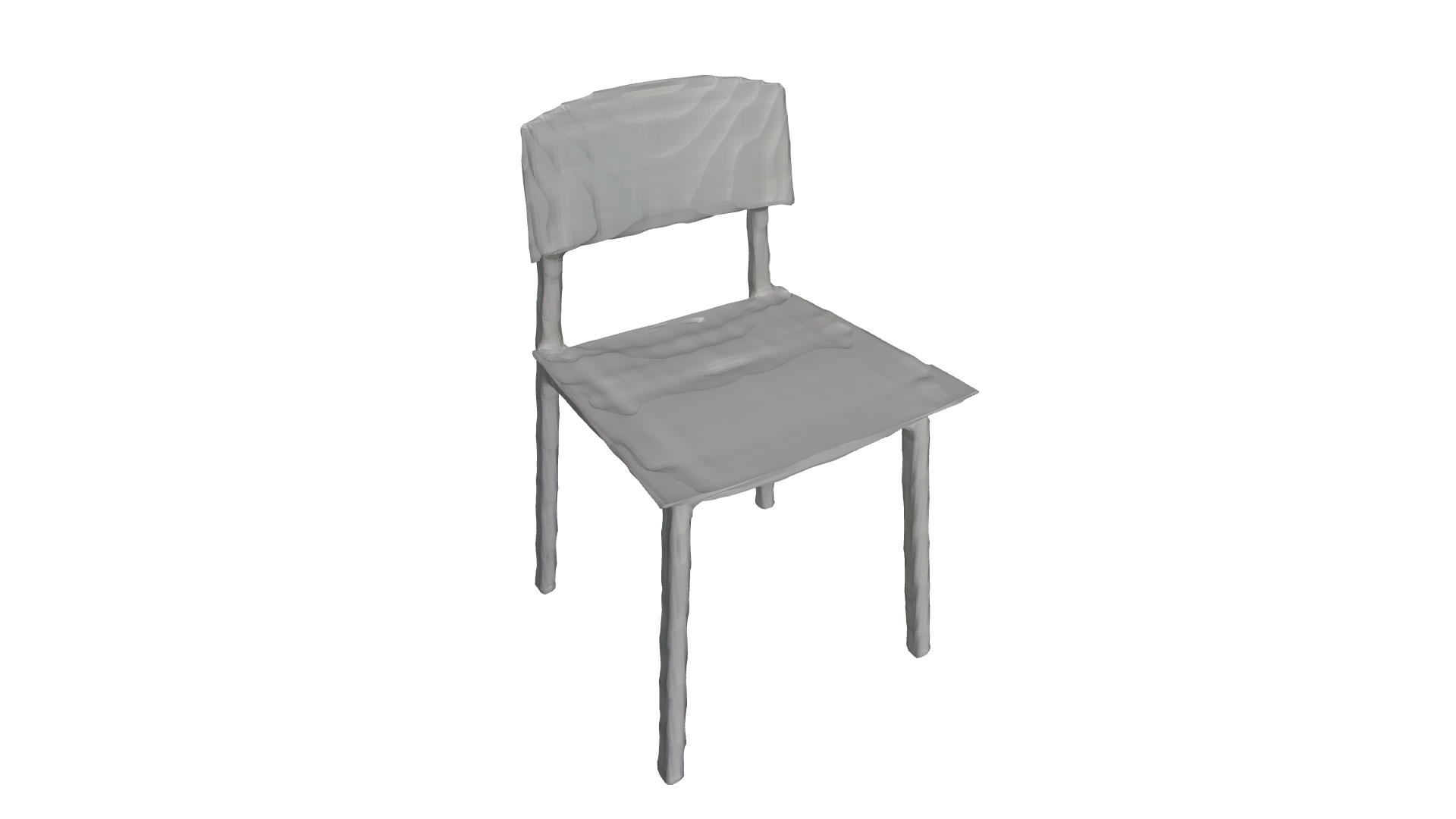}}{}
\jsubfig{\includegraphics[height=2.125cm, trim={2.5cm 1.0cm 2.5cm 4.0cm}, clip]{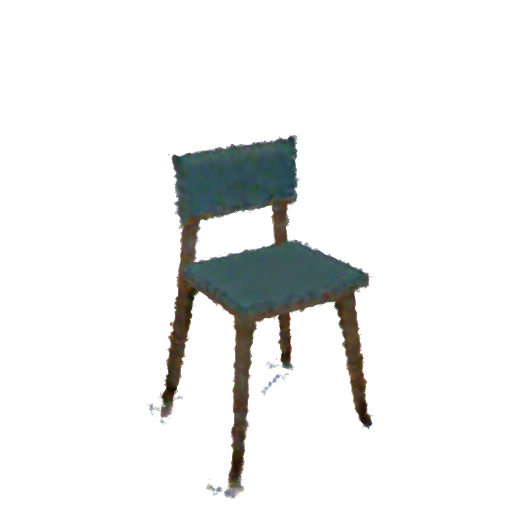}}{}
\jsubfig{\includegraphics[height=2.125cm, trim={15.0cm 0.0cm 15.0cm 0.0cm}, clip]{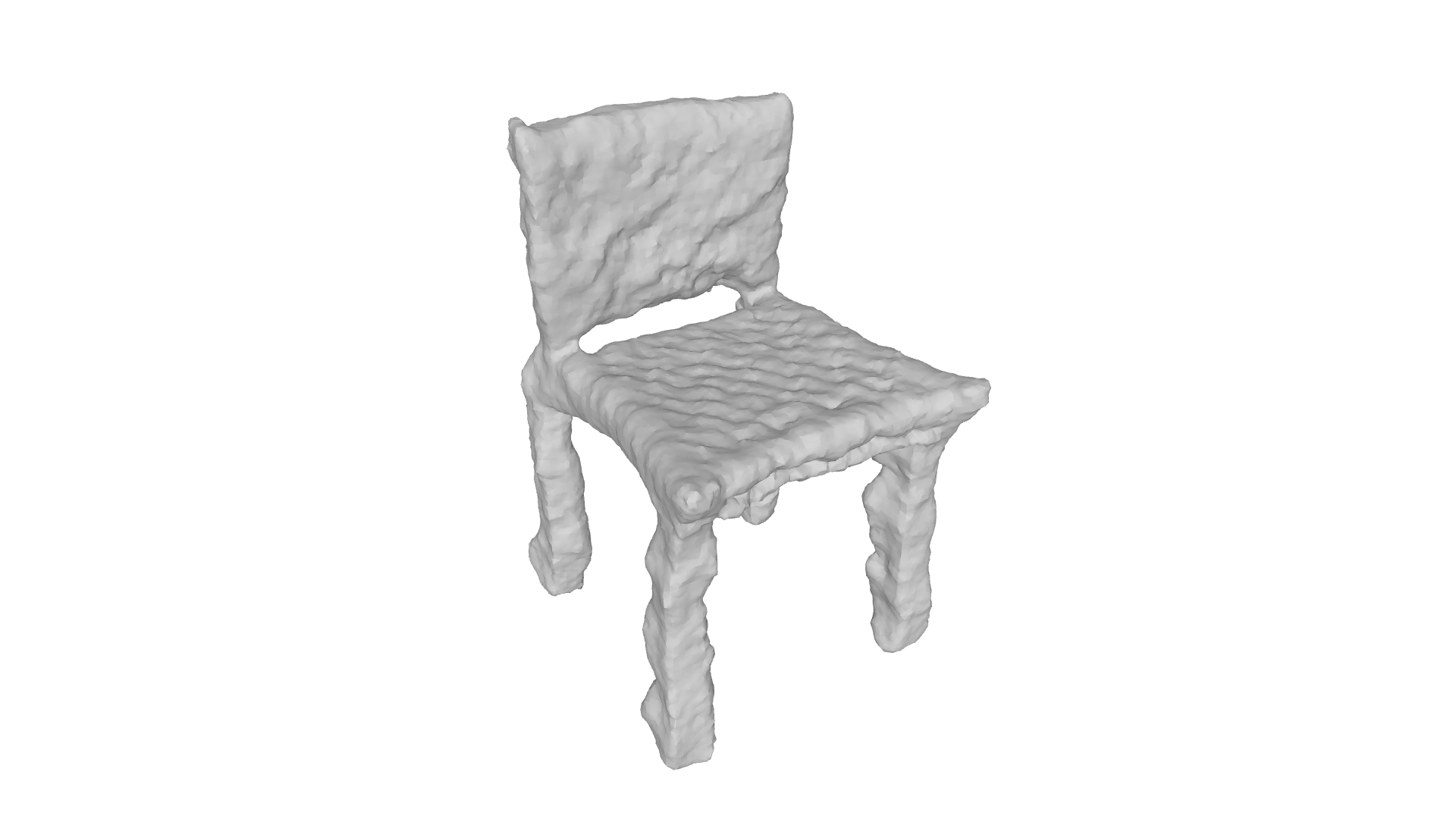}}{}}{} 

\jsubfig{\jsubfig{\includegraphics[height=2.125cm, trim={4.0cm 4.0cm 4.0cm 4.0cm}, clip]{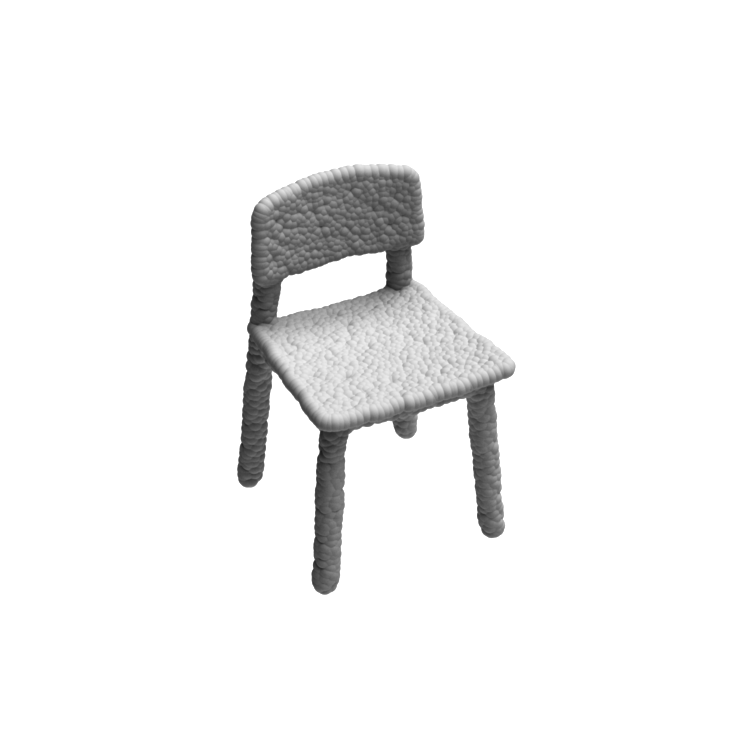}}{}
\jsubfig{\includegraphics[height=2.125cm, trim={4.0cm 4.0cm 4.0cm 4.0cm}, clip]{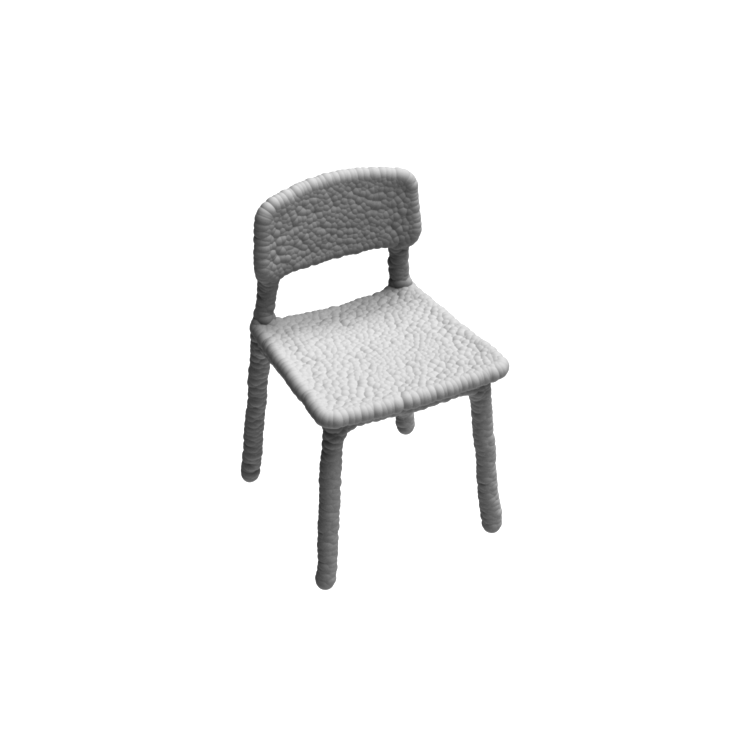}}{}}{}}{}

\rotatebox{90}{\hspace{0.3cm}\footnotesize{\emph{Taller backrest}}}
\jsubfig{\jsubfig{\jsubfig{\includegraphics[height=2.125cm, trim={0.0cm 0.0cm 0.0cm 0.0cm}, clip]{images/baselines/chair_ShapeNet_10c08a28cae054e53a762233fffc49ea/pix2pix_source.png}}{\footnotesize {Input \hspace{0.6cm} (Image)}}
\jsubfig{\includegraphics[height=2.125cm, trim={0.0cm 0.0cm 0.0cm 0.0cm}, clip]{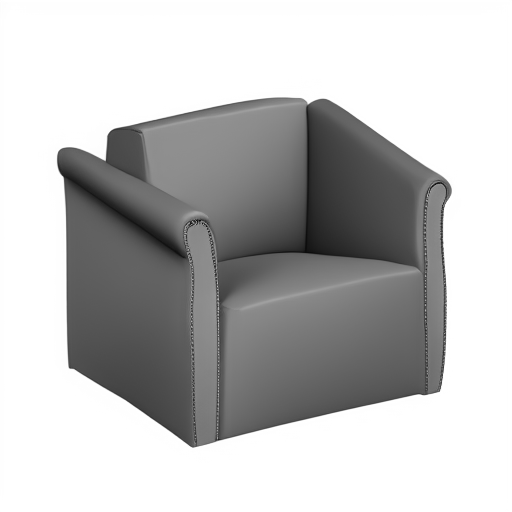}}{\footnotesize {InstructPix2Pix}}
\jsubfig{\includegraphics[height=2.125cm, trim={0.0cm 0.0cm 0.0cm 0.0cm}, clip]{images/baselines/chair_ShapeNet_10c08a28cae054e53a762233fffc49ea/pix2pix_output.png}}{\footnotesize InstructPix2Pix (fine-tuned)}}{\vspace{5pt}2D Editing}

\jsubfig{\jsubfig{\includegraphics[height=2.125cm, trim={15.0cm 0.0cm 15.0cm 0.0cm}, clip]{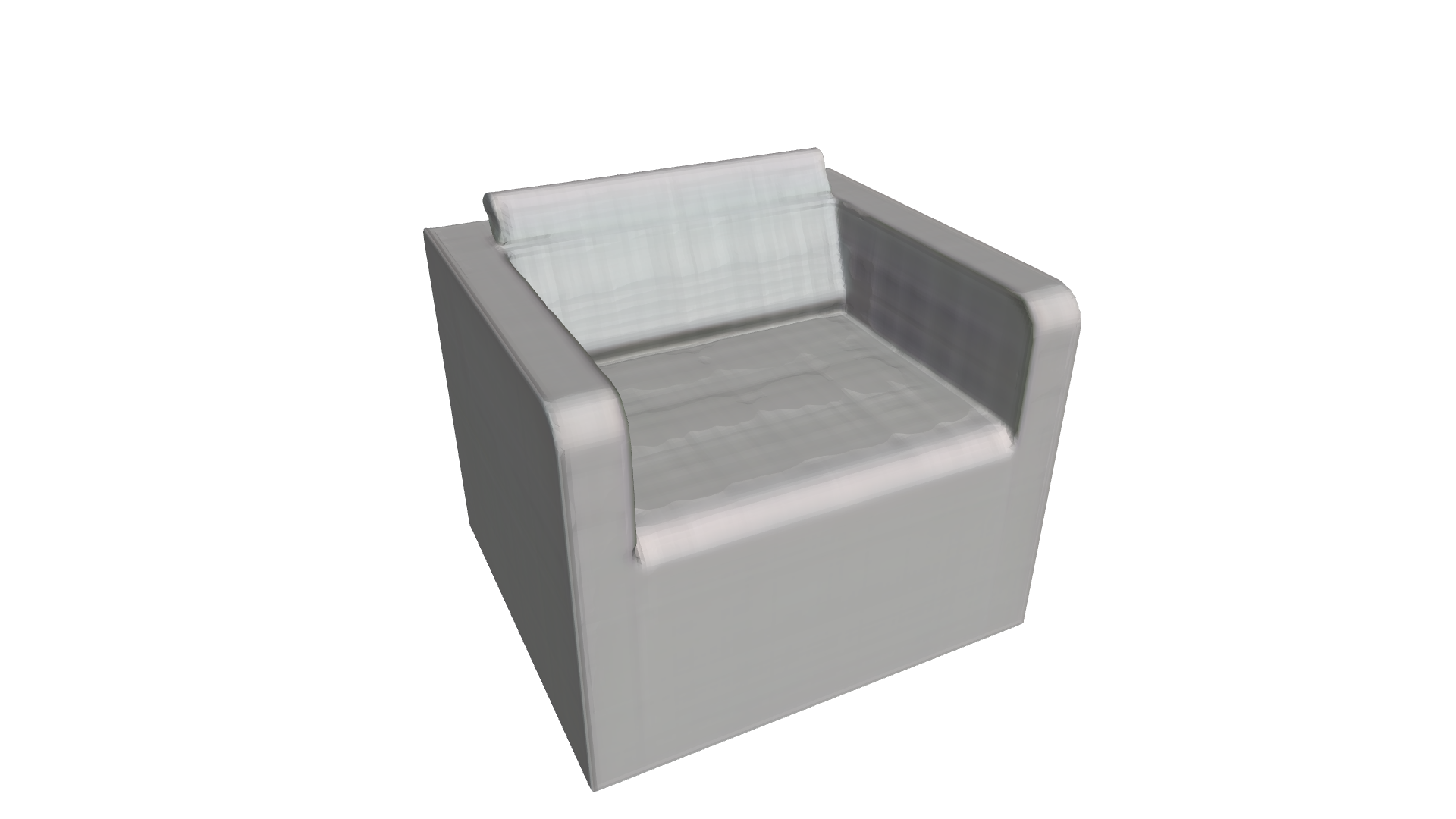}}{\footnotesize {Input \hspace{1.2cm} (3D Mesh)}}
\jsubfig{\includegraphics[height=2.125cm, trim={2.5cm 1.0cm 2.5cm 4.0cm}, clip]{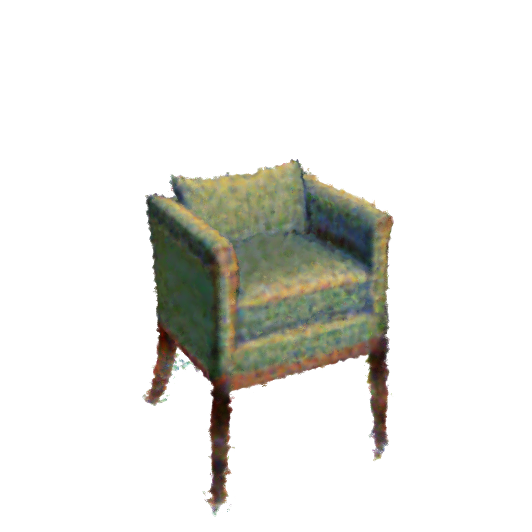}}{\footnotesize {Vox-E}}
\jsubfig{\includegraphics[height=2.125cm, trim={15.0cm 0.0cm 15.0cm 0.0cm}, clip]{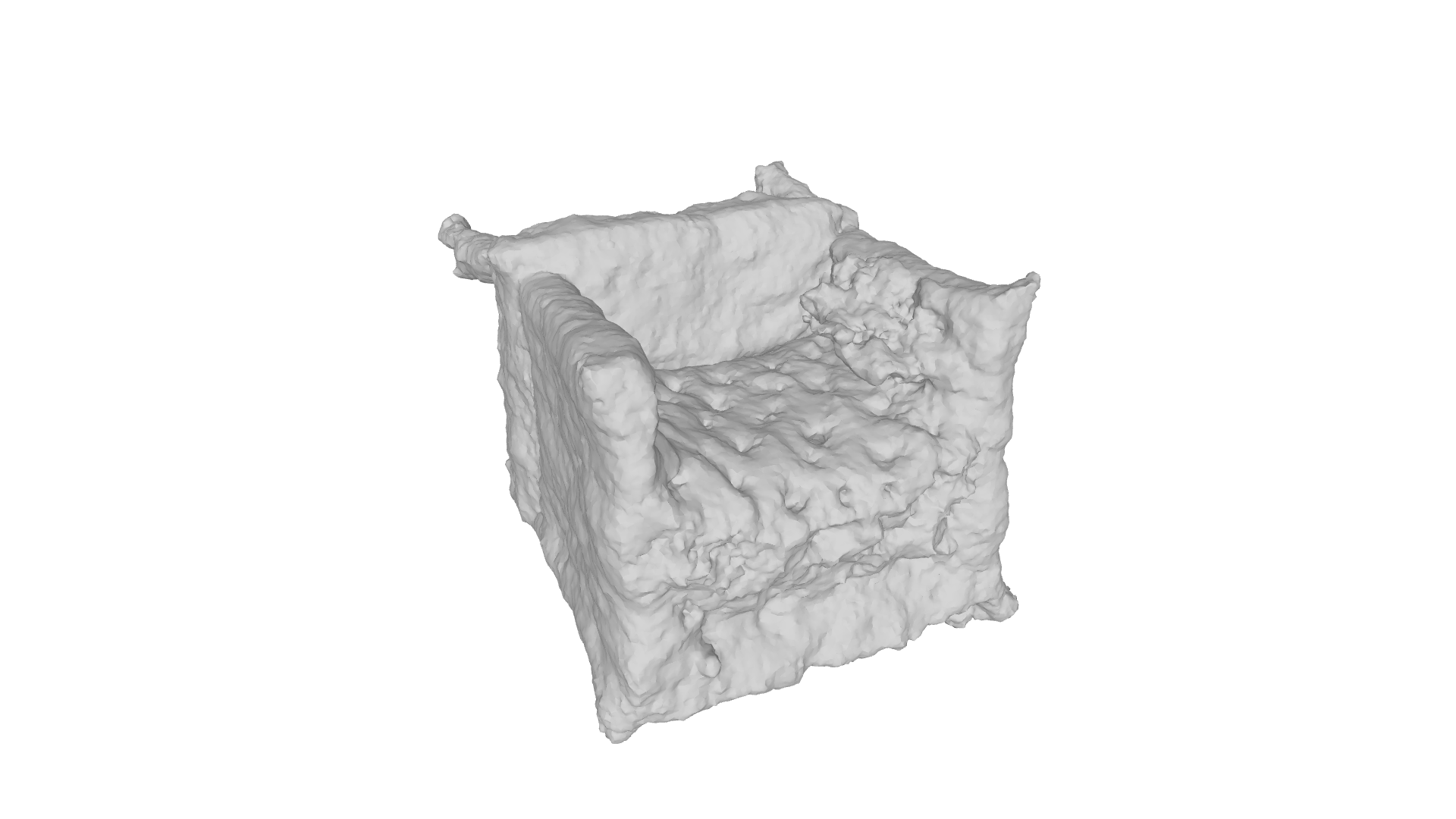}}{\footnotesize {Fantasia 3D}}}{\vspace{5pt}Score Distillation Based Editing} 

\jsubfig{\jsubfig{\includegraphics[height=2.125cm, trim={4.0cm 4.0cm 4.0cm 4.0cm}, clip]{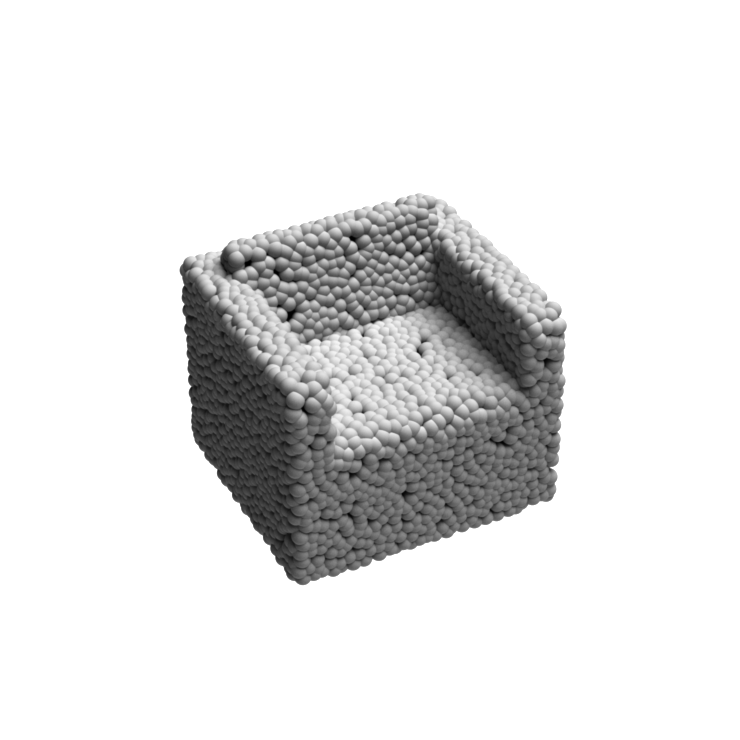}}{\footnotesize {Input \hspace{0.9cm} (Point Cloud)}}
\jsubfig{\includegraphics[height=2.125cm, trim={4.0cm 4.0cm 4.0cm 4.0cm}, clip]{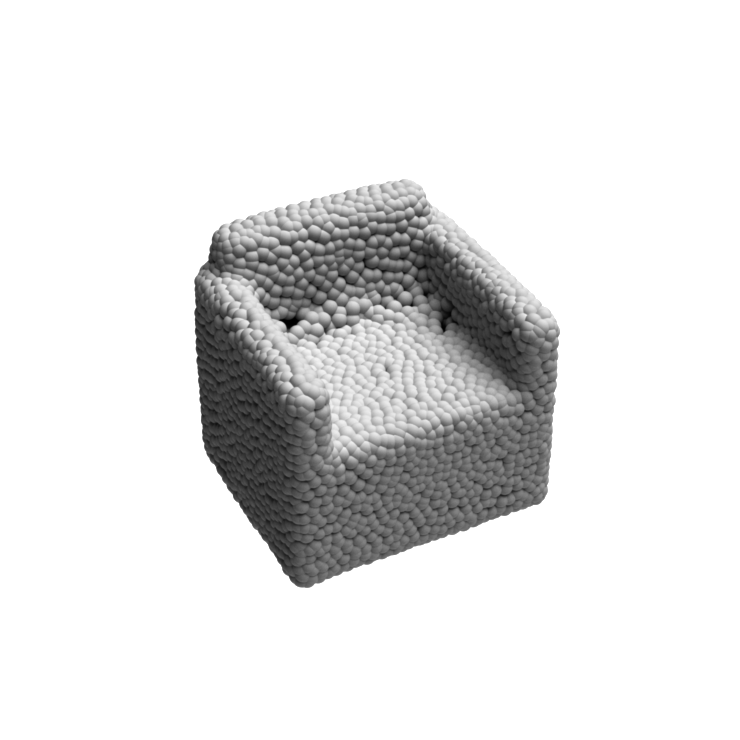}}{\footnotesize {Output}}}{\vspace{5pt}Ours}}{}

\vspace{-1pt} 
\caption{
\textbf{Qualitative comparison.} We compare our method’s
outputs to those of the image editing method InstructPix2Pix \cite{brooks2023instructpix2pix} as well as the score distillation sampling optimization based 3D editing works Fantasia 3D \cite{chen2023fantasia3d} and Vox-E \cite{sella2023vox}. As illustrated above, our method outperforms these baselines in terms of
edit fidelity, identity preservation and overall visual quality.}
\label{fig:supp_qualitative}
\end{figure*}

\medskip \noindent \textbf{Image Editing and Single View Reconstruction.} Recent advancements in image editing and 3D reconstruction methods suggest another potential paradigm: perform image editing on the rendered image of a 3D shape and then use single-view reconstruction to generate an edited 3D shape. We tested this approach by using InstructPix2Pix \cite{xu2023instructp2p} in conjunction with One-2-3-45++ \cite{liu2024one}. Specifically, we rendered an image of the input shape, provided it along with an editing prompt to InstructPix2Pix, and used the edited image as input to One-2-3-45++ to reconstruct the edited 3D shape. As shown in Figure \ref{fig:supp_qualitative}, InstructPix2Pix fail to perform fine-grained shape editing. To address this, we also fine-tuned InstructPix2Pix using images rendered from the ShapeTalk dataset, as also presented in Figure \ref{fig:supp_qualitative}. While the fine-tuned model produces higher-quality results, it still fails to localize edits effectively, especially compared to our method. For single-view reconstruction, shapes generated from InstructPix2Pix results using One-2-3-45++ are shown in Figure \ref{fig:instruct_one}. Beyond InstructPix2Pix's difficulty with precise localization, One-2-3-45++ introduces slight defects in the reconstructed shapes, such as an asymmetrical backrest (second row from the top) and slightly lopsided legs (third row from the top). However, we note that these defects are minor. The overall high visual quality of the results demonstrates promise for future research in this direction.

\begin{figure*} 
\centering %
\rotatebox[origin=l]{90}{\whitetxt{xx}\textit{armrests with}}
\rotatebox[origin=l]{90}{\whitetxt{xxx}\textit{holes at the}}
\rotatebox[origin=l]{90}{\whitetxt{xxx}\textit{at the sides}}
\hspace{16pt}
\includegraphics[height=2.56cm, trim={4.7cm, 4.7cm, 4.7cm, 4.7cm}]{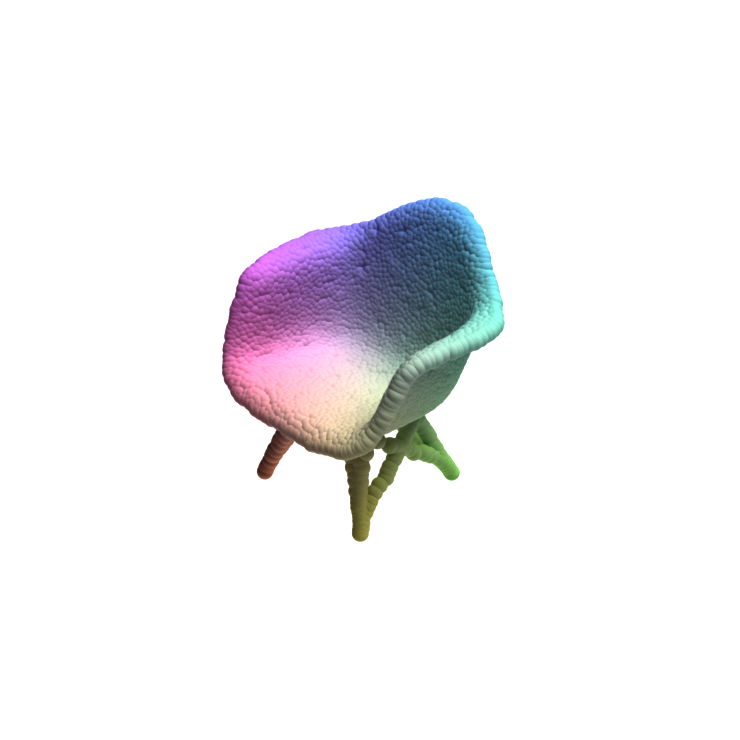}
\hfill
\includegraphics[height=2.56cm, trim={4.7cm, 4.7cm, 4.7cm, 4.7cm}]{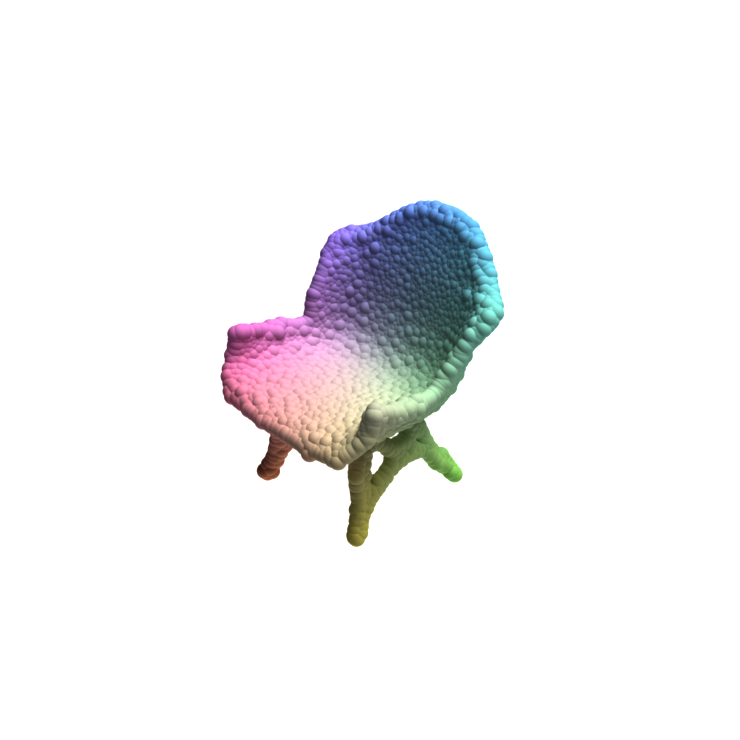}
\hfill
\includegraphics[height=2.56cm, trim={4.7cm, 4.7cm, 4.7cm, 4.7cm}]{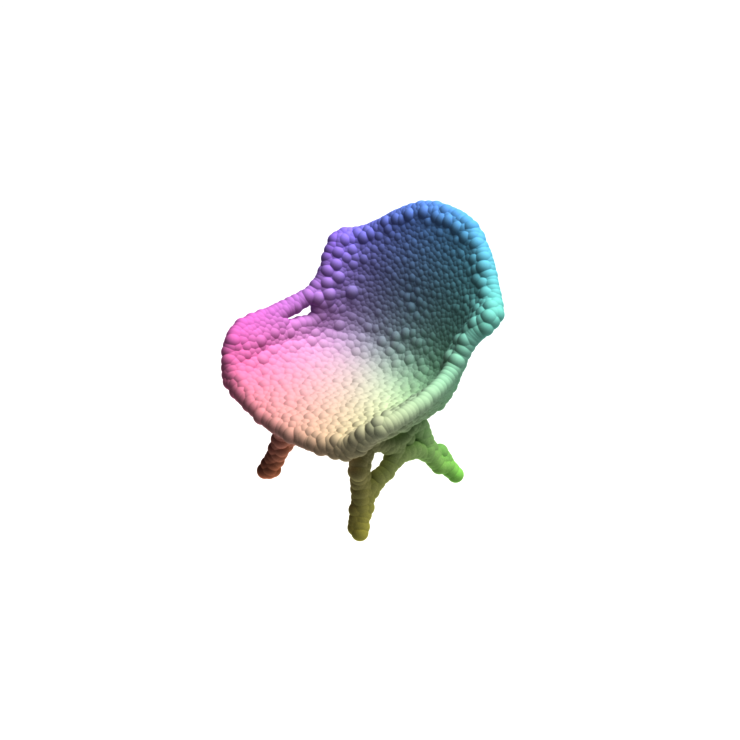}
\hfill
\includegraphics[height=2.56cm, trim={4.7cm, 4.7cm, 4.7cm, 4.7cm}]{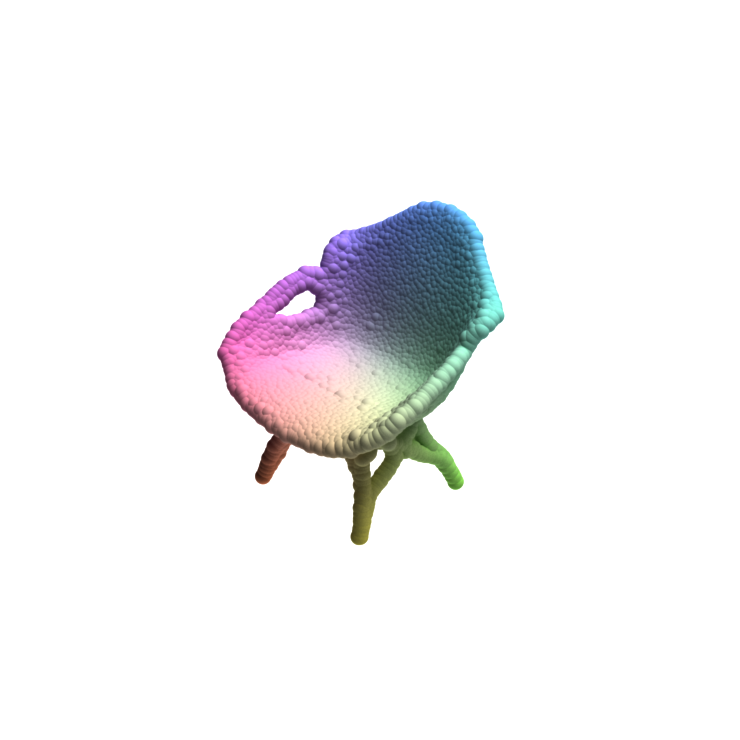}
\hfill
\includegraphics[height=2.56cm, trim={4.7cm, 4.7cm, 4.7cm, 4.7cm}]{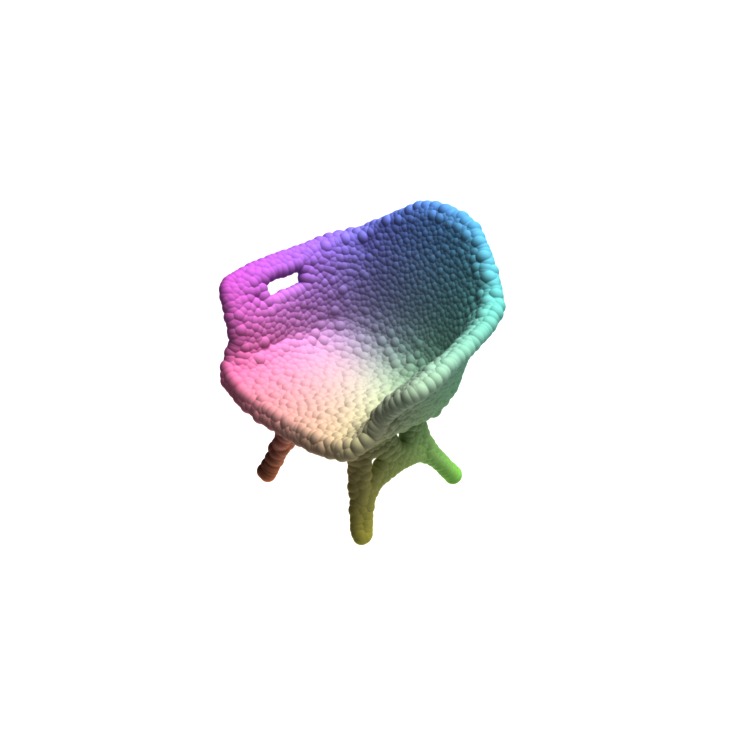}
\hfill
\includegraphics[height=2.56cm, trim={4.7cm, 4.7cm, 4.7cm, 4.7cm}]{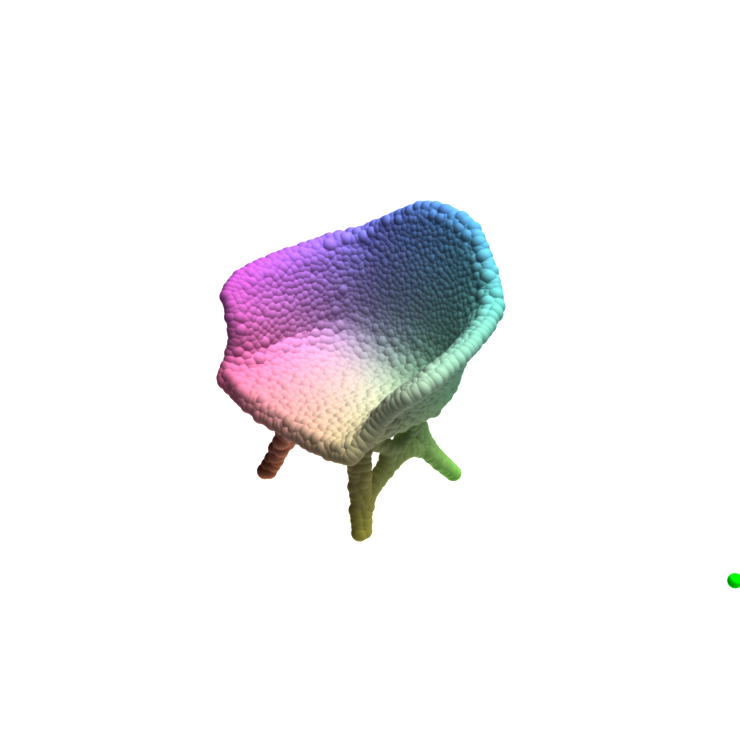} 
\\
\vspace{10pt}
\rotatebox[origin=l]{90}{\whitetxt{xx}}
\rotatebox[origin=l]{90}{\whitetxt{x}\textit{curved backrest}}
\rotatebox[origin=l]{90}{\whitetxt{xxx}}
\hspace{17pt}
\jsubfig{\includegraphics[height=2.5cm, trim={4.5cm, 4.5cm, 4.5cm, 4.5cm}]{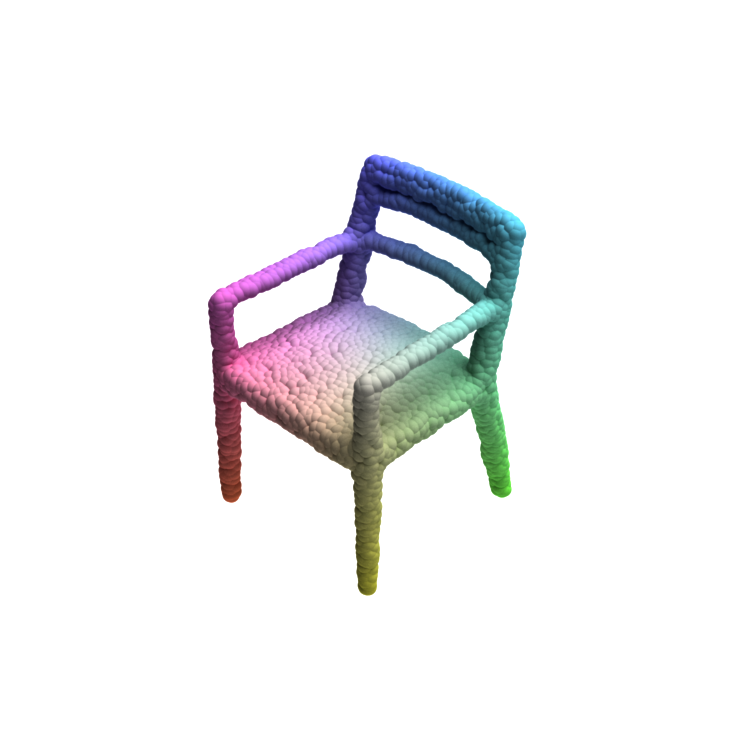}}{Source}
\hfill
\jsubfig{\includegraphics[height=2.5cm, trim={4.5cm, 4.5cm, 4.5cm, 4.5cm}]{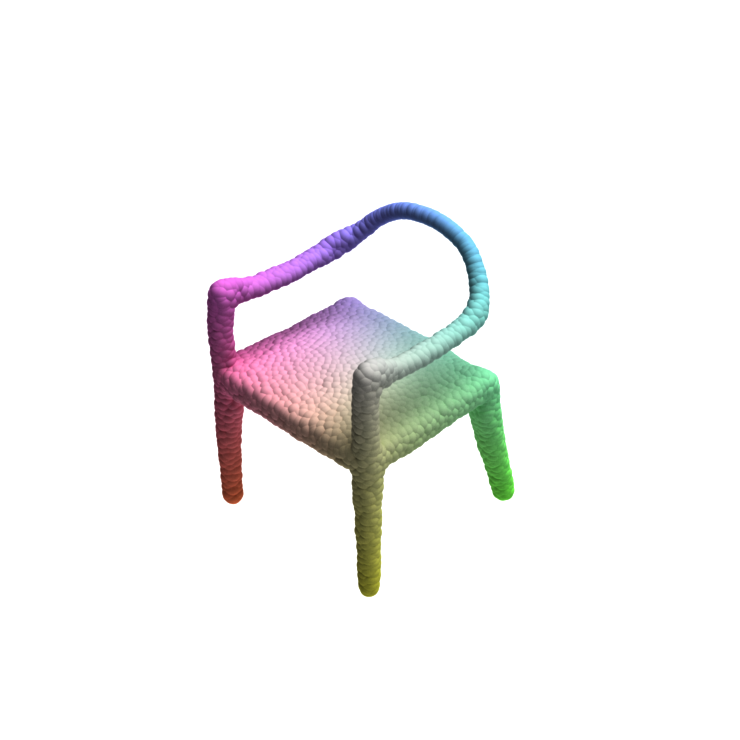}}{$t_r = 65(T)$}
\hfill
\jsubfig{\includegraphics[height=2.5cm, trim={4.5cm, 4.5cm, 4.5cm, 4.5cm}]{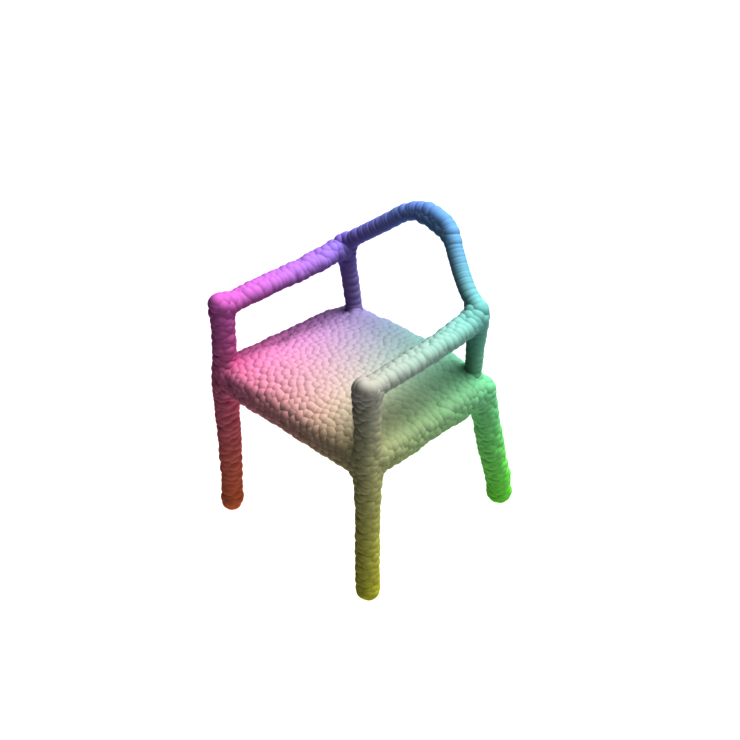}}{$t_r = 40$}
\hfill
\jsubfig{\includegraphics[height=2.5cm, trim={4.5cm, 4.5cm, 4.5cm, 4.5cm}]{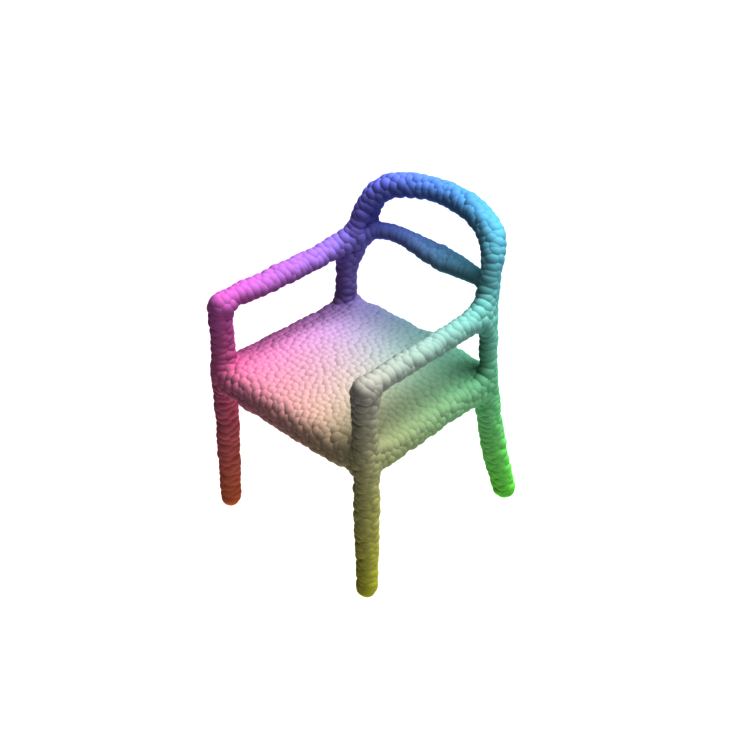}}{\boldmath$t_r = 20$}
\hfill
\jsubfig{\includegraphics[height=2.5cm, trim={4.5cm, 4.5cm, 4.5cm, 4.5cm}]{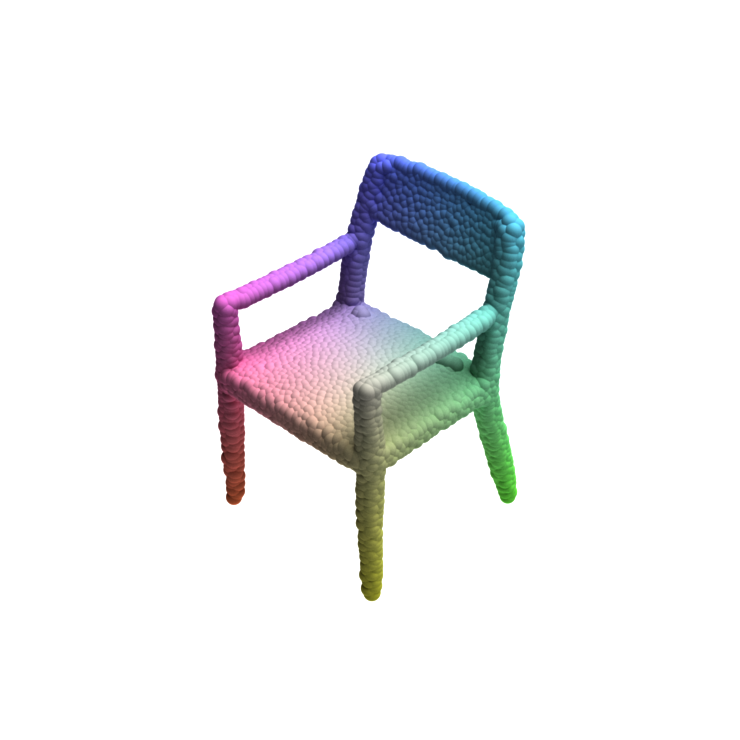}}{$t_r = 10$}
\hfill
\jsubfig{\includegraphics[height=2.5cm, trim={4.5cm, 4.5cm, 4.5cm, 4.5cm}]{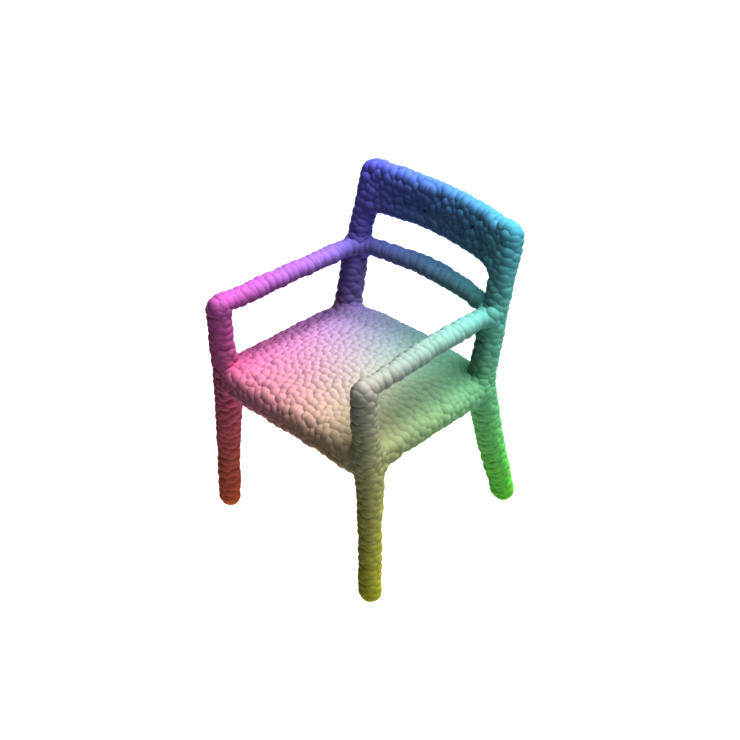}}{$t_r = 0$}

\vspace{-1pt} 
\caption{\textbf{Ablation Study for \boldmath$t_r$ Values.} Higher $t_r$ values increase the number of coordinate blending steps, providing better editing freedom but at the cost of inferior identity preservation. Conversely, lower $t_r$ values improve identity preservation while reducing the model's ability to follow the textual description.}
\label{fig:tr_experiment}
\end{figure*}

\begin{figure*} 
\centering %
\jsubfig{\includegraphics[height=1.75cm, trim={5.5cm, 5.5cm, 5.5cm, 5.5cm}]{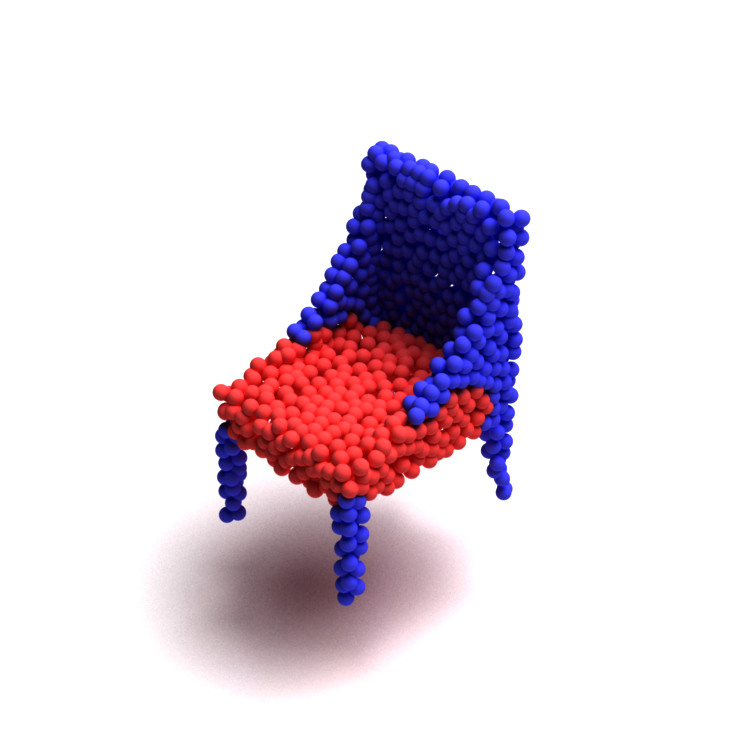}}{\vspace{16pt}Input}
\hfill
\jsubfig{\includegraphics[height=1.75cm, trim={5.5cm, 5.5cm, 5.5cm, 5.5cm}]{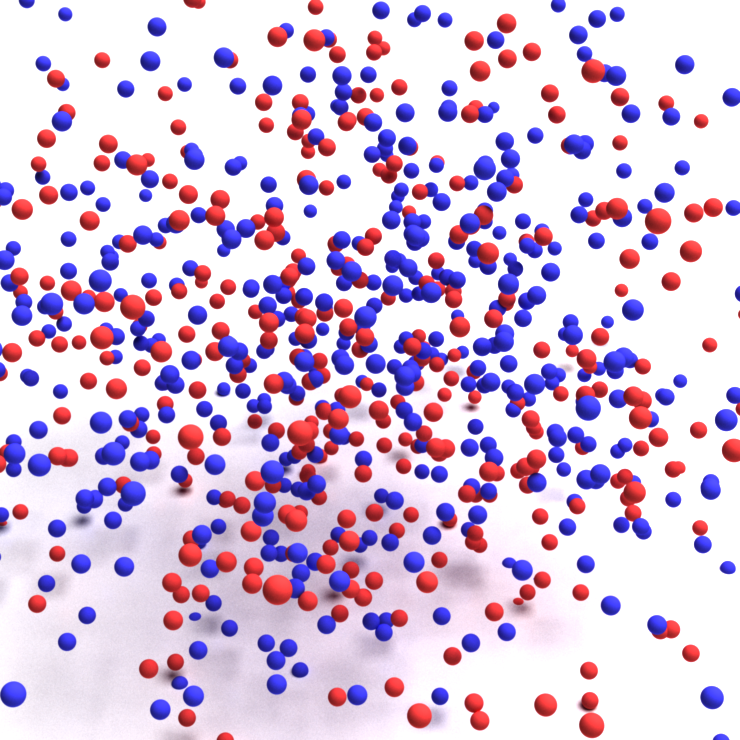}}{\vspace{16pt}$\hat{x}_{recon,T}$}
\hfill
\jsubfig{\includegraphics[height=1.75cm, trim={5.5cm, 5.5cm, 5.5cm, 5.5cm}]{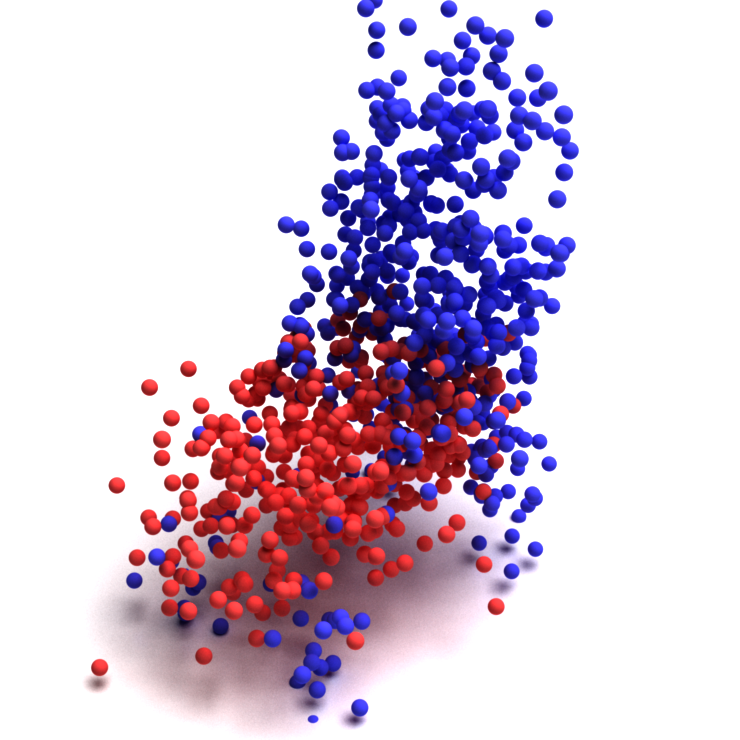}}{\vspace{16pt}$\hat{x}_{recon, 45}$}
\hfill
\jsubfig{\includegraphics[height=1.75cm, trim={5.5cm, 5.5cm, 5.5cm, 5.5cm}]{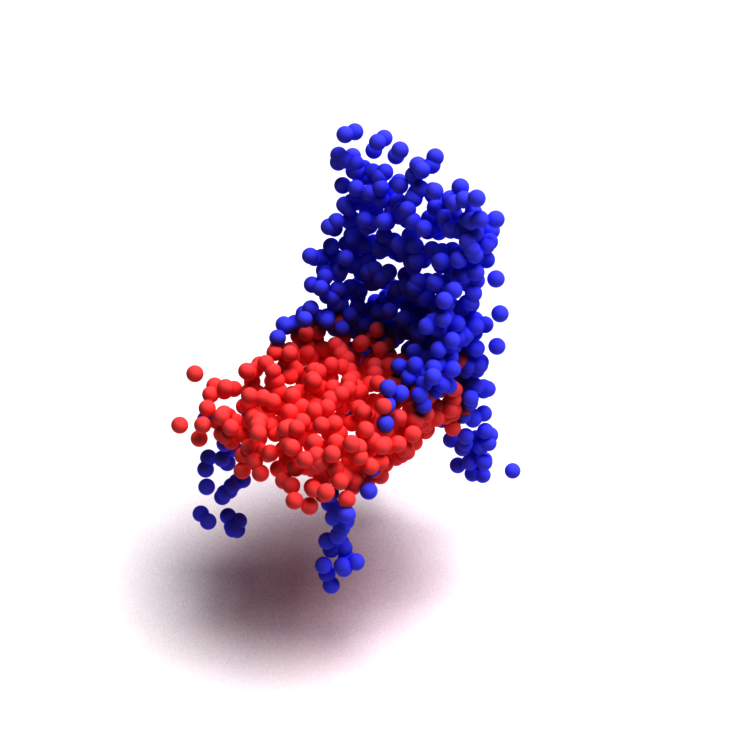}}{\vspace{16pt}$\hat{x}_{recon, 35}$}
\hfill
\jsubfig{\includegraphics[height=1.75cm, trim={5.5cm, 5.5cm, 5.5cm, 5.5cm}]{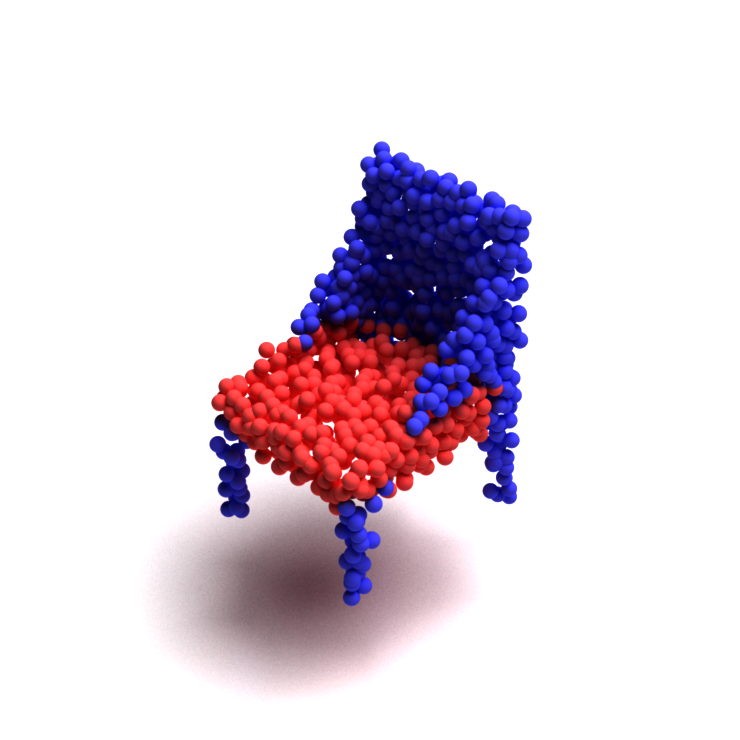}}{\vspace{16pt}$\hat{x}_{recon, 15}$}
\hfill
\jsubfig{\includegraphics[height=1.75cm, trim={5.5cm, 5.5cm, 5.5cm, 5.5cm}]{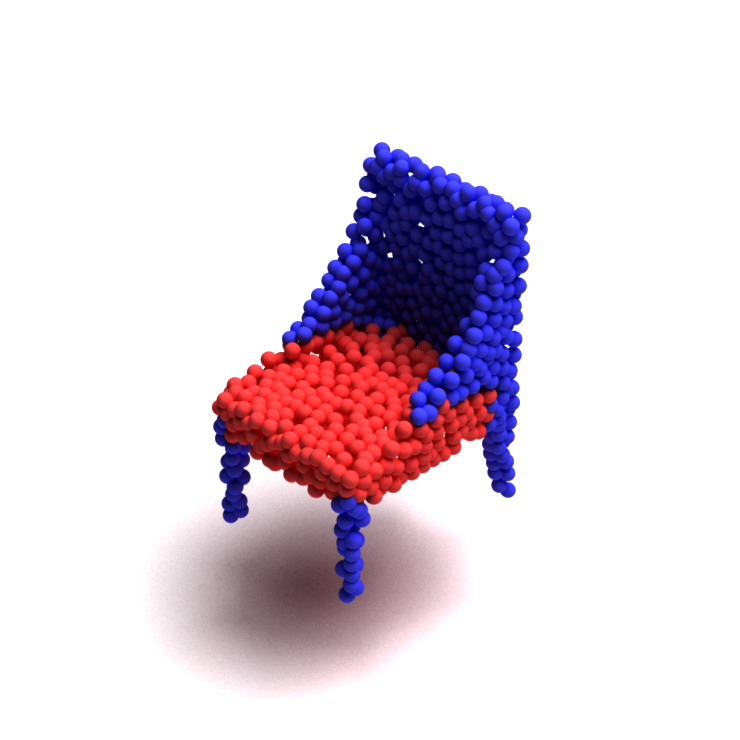}}{\vspace{16pt}$\hat{x}_{recon, 0}$}
\caption{An illustration demonstrating that the  masking module is directly applicable to the intermediate denoised point clouds; additional details are provided in the text. }

\label{fig:mask_consistancy}
\end{figure*}

\medskip \noindent \textbf{Optimization-Based Editing.}
A widely used approach for textual editing of 3D shapes is Score Distillation Sampling (SDS) \cite{poole2022dreamfusion}, which employs inference-time optimization to modify a 3D representation based on a text prompt, using a pre-trained generator as a prior. A qualitative comparison with the SDS based methods Fantasia3D \cite{chen2023fantasia3d} and Vox-E \cite{sella2023vox} is presented in Figure \ref{fig:supp_qualitative}. As shown, while Fantasia3D generates results that resemble the input shapes, it often fails to follow the fine-grained instructions in the editing prompts (e.g., the back is not rounded in the first row, and the chair in the second row lacks four legs). Vox-E also often does not follow the editing instruction (the chair on the third row does not have noticebly thinner legs, chair on the fourth row does not seem to have a taller backrest), but also often fails to correctly maintain the identity of the input objects (adding legs to the chairs on the first and last rows). Additionally, as is common with many SDS-based methods, both methods tend to exhibit noise.

\medskip \noindent \textbf{Visualizing Index Consistency in Coordinate Blending.}
The fact that masked and unmasked indexes remain consistent during the inference process may seem unintuative. For example if the point in index 0 is on the leg of the input chair and the point at index 1 is on the arm, why would these points still be part of their respective parts over the $\hat{x}_{t,recon}$ point clouds, of which the first is pure random noise? While it's true that initially, in $\hat{x}_{T,recon}$ the point indices are random and bear no semantic meaning (e.g., index 0 no longer corresponds to the leg), the reconstruction process in Point-E, and in our fine-tuned Inpaint-E, progressively restores structure in a way that aligns indices with their original semantics. This is due to our use of a permutation-variant loss (MSE) during training, which encourages consistent point ordering. Had we used a loss like Chamfer distance, this behavior would not emerge. To visualize this, in Figure \ref{fig:mask_consistancy} we show a sample input point cloud, with its ``seat'' segmentation mask shown in red. Besides it, we additionally display intermediate $\hat{x}_{recon,t}$ samples from the reconstruction process, starting from $\hat{x}_{recon,T}$ (random noise) to $\hat{x}_{recon,0}$. As the figure shows, points at indexes associated with the seat in the input point cloud gradually converge to the seat of the reconstructed point cloud.

\subsection{Ablation Study for \boldmath$t_r$ Values}
The parameter $t_r$ controls the balance between coordinate blending steps and steps dedicated solely to shape reconstruction. Higher $t_r$ values increase the number of coordinate blending steps, providing greater editing freedom but at the cost of identity preservation. Conversely, lower $t_r$ values improve identity preservation while reducing the model's ability to adhere closely to the text prompt.

Figure \ref{fig:tr_experiment} illustrates the effect of different $t_r$ values. As $t_r$ decreases, the output aligns more closely with the input point cloud. While higher $t_r$ values allow greater editing freedom, they can result in inconsistencies with the masked guidance point cloud.
We set $t_r = 20$ in our experiments as it offers a good balance between identity preservation and edit fidelity.

\begin{table*}[t]
\centering
\setlength{\tabcolsep}{2.9pt}
\def\arraystretch{1.0}
\begin{tabularx}{\textwidth}{lcccccccccccccc}
\toprule
 &           & \multicolumn{6}{c}{Shapetalk} &  & \multicolumn{6}{c}{l-Shapetalk} 
                          \\ 
\cmidrule(lr){3-8} \cmidrule(lr){10-15} 
Metric && $\text{CLIP}_{Sim}\uparrow$& $\text{CLIP}_{Dir}\downarrow$ & 
GD$\downarrow$ & CD$\downarrow$ & FPD$\downarrow$ & l-GD$\downarrow$ && $\text{CLIP}_{Sim}\uparrow$& $\text{CLIP}_{Dir}\downarrow$ & GD$\downarrow$ & CD$\downarrow$ & FPD$\downarrow$ & l-GD$\downarrow$ \\ \midrule
Unified Model && \textbf{0.26} &	{1.00} &	0.32 &	0.03 & 43.05&	0.10
 && 0.26 &	{1.00} &	0.36&	0.09&	138.26&	0.11
\\  
Ours && \textbf{0.26}&	{\textbf{0.99}} & \textbf{0.34} & \textbf{0.05}&	\textbf{33.64}	&\textbf{0.07}
 && \textbf{0.27}	& {\textbf{0.99}} &	\textbf{0.29}&	\textbf{0.04}	&\textbf{13.51}	&\textbf{0.05}
\\ 
\bottomrule
\end{tabularx}
\vspace{-8pt}
\caption{\textbf{Comparison to Unified Model.} Our technique performs effectively across multiple categories training, as evidenced by the comparable results.}
\label{tab:metrics_ablations}
\end{table*}

\subsection{Multi Category Training}

Table \ref{tab:metrics_ablations} demonstrates that our technique is not restricted to single-category training. We trained our model on a unified dataset comprising all three categories combined. The results show that our method handles multiple categories effectively, with the unified model producing results comparable to those of the single-category models.

\begin{table*}[t]
\centering
\setlength{\tabcolsep}{2.6pt}
\def\arraystretch{1.0}
\begin{tabularx}{\textwidth}{llcccccccccccccc}
\toprule
 &&           & \multicolumn{6}{c}{Shapetalk} &  & \multicolumn{6}{c}{l-Shapetalk} 
                          \\ 
\cmidrule(lr){4-9} \cmidrule(lr){11-16} 
& Metric && $\text{CLIP}_{Sim}\uparrow$& $\text{CLIP}_{Dir}\uparrow$ & 
GD$\downarrow$ & CD$\downarrow$ & FPD$\downarrow$ & l-GD$\downarrow$ && $\text{CLIP}_{Sim}\uparrow$& $\text{CLIP}_{Dir}\uparrow$ &  GD$\downarrow$ & CD$\downarrow$ & FPD$\downarrow$ & l-GD$\downarrow$ \\ \midrule
\multirow{3}{*}{\rotatebox[origin=c]{90}{\emph{Chair}}}
& Changeit3D && 0.22&	-0.02&	0.48&	0.05&	127.35&	0.18
 && 0.22&	-0.02&	0.68&	0.07	&156.11&	0.21
\\ 
& Spice-E && 0.26&	0.00&	1.78&	0.18	&527.33&	0.27
 && 0.26&	0.00&	1.78&	0.22&	653.16&	0.32
\\  
& Ours && \textbf{0.28}&	\textbf{0.01}	&\textbf{0.27}	&\textbf{0.02}&	\textbf{14.13}	&\textbf{0.07}
 && \textbf{0.28}	&\textbf{0.01}&	\textbf{0.24}&	\textbf{0.01}	&\textbf{5.87}	& \textbf{0.05}
\\ 
\midrule
\multirow{3}{*}{\rotatebox[origin=c]{90}{\emph{Table}}}
& Changeit3D && 0.22&	-0.03&	0.55&	0.16&	111.27&	0.21
 && 0.22&	-0.04&	0.66&	0.13	&141.12&	0.21
\\ 
& Spice-E && 0.25&	-0.01&	1.85&	0.40	&489.32&	0.30
 && \textbf{0.26}&	-0.04&	2.85&	0.40&	621.43&	0.38
\\  
& Ours && \textbf{0.27}&	\textbf{0.01}	&\textbf{0.33}	&\textbf{0.03}&	\textbf{18.96}	&\textbf{0.06}
 && \textbf{0.26}	&\textbf{0.01}&	\textbf{0.33}&	\textbf{0.03}	&\textbf{11.50}	&\textbf{0.05}
\\
\midrule
\multirow{3}{*}{\rotatebox[origin=c]{90}{\emph{Lamp}}}
& Changeit3D && 0.19&	-0.01&	0.93&	0.33&	310.44&	0.20
 && 0.19&	-0.02&	1.01&	0.38	&354.63&	0.19
\\ 
& Spice-E && 0.23&	-0.02&	1.88&	0.14	&153.41&	0.40
 && 0.23&	-0.02&	2.16&	0.16&	188.56&	0.45
\\  
& Ours && \textbf{0.24}&	\textbf{0.00}	&\textbf{0.41}	&\textbf{0.09}&	\textbf{67.82}	&\textbf{0.09}
 && \textbf{0.26}	&\textbf{0.01}&	\textbf{0.31}&	\textbf{0.07}	&\textbf{23.18}	&\textbf{0.05}
\\ 
\bottomrule
\end{tabularx}
\vspace{-8pt}
    \caption{\textbf{Per-Category Evaluation }. We compare the performance of ChangeIt3D \cite{achlioptas2022changeit3d} and Spice-E \cite{sella2024spice} against ours over the three object categories in the ShapeTalk and l-ShapeTalk datasts.}
\label{tab:metrics_breakdown}
\end{table*}

\subsection{Per-Category Evaluation}
Table \ref{tab:metrics_breakdown} provides the evaluation results for each of the three object categories used for quantitative evaluation. The per-category scores align closely with the overall average scores reported in the main paper.

\begin{figure*} 
\rotatebox{90}{%
  \parbox{3.0cm}{%
    \centering
    \emph{Rectangular wings}%
  }%
}
\jsubfig{\jsubfig{\includegraphics[height=2.7cm, trim={2cm 2cm 2cm 2cm}, clip]{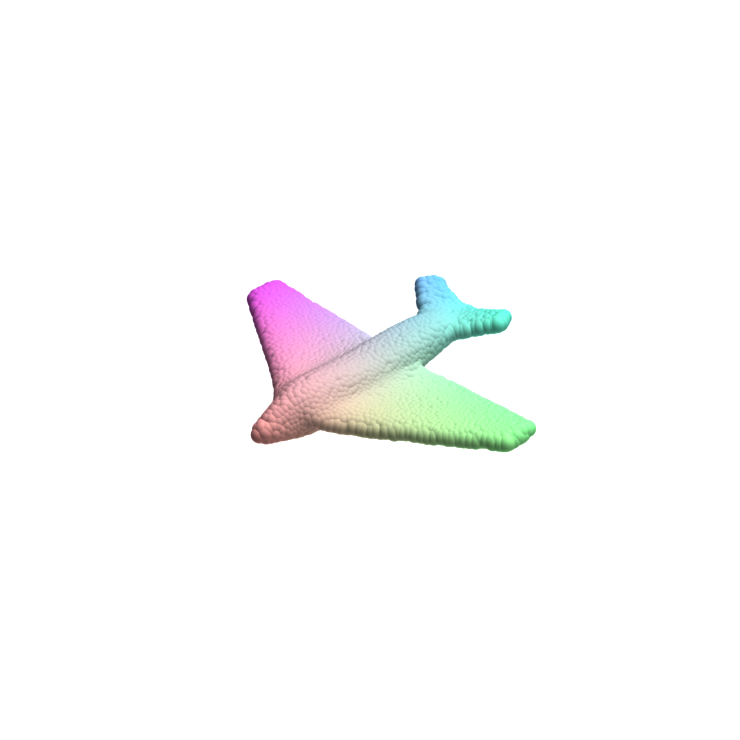}}{}%
\hspace{-0.7cm}
\jsubfig{\includegraphics[height=2.7cm, trim={2cm 2cm 2cm 2cm}, clip]{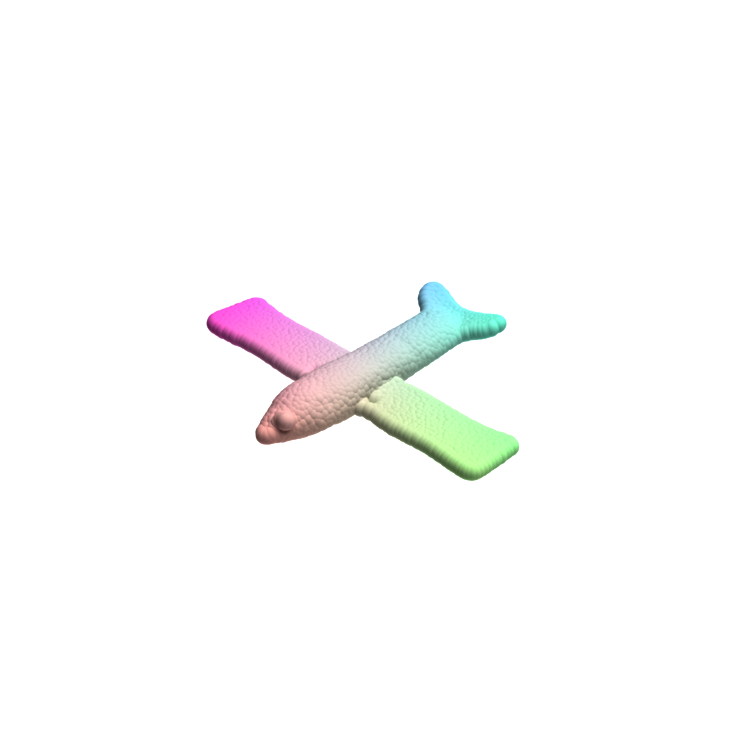}}{}}%

\hspace{0.5cm}
\rotatebox{90}{%
  \parbox{3.0cm}{%
    \centering
    \emph{There are four legs}%
  }%
}
\jsubfig{\jsubfig{\includegraphics[height=2.7cm, trim={3cm 3cm 3cm 3cm}, clip]{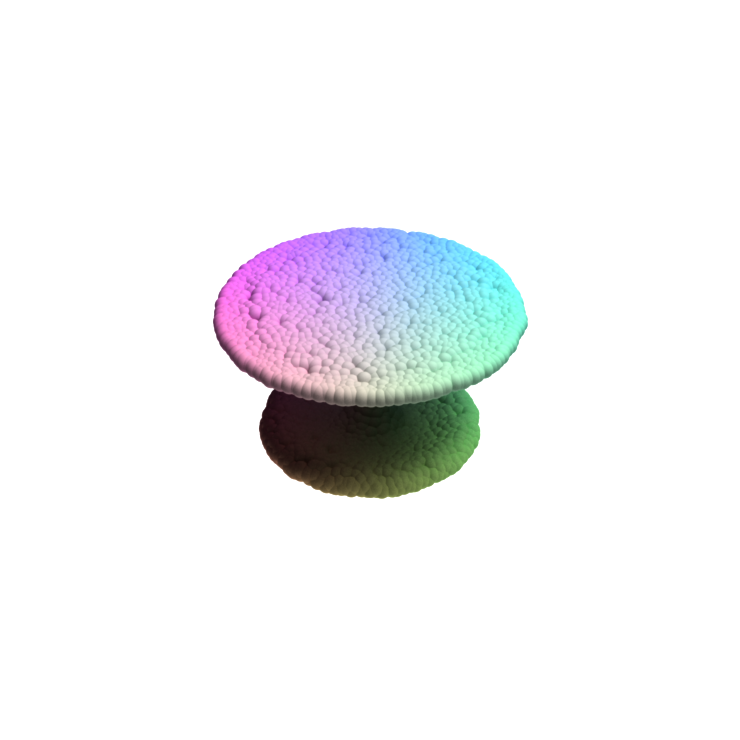}}{}%
\hspace{-0.7cm}
\jsubfig{\includegraphics[height=2.7cm, trim={3cm 3cm 3cm 3cm}, clip]{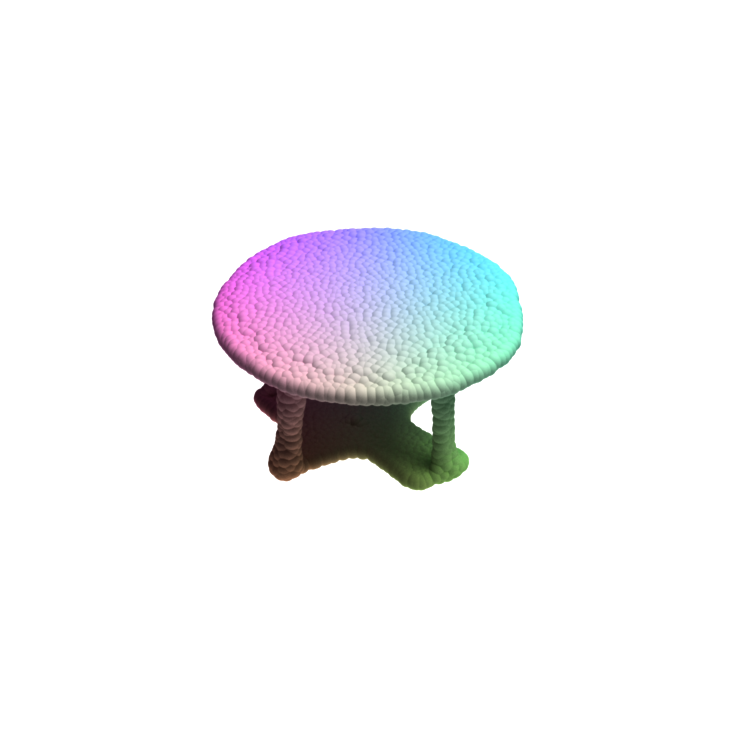}}{}}%

\hspace{0.5cm}
\rotatebox{90}{%
  \parbox{3.0cm}{%
    \centering
    \emph{Square body}%
  }%
}
\jsubfig{\jsubfig{\includegraphics[height=2.7cm, trim={3cm 3cm 3cm 3cm}, clip]{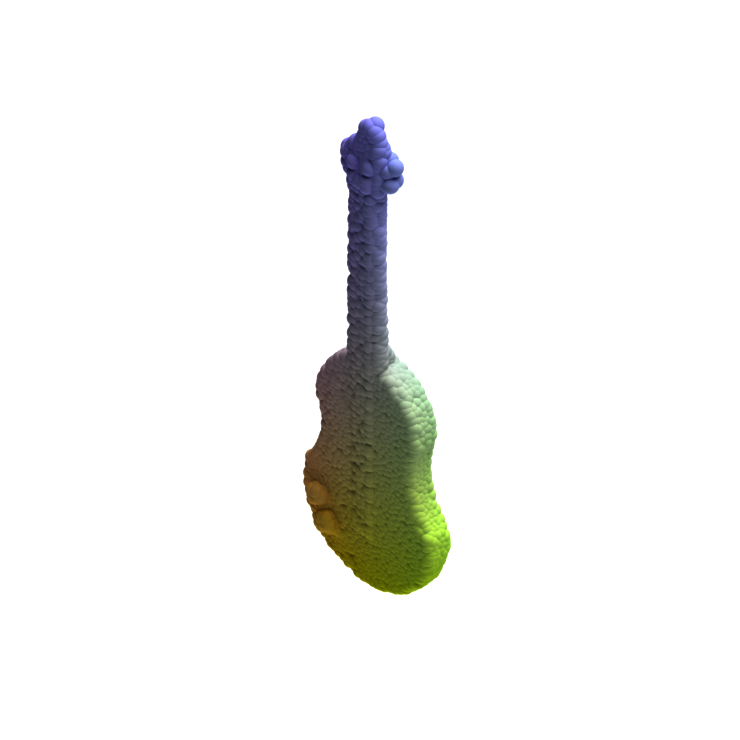}}{}%
\hspace{-0.7cm}
\jsubfig{\includegraphics[height=2.7cm, trim={3cm 3cm 3cm 3cm}, clip]{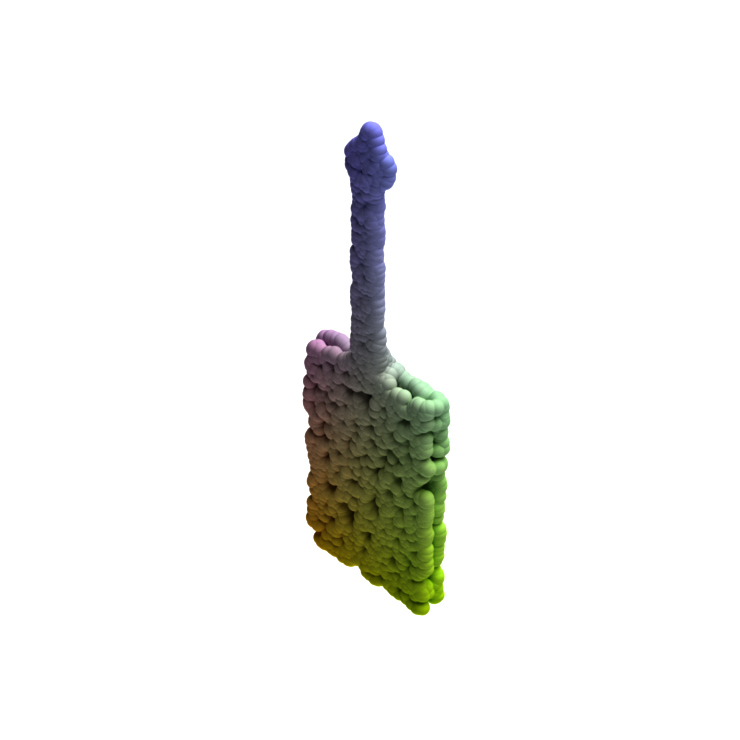}}{}}

\\  

\vspace{0.1cm} 
\rotatebox{90}{%
  \parbox{3.0cm}{%
    \centering
    \emph{Less rolled top}%
  }%
}
\jsubfig{\jsubfig{\includegraphics[height=2.7cm, trim={1cm 1cm 1cm 1cm}, clip]{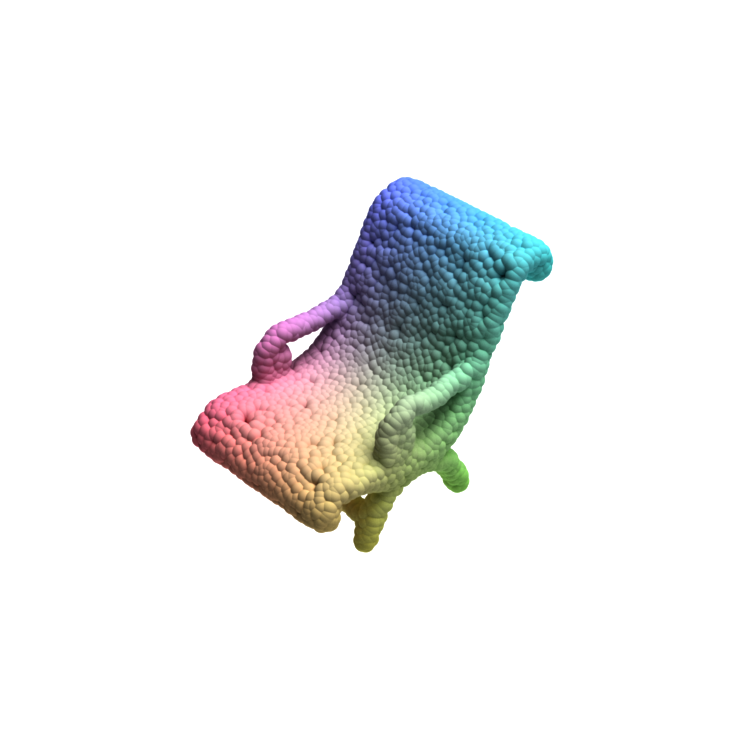}}{}%
\hspace{-0.7cm}
\jsubfig{\includegraphics[height=2.7cm, trim={1cm 1cm 1cm 1cm}, clip]{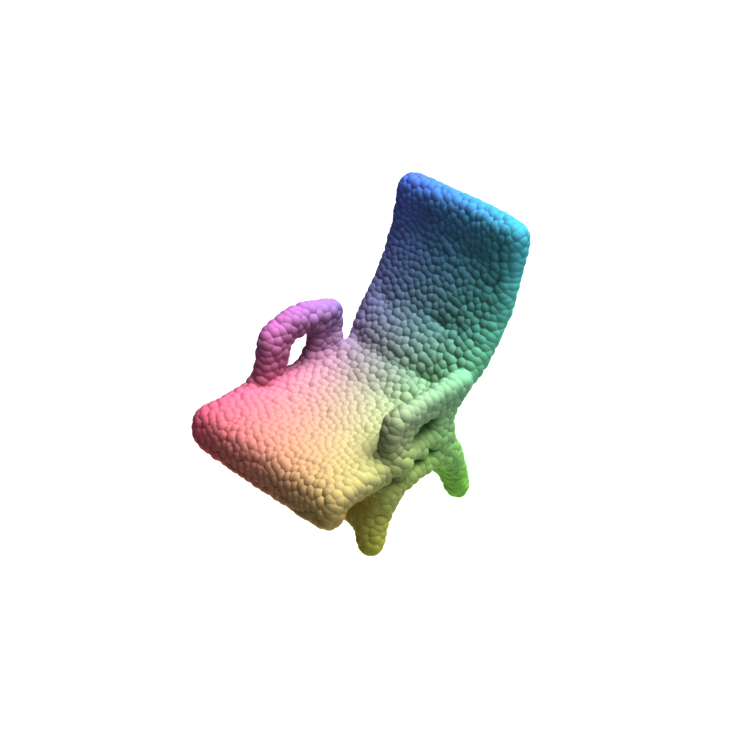}}{}}%

\hspace{0.5cm}
\rotatebox{90}{%
  \parbox{3.0cm}{%
    \centering
    \emph{Thicker board}%
  }%
}
\jsubfig{\jsubfig{\includegraphics[height=2.7cm, trim={2cm 2cm 2cm 2cm}, clip]{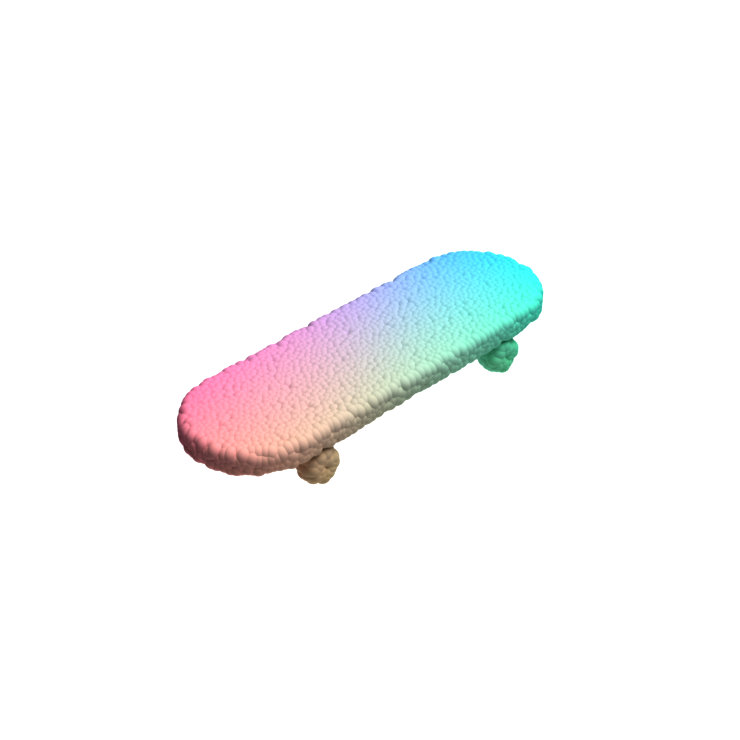}}{}%
\hspace{-0.7cm}
\jsubfig{\includegraphics[height=2.7cm, trim={2cm 2cm 2cm 2cm}, clip]{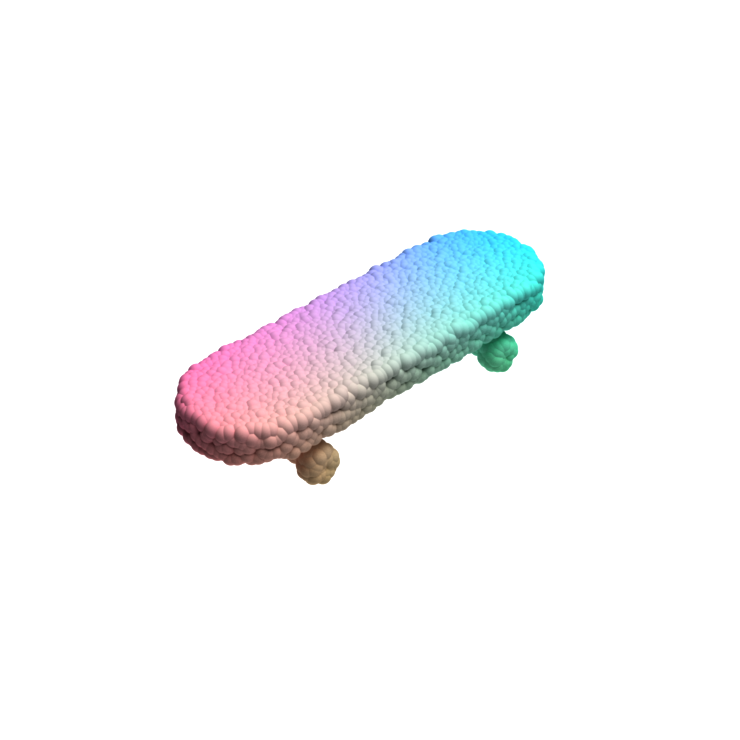}}{}}%

\hspace{0.5cm}
\rotatebox{90}{%
  \parbox{3.0cm}{%
    \centering
    \emph{Smooth blade}%
  }%
}
\jsubfig{\jsubfig{\includegraphics[height=2.7cm, trim={3cm 3cm 3cm 3cm}, clip]{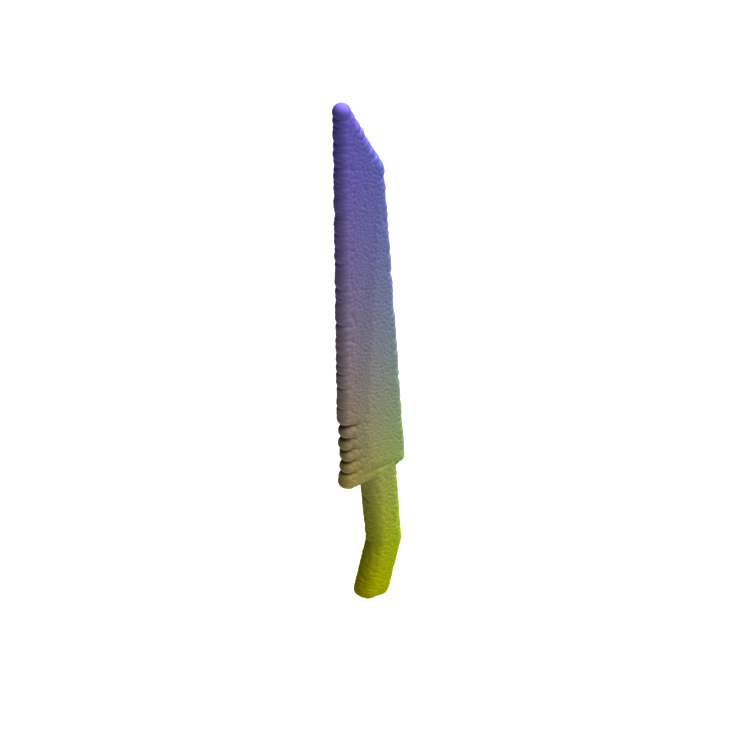}}{}%
\hspace{-0.7cm}
\jsubfig{\includegraphics[height=2.7cm, trim={3cm 3cm 3cm 3cm}, clip]{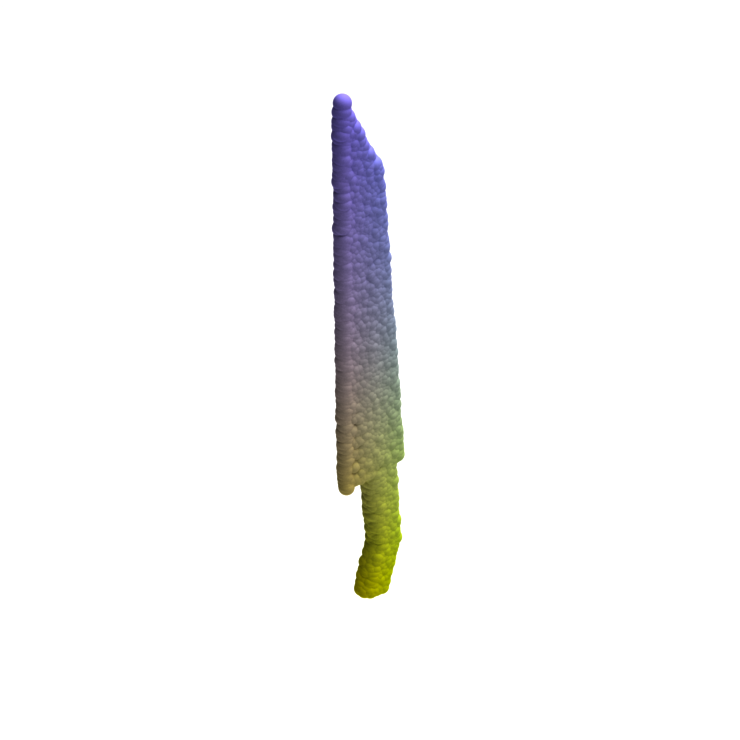}}{}}%

\\

\vspace{0.1cm} 
\rotatebox{90}{%
  \parbox{3.0cm}{%
    \centering
    \emph{The handle is shorter}%
  }%
}
\jsubfig{\jsubfig{\includegraphics[height=2.7cm, trim={3cm 3cm 3cm 3cm}, clip]{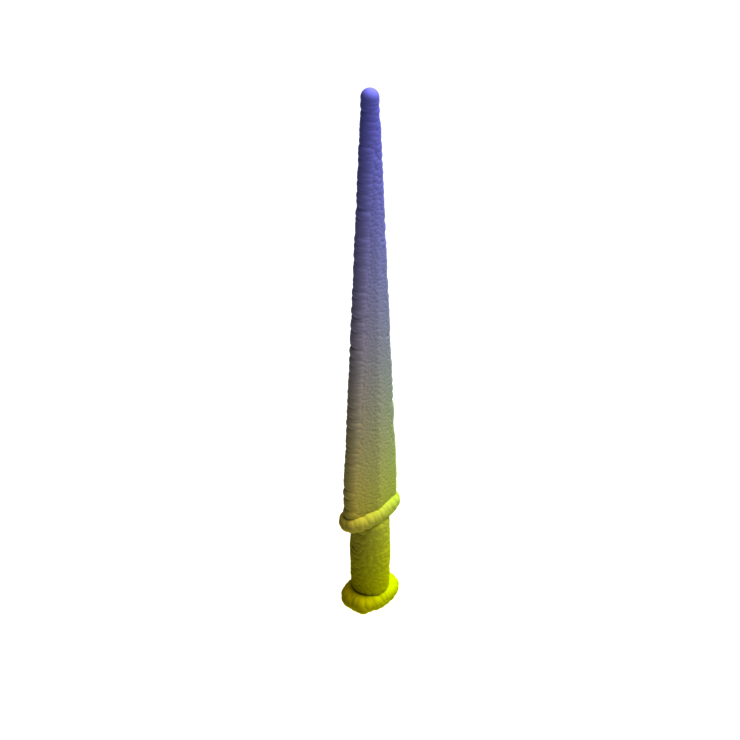}}{}%
\hspace{-0.7cm}
\jsubfig{\includegraphics[height=2.7cm, trim={3cm 3cm 3cm 3cm}, clip]{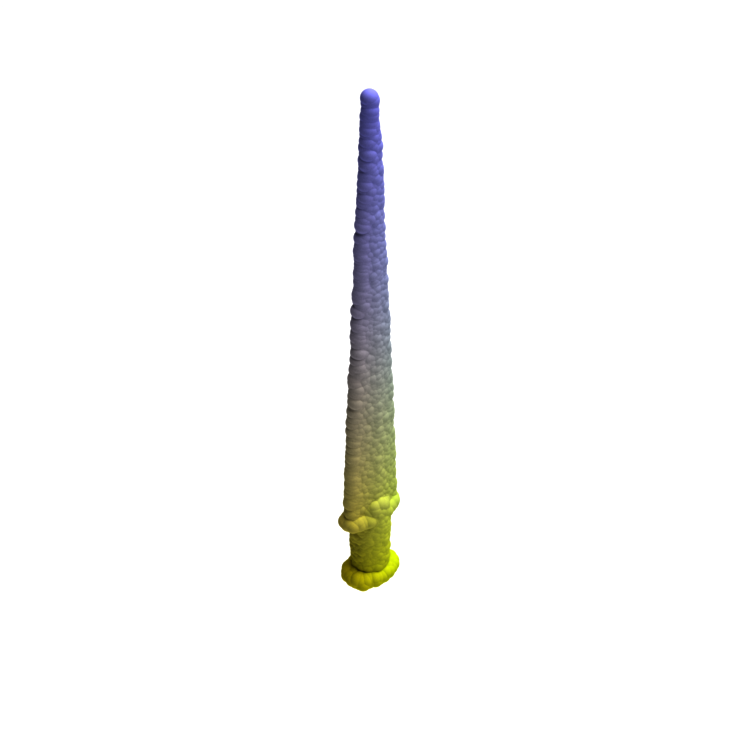}}{}}%

\hspace{0.5cm}
\rotatebox{90}{%
  \parbox{3.0cm}{%
    \centering
    \emph{It has no engines}%
  }%
}
\jsubfig{\jsubfig{\includegraphics[height=2.7cm, trim={3cm 3cm 3cm 3cm}, clip]{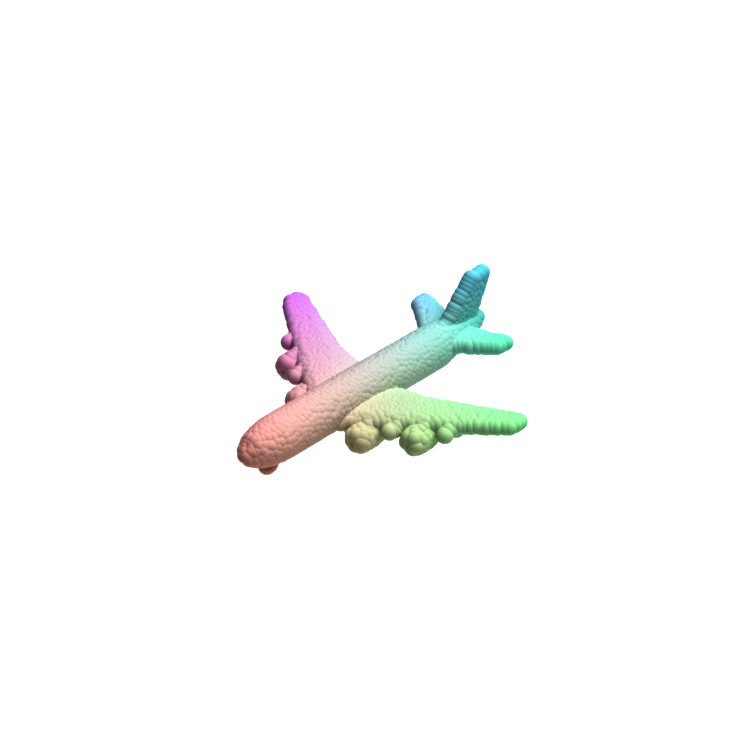}}{}%
\hspace{-0.7cm}
\jsubfig{\includegraphics[height=2.7cm, trim={3cm 3cm 3cm 3cm}, clip]{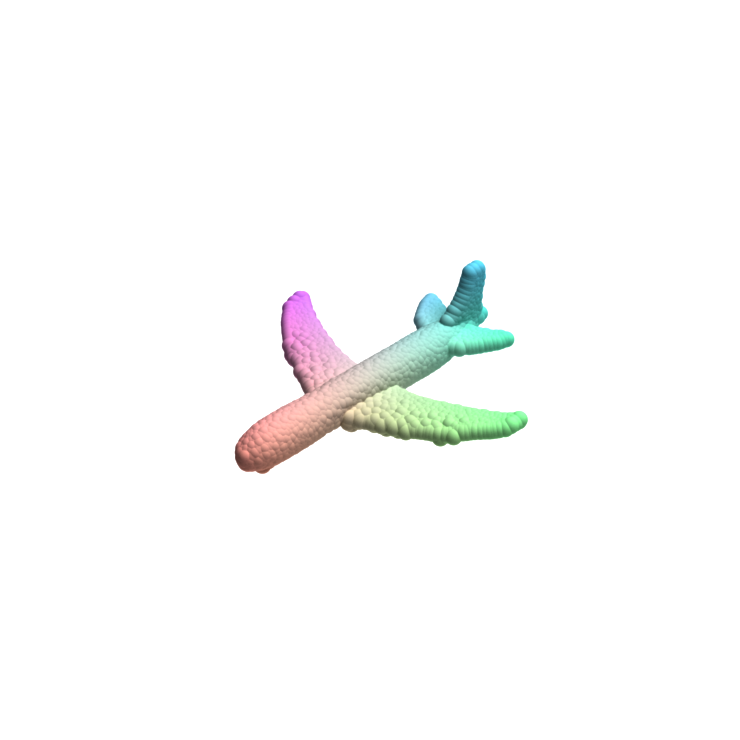}}{}}%

\hspace{0.5cm}
\rotatebox{90}{%
  \parbox{3.0cm}{%
    \centering
    \emph{Thinner pole}%
  }%
}
\jsubfig{\jsubfig{\includegraphics[height=2.7cm, trim={3cm 3cm 3cm 3cm}, clip]{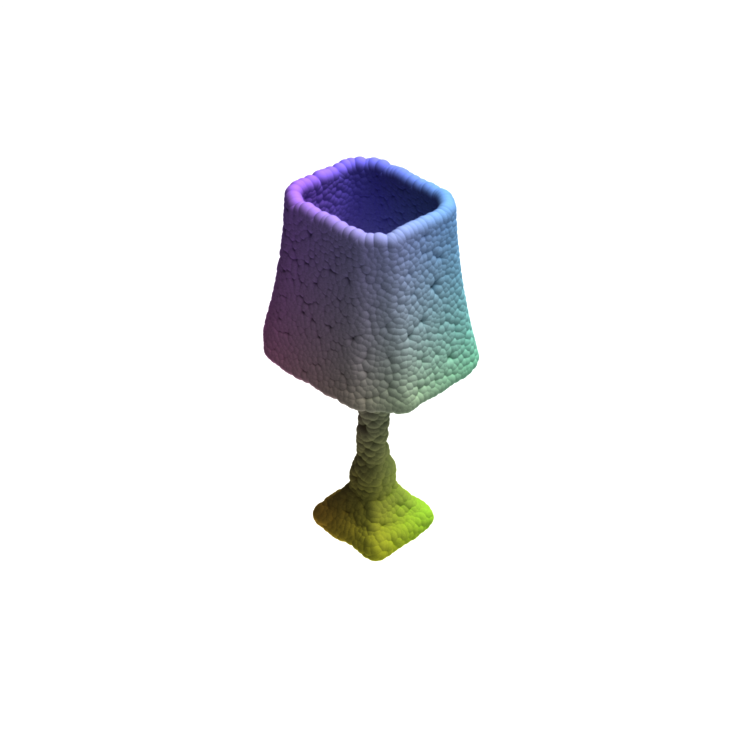}}{}%
\hspace{-0.7cm}
\jsubfig{\includegraphics[height=2.7cm, trim={3cm 3cm 3cm 3cm}, clip]{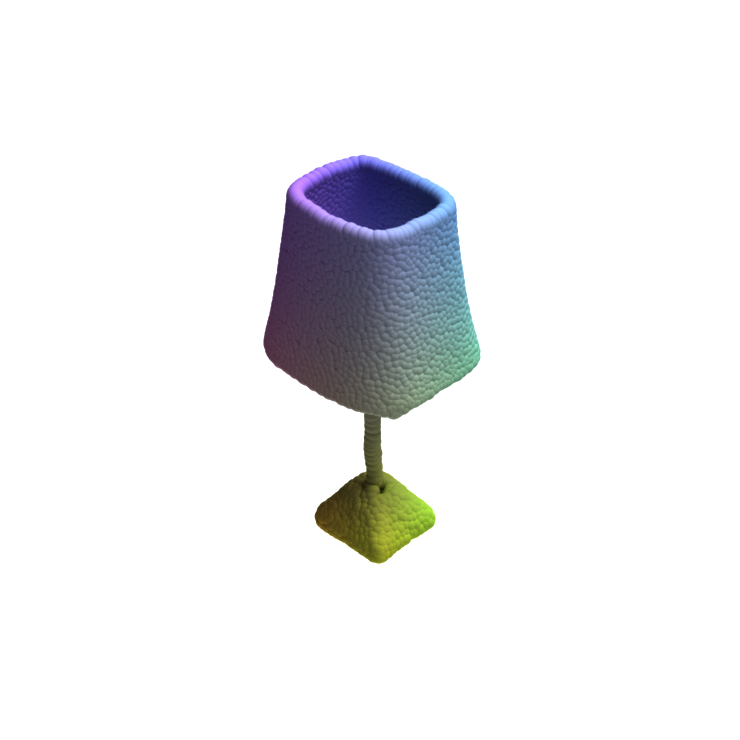}}{}}%

\\

\vspace{0.1cm} 
\rotatebox{90}{%
  \parbox{3.0cm}{%
    \centering
    \emph{Thinner wings}%
  }%
}
\jsubfig{\jsubfig{\includegraphics[height=2.7cm, trim={2cm 2cm 2cm 2cm}, clip]{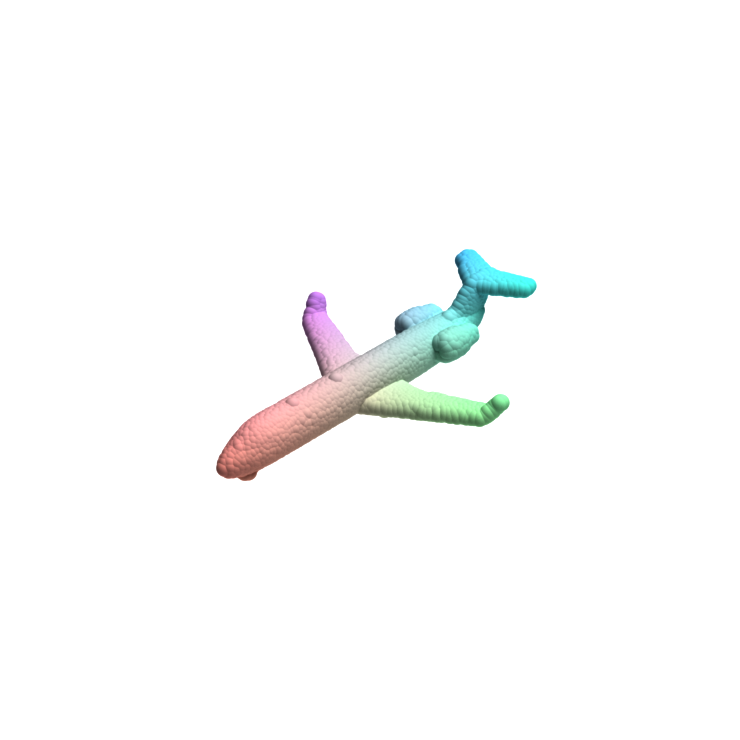}}{}%
\hspace{-0.7cm}
\jsubfig{\includegraphics[height=2.7cm, trim={2cm 2cm 2cm 2cm}, clip]{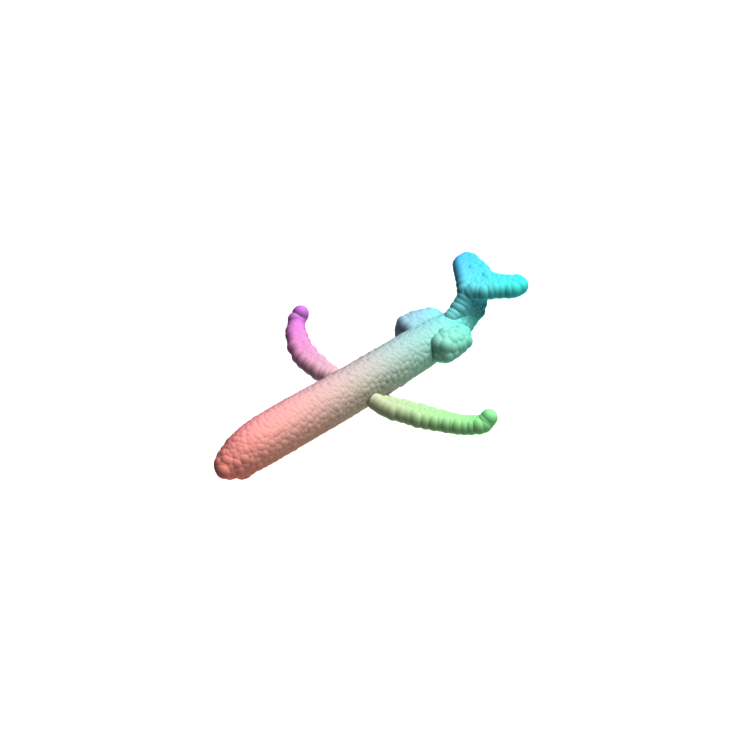}}{}}

\hspace{0.5cm}
\rotatebox{90}{%
  \parbox{3.0cm}{%
    \centering
    \emph{Has a round shade}%
  }%
}
\jsubfig{\jsubfig{\includegraphics[height=2.7cm, trim={3cm 3cm 3cm 3cm}, clip]{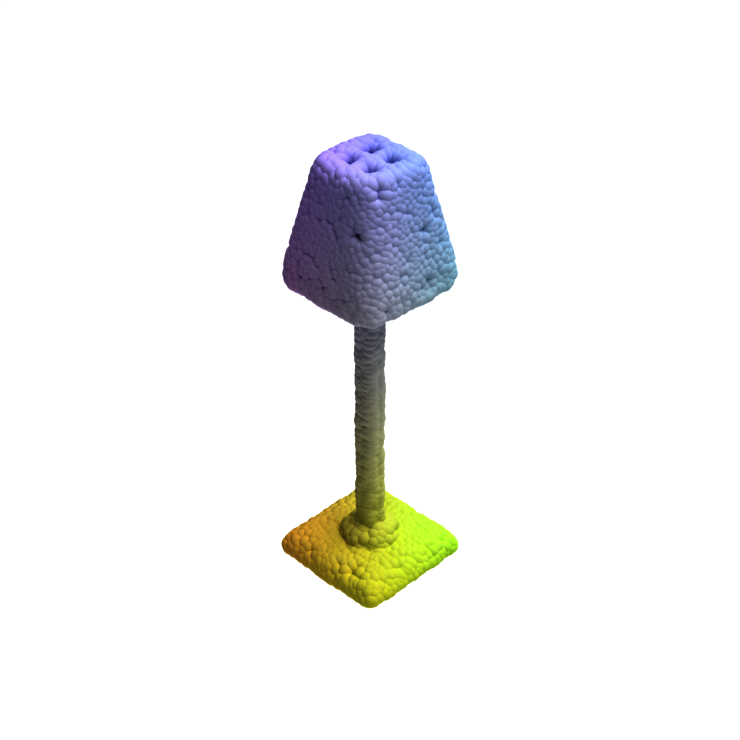}}{}%
\hspace{-0.7cm}
\jsubfig{\includegraphics[height=2.7cm, trim={3cm 3cm 3cm 3cm}, clip]{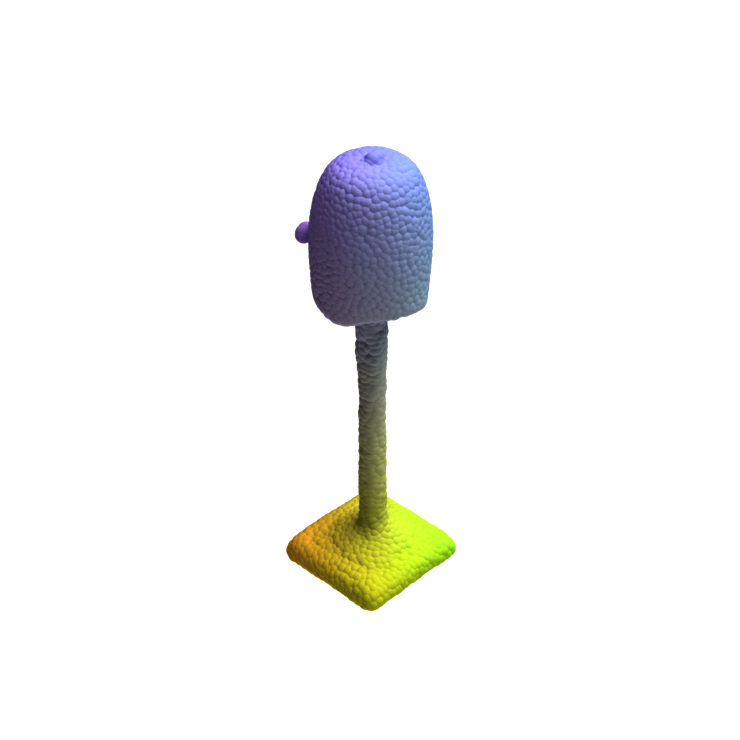}}{}}%

\hspace{0.5cm}
\rotatebox{90}{%
  \parbox{3.0cm}{%
    \centering
    \emph{Longer crown}%
  }%
}
\jsubfig{\jsubfig{\includegraphics[height=2.7cm, trim={2cm 2cm 2cm 2cm}, clip]{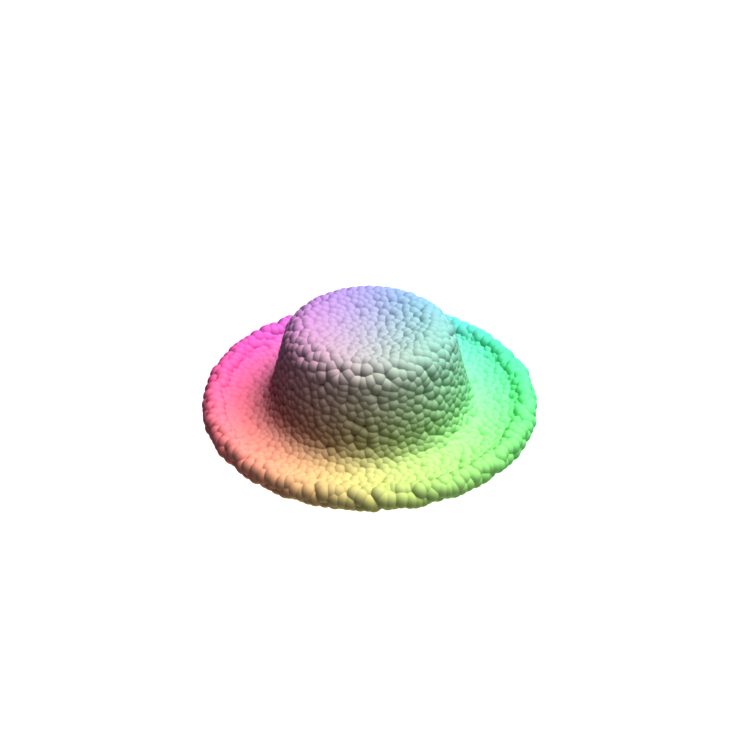}}{}%
\hspace{-0.7cm}
\jsubfig{\includegraphics[height=2.7cm, trim={2cm 2cm 2cm 2cm}, clip]{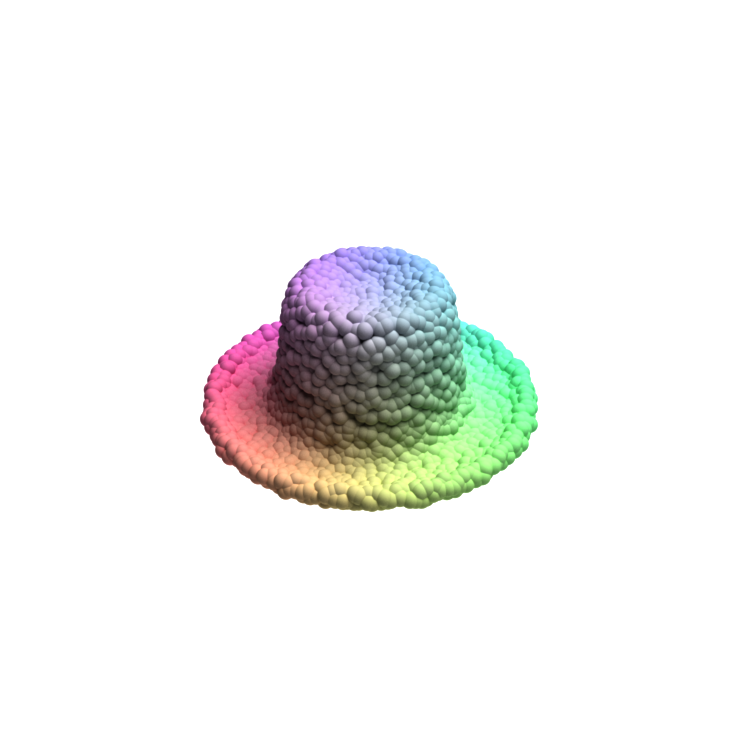}}{}}

\\

\vspace{0.1cm} 
\rotatebox{90}{%
  \parbox{3.0cm}{%
    \centering
    \emph{The head is straight}%
  }%
}
\jsubfig{\jsubfig{\includegraphics[height=2.7cm, trim={3cm 3cm 3cm 3cm}, clip]{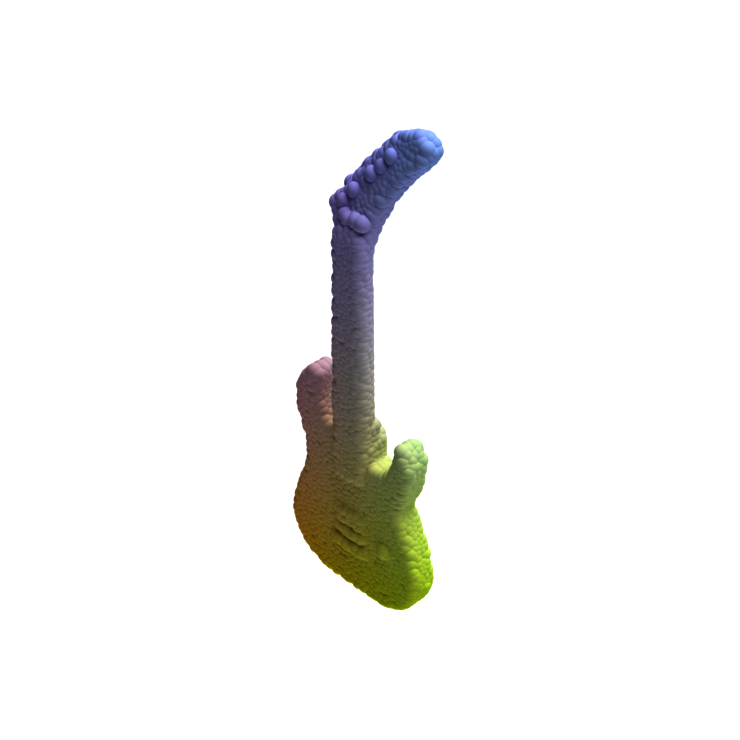}}{}%
\hspace{-0.7cm}
\jsubfig{\includegraphics[height=2.7cm, trim={3cm 3cm 3cm 3cm}, clip]{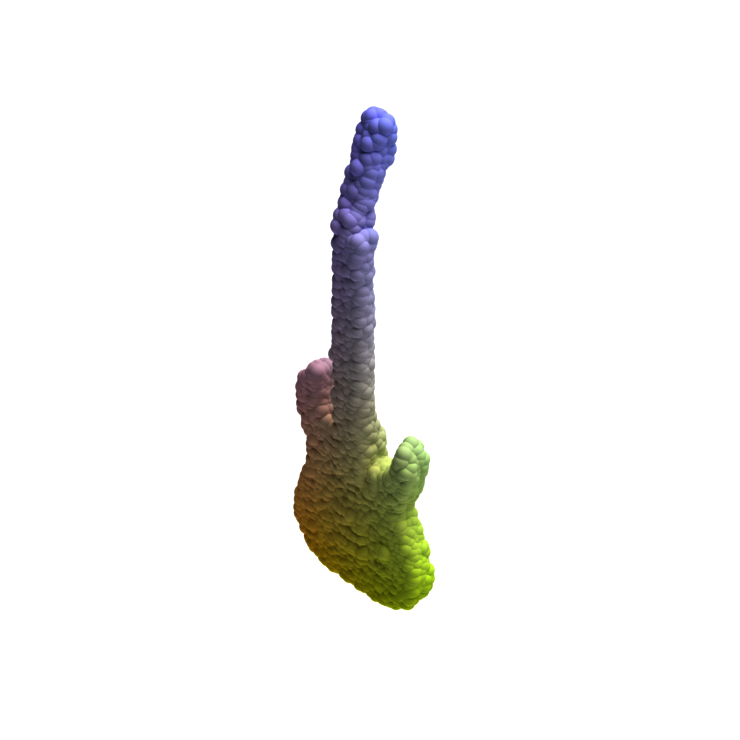}}{}}%

\hspace{0.5cm}
\rotatebox{90}{%
  \parbox{3.0cm}{%
    \centering
    \emph{Legs are tinner}%
  }%
}
\jsubfig{\jsubfig{\includegraphics[height=2.7cm, trim={0cm 0cm 0cm 0cm}, clip]{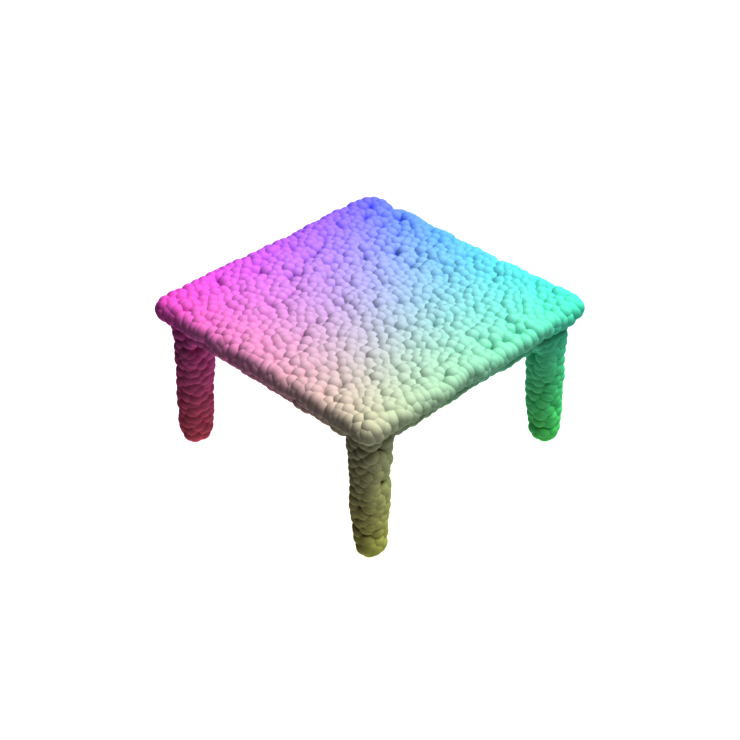}}{}%
\hspace{-0.7cm}
\jsubfig{\includegraphics[height=2.7cm, trim={0cm 0cm 0cm 0cm}, clip]{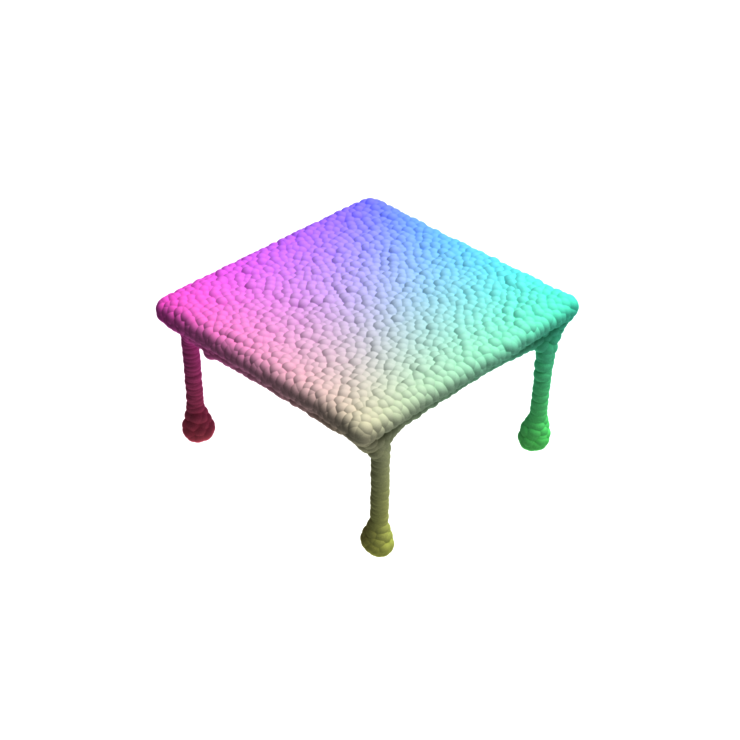}}{}}%

\hspace{0.5cm}
\rotatebox{90}{%
  \parbox{3.0cm}{%
    \centering
    \emph{Armholes on the sides}%
  }%
}
\jsubfig{\jsubfig{\includegraphics[height=2.7cm, trim={3cm 3cm 3cm 3cm}, clip]{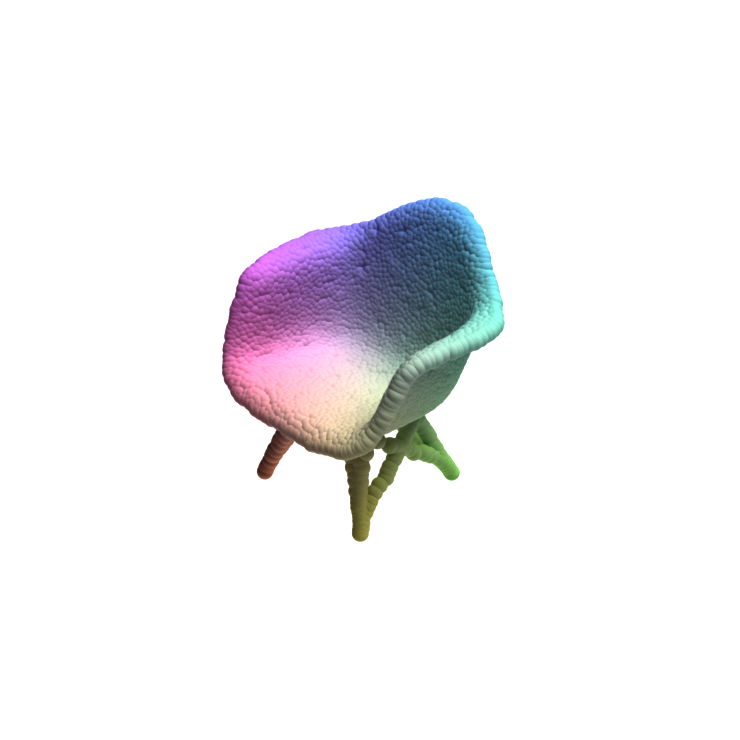}}{}%
\hspace{-0.7cm}
\jsubfig{\includegraphics[height=2.7cm, trim={3cm 3cm 3cm 3cm}, clip]{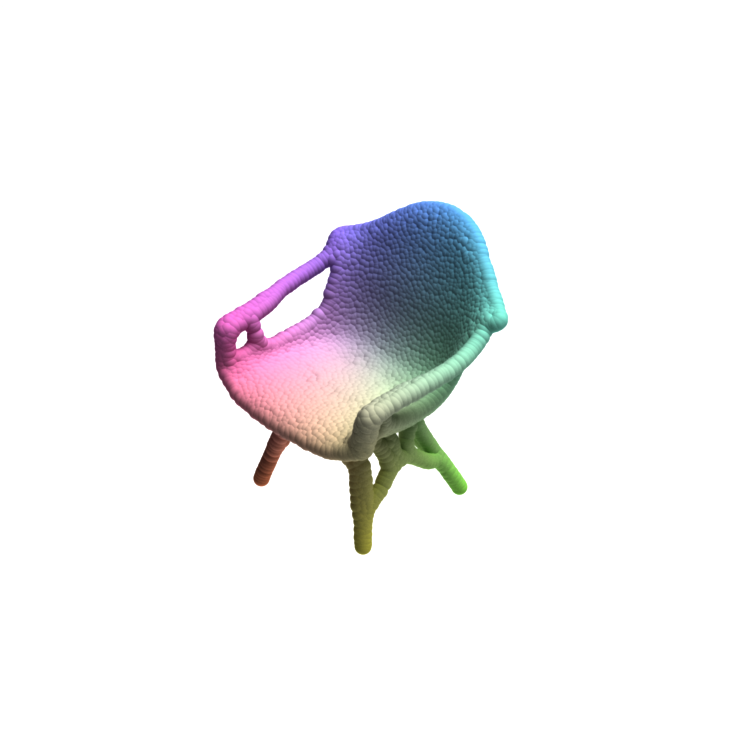}}{}}%

\\

\vspace{0.1cm} 
\rotatebox{90}{%
  \parbox{3.0cm}{%
    \centering
    \emph{Has an H stretcher}%
  }%
}
\jsubfig{\jsubfig{\includegraphics[height=2.7cm, trim={3cm 3cm 3cm 3cm}, clip]{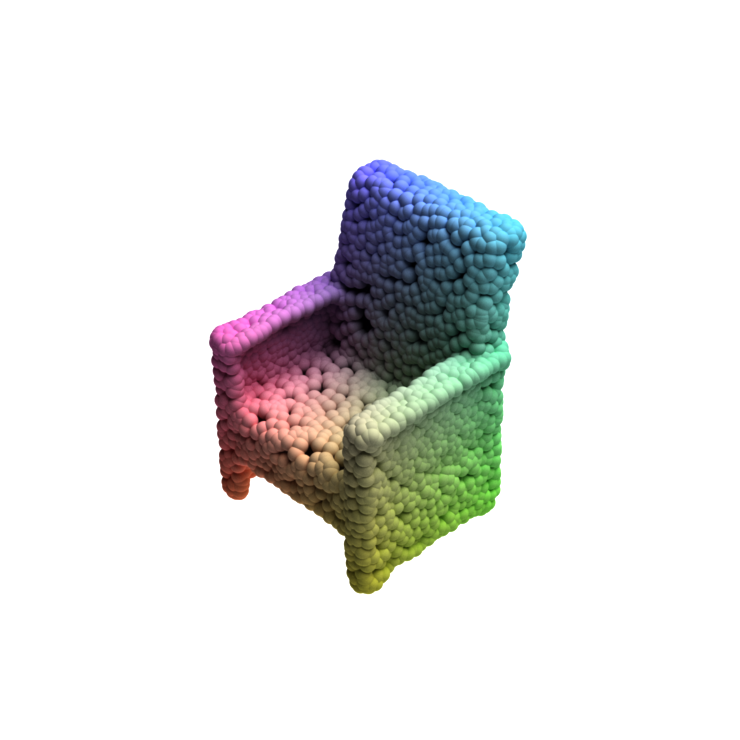}}{\large \emph{Input}}%
\hspace{-0.7cm}
\jsubfig{\includegraphics[height=2.7cm, trim={3cm 3cm 3cm 3cm}, clip]{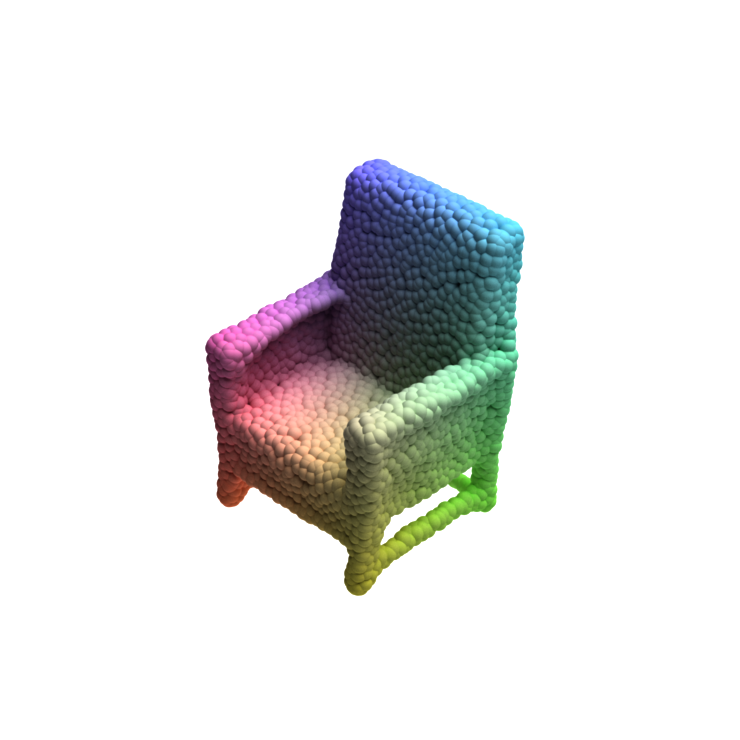}}{\large \emph{Output}}}%

\hspace{0.5cm}
\rotatebox{90}{%
  \parbox{3.0cm}{%
    \centering
    \emph{Rounded brim}%
  }%
}
\jsubfig{\jsubfig{\includegraphics[height=2.7cm, trim={2cm 2cm 2cm 2cm}, clip]{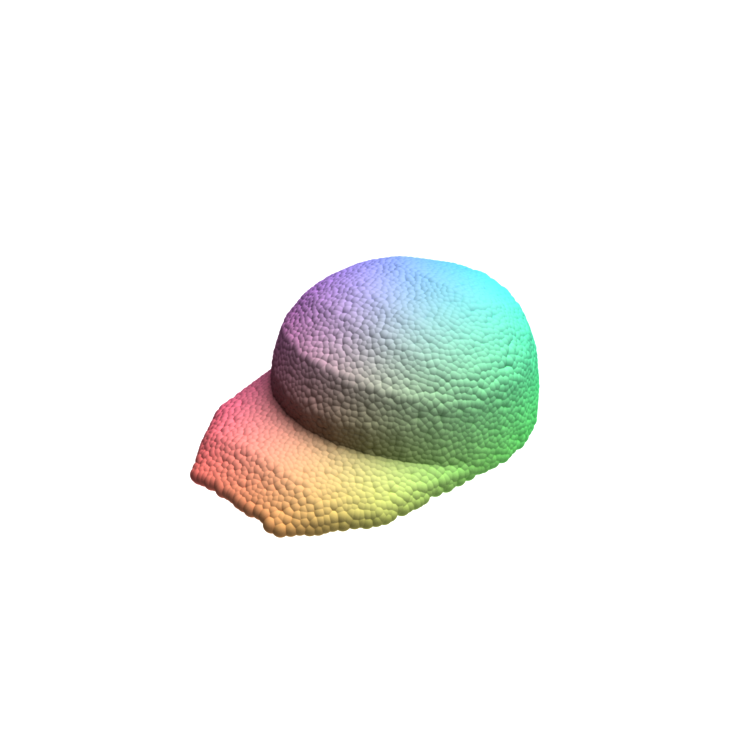}}{\large \emph{Input}}%
\hspace{-0.7cm}
\jsubfig{\includegraphics[height=2.7cm, trim={2cm 2cm 2cm 2cm}, clip]{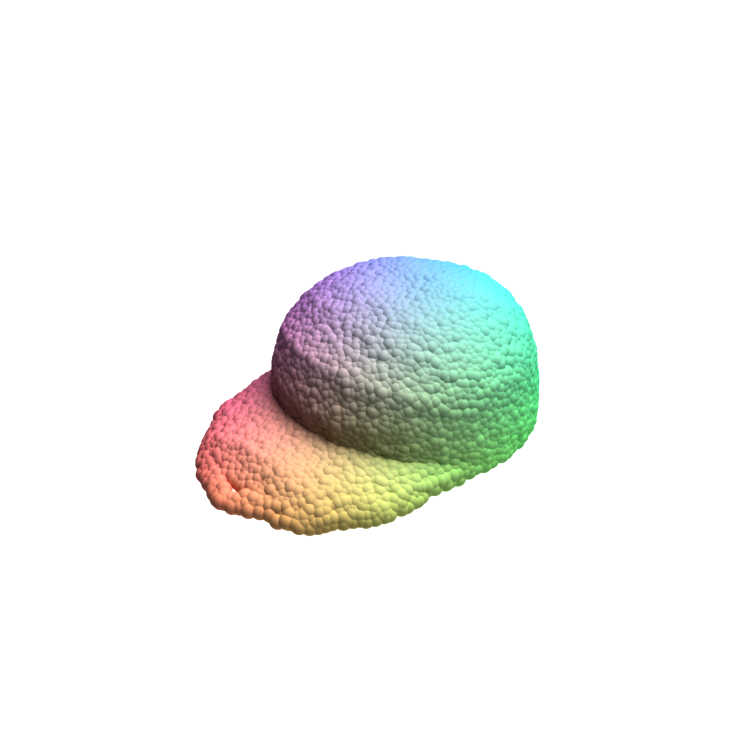}}{\large \emph{Output}}}

\hspace{0.5cm}
\rotatebox{90}{%
  \parbox{3.0cm}{%
    \centering
    \emph{The deck is straight}%
  }%
}
\jsubfig{\jsubfig{\includegraphics[height=2.7cm, trim={2cm 2cm 2cm 2cm}, clip]{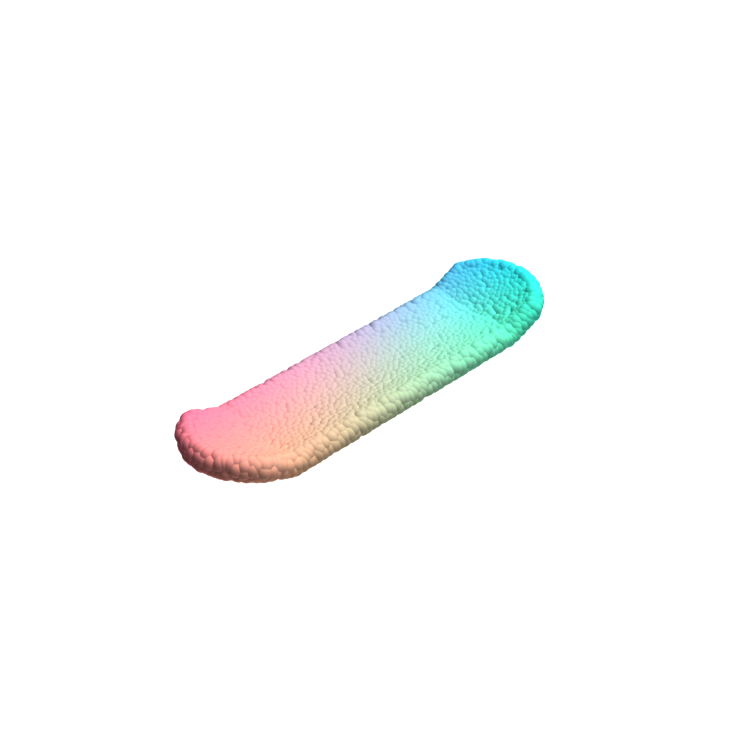}}{\large \emph{Input}}%
\hspace{-0.7cm}
\jsubfig{\includegraphics[height=2.7cm, trim={2cm 2cm 2cm 2cm}, clip]{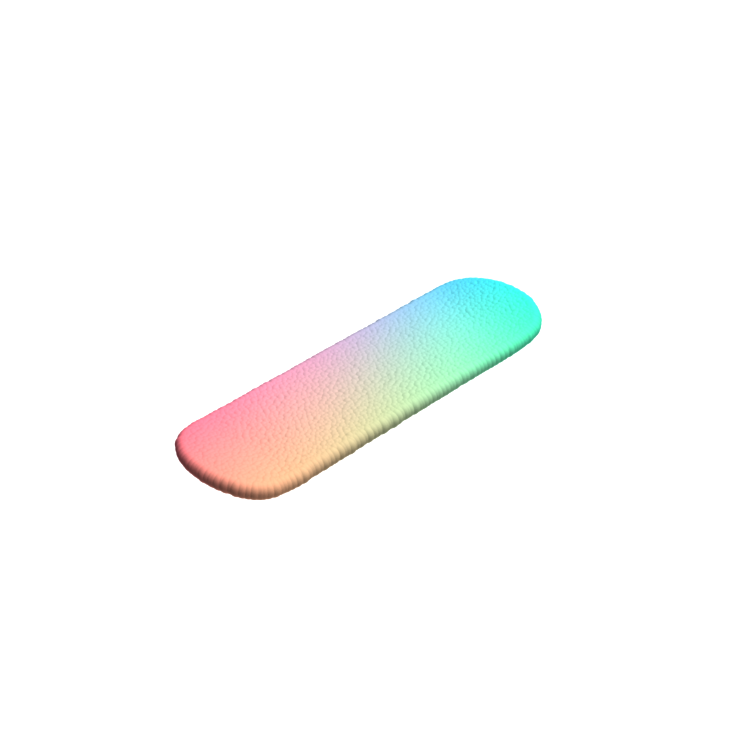}}{\large \emph{Output}}}%

\vspace{-1pt} 
\caption{\textbf{Results Gallery.} Above we show results over various object categories including \emph{chair}, \emph{table}, \emph{lamp}, \emph{airplane}, \emph{cap}, \emph{guitar}, \emph{skateboard} and \emph{knife}.}
\label{fig:results_gallery}
\end{figure*}

\subsection{Extended Qualitative Results}
In Figure \ref{fig:results_gallery} we present qualitative results from multiple shapeNet categories (\textit{Guitar}, \textit{Airplane}, \textit{Chair}, \textit{Lamp}, \textit{Sword}, \textit{Hat}, \textit{Skateboard} and \textit{Table}), demonstrating our method's ability to make meaningful fine grained edits across a wide variety of shape types.

\begin{table}[t]
\centering
\setlength{\tabcolsep}{2.7pt}
\def\arraystretch{0.7}
\begin{tabularx}{\linewidth}{lccccccc}
\toprule
Metric & \footnotesize{$\text{CLIP}_{Sim}$}\small{$\uparrow$} & \footnotesize{$\text{CLIP}_{Dir}$}\small{$\uparrow$} & \small{GD$\downarrow$} & \small{CD$\downarrow$} & \small{FPD$\downarrow$} & \small{l-GD$\downarrow$} \\ \midrule
$\text{RP}_{r=1,j=1}$ & 0.22&	-0.02&	0.57&	0.12&	69.11&	0.72 \\
$\text{RP}_{r=10,j=1}$ & 0.19&	-0.02&	0.62&	0.15&	79.33&	\textbf{0.55} \\
$\text{RP}_{r=1,j=10}$ & 0.19&	-0.03&	0.63&	0.14&	63.48&	0.71 \\
$\text{RP}_{r=10,j=10}$ & 0.21&	-0.03&	0.72 &	0.13&	64.17&	0.57 \\
Ours & \textbf{0.27}	&\textbf{0.01}&	\textbf{0.29}&	\textbf{0.04}	&\textbf{13.51}	&\textbf{0.55} \\
\bottomrule
\end{tabularx}
\vspace{-8pt}
\caption{\textbf{Quantitative Evaluation} against RePaint (RP) with different resampling (r) and jumping (j) values. Note that this implementation uses our InPaint-E model and our reconstructed noise on the non-edit regions as directly using Point-E and random noise resulted in highly noisy (and uninformative) outputs.
}
\vspace{-15pt}
\label{tab:repaint}
\end{table}

\begin{table*}[t]
\centering
\setlength{\tabcolsep}{3.9pt}
\def\arraystretch{1.0}
\begin{tabularx}{\textwidth}{lcccccccccccccc}
\toprule
 &           & \multicolumn{6}{c}{Shapetalk} &  & \multicolumn{6}{c}{l-Shapetalk} \\ 
\cmidrule(lr){3-8} \cmidrule(lr){10-15}
Metric && $\text{CLIP}_{Sim}\uparrow$ & $\text{CLIP}_{Dir}\downarrow$ &
GD$\downarrow$ & CD$\downarrow$ & FPD$\downarrow$ & l-GD$\downarrow$ &&
$\text{CLIP}_{Sim}\uparrow$ & $\text{CLIP}_{Dir}\downarrow$ &
GD$\downarrow$ & CD$\downarrow$ & FPD$\downarrow$ & l-GD$\downarrow$ \\ \midrule

0\%  && 0.24 & 1.01 & 1.00 & 0.11 & 49.12 & 0.48 &&
        0.24 & 1.01 & 1.11 & 0.14 & 69.01 & 0.51 \\

5\%  && \textbf{0.26} & 1.00 & 0.39 & 0.07 & \textbf{31.12} & 0.08 &&
        0.26 & \textbf{0.99} & 0.32 & 0.05 & \textbf{12.19} & 0.06 \\

10\% && \textbf{0.26} & \textbf{0.99} & \textbf{0.34} & \textbf{0.05} &
        33.64 & \textbf{0.07} &&
        \textbf{0.27} & \textbf{0.99} & \textbf{0.29} & \textbf{0.04} &
        13.51 & \textbf{0.05} \\

25\% && 0.25 & 1.00 & \textbf{0.34} & \textbf{0.05} & 36.16 & 0.08 &&
        0.25 & 1.00 & 0.30 & \textbf{0.04} & 17.31 & 0.06 \\

50\% && 0.22 & 1.04 & 1.01 & 0.16 & 60.75 & 0.42 &&
        0.23 & 1.03 & 0.95 & 0.18 & 75.38 & 0.45 \\

\bottomrule
\end{tabularx}
\vspace{-8pt}
\caption{\textbf{Ablating reconstruction percentage during training.} 
 We present quantitative evaluation results for different \inpaintingModel{} models, each trained with a different proportion of reconstruction samples (\emph{``Recon $\%$''}) during training.}
\label{tab:rec_steps_ablations}
\end{table*}

\subsection{Comparison to RePaint}
The free form inpainting method RePaint \cite{lugmayr2022repaint} has had a major impact on the field of diffusion based image inpainting. This work introduced an inference time algorithm which, somewhat similarly to our coordinate blending algorithm, blends noisy versions of the ``known'' regions of the image with the predicted denoised versions of the inpainted region according to an input binary mask. RePaint also proposes to resample noise and repeat the process for a given number of iterations to further refine this inpainting process. By contrast, in addition to other core differences our work dedicates a significant portion of the inference process to reconstructing the full input shape before starting the inpainting process, as well as operating on a specific model tailored for this task instead of a more general text-to-3D model. To test the significance of some of these design choices we conducted a quantitative comparison against a baseline which resembles RePaint in its function. Specifically, in this baseline inpainting is performed at every inference step ($t_r = T$) and the edit region is initialized with random noise. This baseline also incorporates the ``resampling'' and ``jumping'' mechanisms introduced in RePaint. Unlike RePaint however, we used our reconstructed noise for the non-edit region and operated on \inpaintingModel{}, as directly using Point-E and random noise resulted in highly noisy (and uninformative) outputs. 

The results of this comparison are presented in Table \ref{tab:repaint} and clearly shows that our method outperforms this baseline across all metrics.

\begin{figure} 
\rotatebox{90}{\hspace{0.8cm}\footnotesize{\emph{Legs}}}
\jsubfig{
\jsubfig{\jsubfig{\includegraphics[height=2cm, trim={2.0cm 2.0cm 2.0cm 2.0cm}, clip]{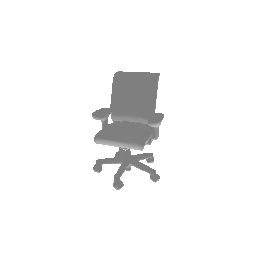}}{}
\jsubfig{\includegraphics[height=2cm, trim={2.0cm 2.0cm 2.0cm 2.0cm}, clip]{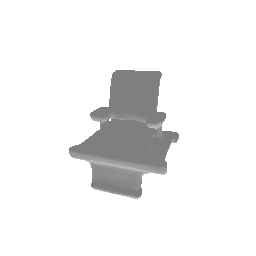}}{}}{} 

\jsubfig{\jsubfig{\includegraphics[height=2.125cm, trim={4.0cm 4.0cm 4.0cm 4.0cm}, clip]{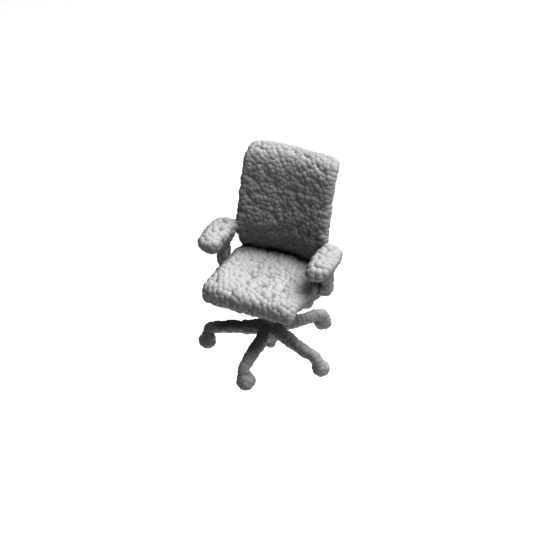}}{}
\jsubfig{\includegraphics[height=2.125cm, trim={4.0cm 4.0cm 4.0cm 4.0cm}, clip]{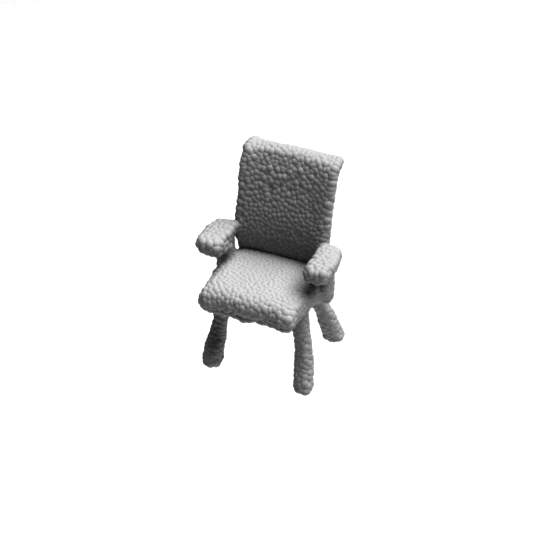}}{}}{}}{}

\rotatebox{90}{\hspace{0.7cm}\footnotesize{\emph{Back}}}
\jsubfig{

\jsubfig{\jsubfig{\includegraphics[height=2cm, trim={2.0cm 2.0cm 2.0cm 2.0cm}, clip]{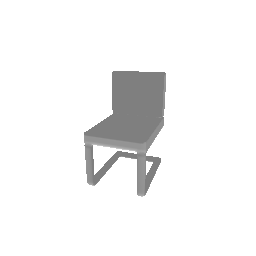}}{}
\jsubfig{\includegraphics[height=2cm, trim={2.0cm 2.0cm 2.0cm 2.0cm}, clip]{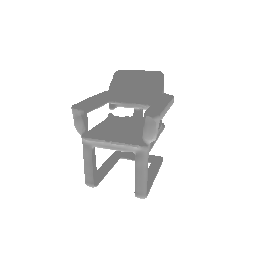}}{}}{} 

\jsubfig{\jsubfig{\includegraphics[height=2.125cm, trim={4.0cm 4.0cm 4.0cm 4.0cm}, clip]{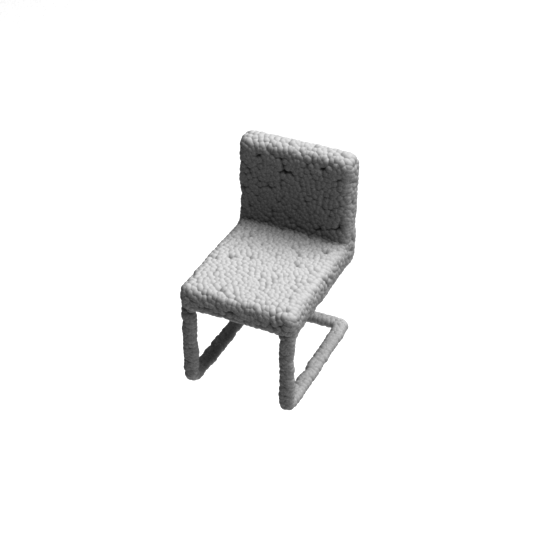}}{}
\jsubfig{\includegraphics[height=2.125cm, trim={4.0cm 4.0cm 4.0cm 4.0cm}, clip]{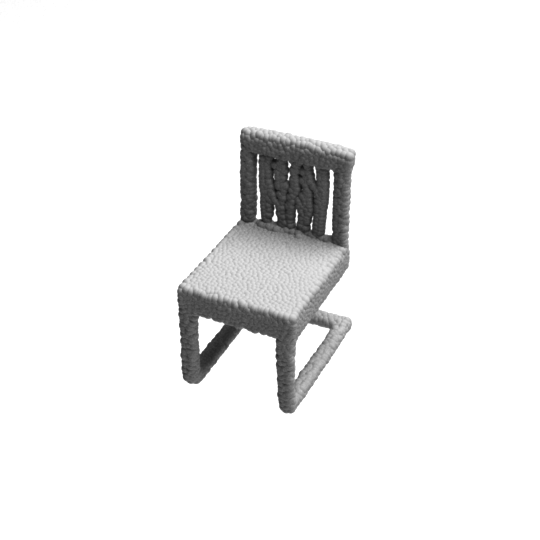}}{}}{}}{}

\rotatebox{90}{\hspace{0.3cm}\footnotesize{\emph{Legs}}}
\jsubfig{
\jsubfig{\jsubfig{\includegraphics[height=2cm, trim={2.0cm 2.0cm 2.0cm 2.0cm}, clip]{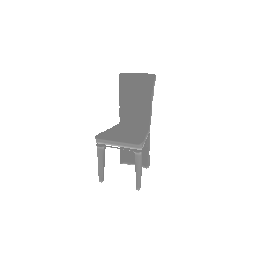}}{\footnotesize {Input \hspace{1.2cm} (3D Mesh)}}
\jsubfig{\includegraphics[height=2cm, trim={2.0cm 2.0cm 2.0cm 2.0cm}, clip]{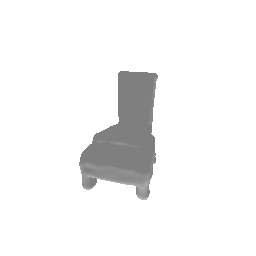}}{\footnotesize {Output}}}{\vspace{5pt}SDFusion} 

\jsubfig{\jsubfig{\includegraphics[height=2.125cm, trim={4.0cm 4.0cm 4.0cm 4.0cm}, clip]{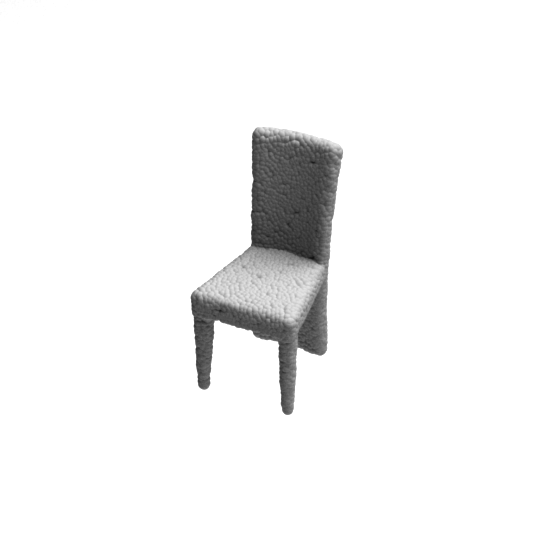}}{\footnotesize {Input \hspace{0.9cm} (Point Cloud)}}
\jsubfig{\includegraphics[height=2.125cm, trim={4.0cm 4.0cm 4.0cm 4.0cm}, clip]{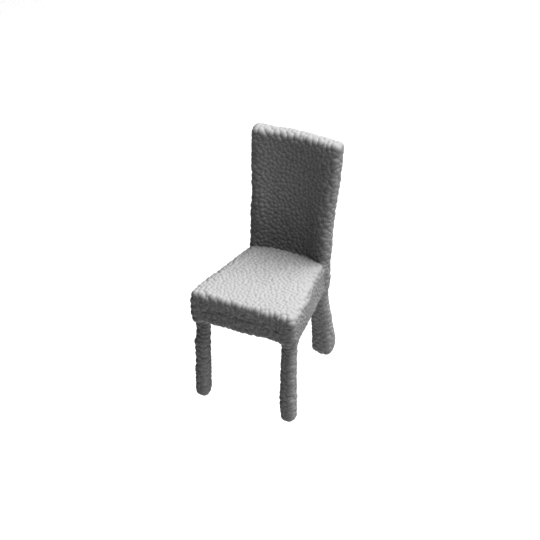}}{\footnotesize {Output}}}{\vspace{5pt}Ours}}{}

\vspace{-1pt} 
\caption{
\textbf{Qualitative comparison against SDFusion \cite{cheng2023sdfusion} unconditional part completion.} We compare our method's outputs against the unconditional part completion results of the SDFusion \cite{cheng2023sdfusion} baseline. In each row we task the methods with completing (in SDFusion's case) or editing\setcounter{footnote}{0}\protect\footnotemark{} (in our case) a different part. As these results show, SDFusion's ability to preserve identity across all regions of the shape is limited, in comparison to our approach that performs localized fine-grained editing of 3D shapes.}
\label{fig:sdfusion}
\end{figure}

\subsection{Shape Completion Comparison}
Shape, or part, completion is a longstanding task involving completing a shape that is missing one or more parts in a way that maintains plausibility and optionally aligns with a text prompt. In Figure \ref{fig:sdfusion}, we illustrate that these methods are not well suited for fine-grained shape editing as they are inherently blind to the missing parts, focusing on a comparison with SDFusion~\cite{cheng2023sdfusion}. Note that we demonstrate this in a slightly different setting in comparison to the one addressed in our work---\emph{unconditional} generation, and not \emph{text-guided} part completion---as their official codebase does not include text-guided part completion.

Nonetheless, we performed a qualitative comparison in which we compared our results (text-guided) against unconditional part completion results of SDFusion. These results show that this method's ability to maintain identity is somewhat limited, even outside of the edit region (thicker legs on the chair in the second row, as well as adding arms to it). As SDFusion's part completion in this case is not guided by text it is somewhat hard to judge its quality. However, it is evident that the completed parts often don't match the general identity of the original shape particularly well (office chair type backrest in row 2, huge elaborate legs in row 1) compared to our method.

\footnotetext{The text guidance provided for our method (from top to bottom): \emph{four legs}, \emph{spindle backrest}, \emph{four legs}.}

\subsection{Ablating  the proportion of reconstruction training samples} As mentioned in Section \ref{sec:inference_time}, to support our inference time algorithm, we occasionally replace during training the masked point cloud $x_{M}$ and the text condition $\mathcal{C}$ with the full point cloud $x$ and an empty prompt $\mathcal{C}_{0} = ``\,"$. In essence, this replaces the inpainting objective for that specific iteration with full shape reconstruction. This strengthens the model's ability to reconstruct full shapes, which supports our inversion free algorithm. In our finalized pipeline we replace the inpainting objective with reconstruction in $10\%$ of all training iterations.

To evaluate the effect of this design decision we conducted an experiment in which we trained several \inpaintingModel{} models, each trained with a different \emph{``Recon $\%$''} values and evaluated their performance. The quantitative results of this experiment are presented in table \ref{tab:rec_steps_ablations}. These results show that when we don't train for reconstruction ($0$ \emph{``Recon $\%$''}) our identity preservation capabilities are significantly hampered, yielding high GD and l-GD scores. Conversely, high \emph{``Recon $\%$''} values hamper the model's inpainting capabilities, yielding low $\text{CLIP}_{Sim}$ and high $\text{CLIP}_{Dir}$ values.

\begin{figure} 
\rotatebox{90}{%
  \parbox{2.0cm}{%
    \centering
    \emph{Thick arms}%
  }%
}
\jsubfig{\jsubfig{\includegraphics[height=2.1cm, trim={2cm 2cm 2cm 2cm}, clip]{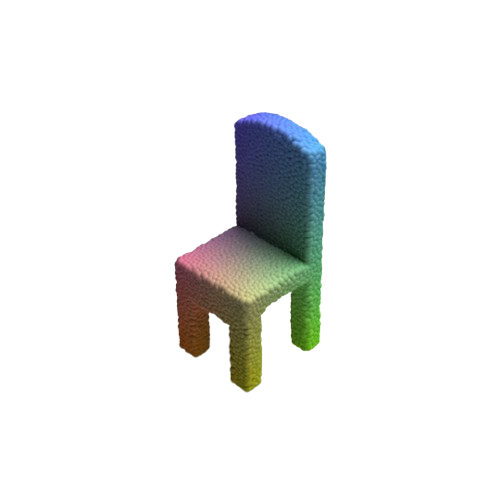}}{\large \emph{Input}}%
\hspace{-0.7cm}
\jsubfig{\includegraphics[height=2.1cm, trim={2cm 2cm 2cm 2cm}, clip]{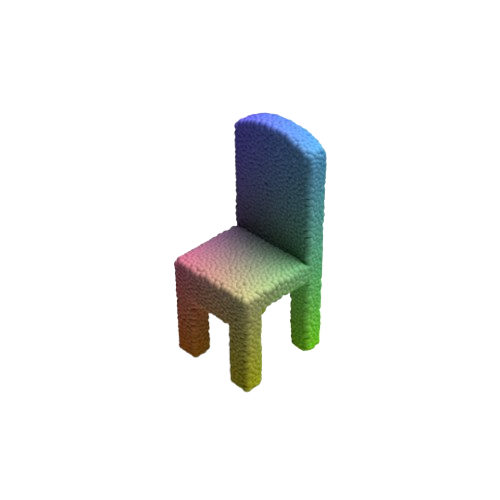}}{\large \emph{Output}}}

\hspace{0.0cm}
\rotatebox{90}{%
  \parbox{2.0cm}{%
    \centering
    \emph{A royal chair}%
  }%
}
\jsubfig{\jsubfig{\includegraphics[height=2.1cm, trim={2cm 2cm 2cm 2cm}, clip]{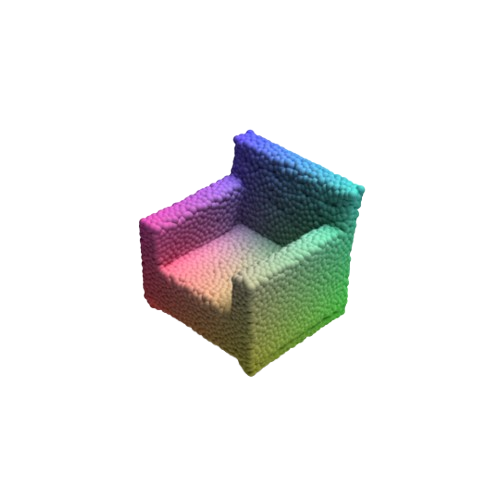}}{\large \emph{Input}}%
\hspace{-0.7cm}
\jsubfig{\includegraphics[height=2.1cm, trim={2cm 2cm 2cm 2cm}, clip]{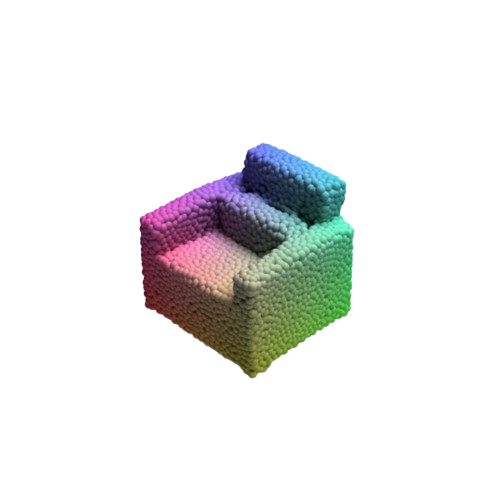}}{\large \emph{Output}}}%

\vspace{-1pt} 
\caption{\textbf{Limitations.} As illustrated above, our approach cannot add new parts (such as arms on the left) or perform global edits that involve multiple parts (such as turning the chair into a royal chair on the right).  }
\label{fig:limitations}
\end{figure}

\section{Limitations}
\label{sec:limitations}
While our method performs well in most scenarios, it has certain limitations; see Figure \ref{fig:limitations}. As illustrated in the figure, our inpainting-based approach restricts the ability to generate entirely new objects or parts. BlendedPC is specifically designed for localized editing and struggles with global shape transformations. Furthermore, as a supervised learning approach, the method's generalizability is constrained by the object categories in the training dataset. This limits performance when editing shape categories not encountered during training. Furthermore, as our method uses off the shelf point cloud segmentation methods, it is currently unable to segment sub-parts or part instances, an issue which could be solved with more powerful segmentation methods. On a whole, these limitations present exciting opportunities for future research, such as developing techniques for generating object parts, enabling global shape modifications, enabling instance level editing and improving cross-category generalization.

\end{document}